\documentclass[12pt]{article}
\usepackage{amsmath,amssymb,amsthm,amsxtra,overpic,bbm,bm,epsfig,subfigure}
\usepackage{hyperref}
\usepackage{mathrsfs}
\usepackage{enumitem}
\usepackage{graphicx}
\usepackage{color}
\usepackage[table]{xcolor}
\usepackage{comment}
\usepackage{epstopdf}
\usepackage{float}
\usepackage{cite}
\usepackage{slashed,stmaryrd}

\usepackage[title]{appendix}

\allowdisplaybreaks[1]

\def\thefootnote{\fnsymbol{footnote}}

\usepackage{multirow}
\newcommand{\Dlr}{\mbox{$\raisebox{2mm}{\boldmath ${}^\leftrightarrow$}\hspace{-4mm}D^{}_\mu$}}
\newcommand{\Dilr}{\mbox{$\raisebox{2mm}{\boldmath ${}^\leftrightarrow$}\hspace{-4mm}D^I_\mu$}}
\newcommand{\Dilrs}{\mbox{$\raisebox{2mm}{\boldmath ${}^\leftrightarrow$}\hspace{-4mm}\slashed{D}^I$}}
\newcommand{\Dlrs}{\mbox{$\raisebox{2mm}{\boldmath ${}^\leftrightarrow$}\hspace{-4mm}\slashed{D}$}}

\newcommand{\Dlrn}{\mbox{$\raisebox{2mm}{\boldmath ${}^\leftrightarrow$}\hspace{-4mm}D^\nu$}}

\newcommand{\Yn}{Y^{}_\nu}
\newcommand{\DYn}{Y^\dagger_\nu}
\newcommand{\SYn}{Y^\ast_\nu}
\newcommand{\TYn}{Y^{\rm T}_\nu}
\newcommand{\Yl}{Y^{}_l}
\newcommand{\DYl}{Y^\dagger_l}
\newcommand{\Yu}{Y^{}_{\rm u}}

\newcommand{\DYd}{Y^\dagger_{\rm d}}
\newcommand{\Ydel}{Y^{}_\Delta}
\newcommand{\DYdel}{Y^\dagger_\Delta}

\newcommand{\DDYni}[2][i]{\left(Y^\dagger_\nu Y^{}_\nu\right)^{}_{#1 #2}} 
\newcommand{\Yni}[2][\alpha]{\left(Y^{}_\nu\right)^{}_{#1 #2}} 
\newcommand{\DYni}[2][\beta]{\left(Y^\dagger_\nu \right)^{}_{#2 #1}} 
\newcommand{\TYni}[2][\beta]{\left(Y^{\rm T}_\nu \right)^{}_{#2 #1}}
\newcommand{\Yli}[2][\alpha]{\left(Y^{}_l\right)^{}_{#1 #2}}

\newcommand{\Yui}[2][\alpha]{\left(Y^{}_{\rm u}\right)^{}_{#1 #2}}

\newcommand{\Ydi}[2][\alpha]{\left(Y^{}_{\rm d}\right)^{}_{#1 #2}}

\newcommand{\Lelli}[1][\beta]{\ell^{}_{#1 \rm L}}
\newcommand{\BLelli}[1][\alpha]{\overline{\ell^{}_{#1 \rm L}}}
\newcommand{\REi}[1][\beta]{E^{}_{#1 \rm R}}
\newcommand{\BREi}[1][\alpha]{\overline{E^{}_{#1 \rm R}}}
\newcommand{\LQi}[1][\beta]{Q^{}_{#1 \rm L}}
\newcommand{\BLQi}[1][\alpha]{\overline{Q^{}_{#1 \rm L}}}
\newcommand{\RUi}[1][\beta]{U^{}_{#1 \rm R}}
\newcommand{\BRUi}[1][\alpha]{\overline{U^{}_{#1 \rm R}}}
\newcommand{\RDi}[1][\beta]{D^{}_{#1 \rm R}}
\newcommand{\BRDi}[1][\alpha]{\overline{D^{}_{#1 \rm R}}}

\newcommand{\rmI}{{\rm i}}
\newcommand{\lnmd}{ L^{}_\Delta }
\newcommand{\lnmi}{ L^{}_i }
\newcommand{\lnmj}{ L^{}_j }
\newcommand{\lnij}{ L^{}_{ij} }

\newcommand{\pM}[2][i]{M^{#2}_#1}
\newcommand{\Op}{\mathcal{O}}
\newcommand{\Opr}{\mathcal{R}}
\newcommand{\Mds}[1][2]{M^{#1}_\Delta}

\newcommand{\lint}{\int \frac{{\rm d}^d k}{\left( 2\pi \right)^d}}
\newcommand{\del}[1]{\delta^{}_{#1}}
\newcommand{\lamd}[1][]{\lambda^{#1}_\Delta}
\newcommand{\momp}[2][]{p^{#1}_{#2}}
\newcommand{\momps}[2][]{\slashed{p}^{#1}_{#2}}
\newcommand{\amp}[1]{\rmI \mathcal{M}^{\rm ct}_{#1} }
\newcommand{\ampd}[1]{\left. \rmI \mathcal{M}^{\rm ct}_{#1} \right|^{}_{\rm hard}}

\newcommand{\pl}{P^{}_{\rm L}}
\newcommand{\pr}{P^{}_{\rm R}}

\usepackage{ulem}

\begin{document}

\begin{center}
{\Large\bf Complete One-loop Structure of the Type-(I+II) Seesaw Effective Field Theory}
\end{center}

\vspace{0.2cm}

\begin{center}
{\bf Di Zhang~$^{a,~b,~c}$}~\footnote{E-mail: zhangdi@ihep.ac.cn}
\\
\vspace{0.2cm}
{\small
$^a$Institute of High Energy Physics, Chinese Academy of Sciences, Beijing 100049, China\\
\small$^b$School of Physical Sciences, University of Chinese Academy of Sciences, Beijing 100049, China}\\
\small$^c$Physik-Department, Technische Universität München, James-Franck-Straße, 85748 Garching, Germany
\end{center}

\vspace{1.5cm}

\begin{abstract}
Besides the three canonical seesaw mechanisms, the hybrid scenario, i.e., the so-called type-(I+II) seesaw mechanism containing both the right-handed neutrinos $N^{}_{\rm R}$ and the triplet Higgs $\Phi$ is also an appealing extension of the Standard Model (SM) to account for tiny neutrino masses. Recently, the seesaw effective field theories (SEFTs) of the three canonical seesaw mechanisms have already been completely constructed up to one-loop level. In this work, we carry out the one-loop matching of the type-(I+II) seesaw mechanism onto the corresponding type-(I+II) SEFT, which is by no means the trivial combination of the type-I and type-II SEFTs and contains additional contributions even though the right-handed neutrinos and the triplet Higgs have no direct interactions. Employing the Feynman diagrammatic approach, we calculate all those additional contributions from the entangled effects of $N^{}_{\rm R}$ and $\Phi$, and finally achieve the complete one-loop structure of the type-(I+II) SEFT. In the type-(I+II) SEFT,  the number and content of dim-6 operators are exactly the same as those in the type-II SEFT, but the Wilson coefficients of the unique dim-5 and nine dim-6 operators as well as the quartic coupling constant of the SM Higgs gain some additional contributions, which are absent in the type-I and type-II SEFTs.
\end{abstract}


\def\thefootnote{\arabic{footnote}}
\setcounter{footnote}{0}
\newpage

\section{Introduction}

Indisputable evidences from the atmospheric, solar, reactor and accelerator neutrino (or antineutrino) oscillation experiments demonstrate that at least two of three neutrinos have nonzero but tiny masses and lepton flavors are significantly mixed \cite{Workman:2022ynf}. However, neutrino masses and the lepton flavor mixing can not be accommodated in the SM, which infers that the SM is not complete and only serves as an effective field theory (EFT) under a certain energy scale \cite{Xing:2020ijf}. To account for neutrino masses, one needs to extend the SM with some new ingredients. Among various extensions, seesaw mechanisms \cite{Minkowski:1977sc,Yanagida:1979as,Gell-Mann:1979vob,Glashow:1979nm,Mohapatra:1979ia,Konetschny:1977bn,Magg:1980ut,Schechter:1980gr,Cheng:1980qt,Mohapatra:1980yp,Lazarides:1980nt,Foot:1988aq} are the most natural and popular ones. Their main spirit is that active neutrino masses are extremely suppressed by masses of the introduced heavy degrees of freedom, such as right-handed neutrinos \cite{Minkowski:1977sc,Yanagida:1979as,Gell-Mann:1979vob,Glashow:1979nm,Mohapatra:1979ia}, the triplet Higgs \cite{Konetschny:1977bn,Magg:1980ut,Schechter:1980gr,Cheng:1980qt,Mohapatra:1980yp,Lazarides:1980nt} and triplet fermions \cite{Foot:1988aq} in the type-I, type-II and type-III seesaw mechanisms, respectively. Therefore, seesaw mechanisms can naturally explain the smallness of neutrino masses. Moreover, the other bonus of seesaw mechanisms is that they may provide an elegant explanation for the matter-antimatter asymmetry of the Universe via the so-called leptogenesis \cite{Fukugita:1986hr}~\footnote{The type-II seesaw mechanism with one triplet Higgs can not lead to successful leptogenesis that actually requires at least two triplet scalars in this framework \cite{Ma:1998dx}.}. Besides the three canonical seesaw mechanisms, there is an appealing hybrid scenario, namely the type-(I+II) seesaw mechanism. This hybrid scenario contains both right-handed neutrinos and the triplet Higgs, and can be naturally embedded into more fundamental frameworks, e.g., the $\rm SO(10)$ grand unified theories \cite{Fritzsch:1974nn,Georgi:1974my}. The type-I and type-II seesaw mechanisms can be regarded as two special cases of the type-(I+II) seesaw mechanism where the active neutrinos acquire their tiny masses via contributions from both right-handed neutrinos and the triplet Higgs. The latter one not only has all properties of the former two, but also contains additional interfering effects of them. For example, the type-(I+II) seesaw mechanism has very rich phenomena induced by the triplet Higgs but with new interfering effects of right-handed neutrinos and the triplet Higgs, which can be searched and probed at colliders if their masses are at the TeV scale. Another advantage is that it can improve the realization of the leptogenesis to account for the matter-antimatter asymmetry of the Universe, where the interfering contributions may play an important role and can reduce the required mass scale of right-handed neutrinos (see e.g., Ref. \cite{Hambye:2012fh} for a review).


Barring unnatural small Yukawa couplings or large cancellations, the mass scale of heavy degrees of freedom in seesaw mechanisms needs to be considerably heavier than the electroweak scale. This means that these heavy particles accounting for tiny neutrino masses can hardly be probed directly at current or near future collider experiments. One way out of this issue is to indirectly explore their low-energy consequences. In this case, one can integrate out those heavy degrees of freedom in seesaw mechanisms and then focus on the obtained seesaw effective field theories where some effective operators of mass dimension higher than four appear, i.e.,
\begin{eqnarray}\label{eq:seft}
\mathcal{L}^{}_{\rm SEFT} = \mathcal{L}^{}_{\rm SM} + \sum^{n^{}_d}_{i=1} C^{(d)}_i \Op^{(d)}_i \;,
\end{eqnarray}
where $\mathcal{L}^{}_{\rm SM}$ denotes the SM Lagrangian and $\Op^{(d)}_i$ are  effective operators consisting of the SM fields and satisfying the SM gauge symmetry and Lorentz invariance. 
In Eq. \eqref{eq:seft}, $d>4$ is the mass dimension of the operator $\Op^{(d)}_i$ (for $i=1,2,\dots,n^{}_d$ with $n^{}_d$ being the number of $d$-dimensional operators), and $C^{(d)}_i$ stands for the corresponding Wilson coefficient of $\Op^{(d)}_i$ and is suppressed by an $\mathcal{O} \left( M^{4-d} \right)$ factor with $M$ being the typical mass scale of heavy particles in seesaw mechanisms. Usually, effective operators $\Op^{(d)}_i$ in the SEFTs are only a subset of those in the SM effective field theory (SMEFT) \cite{Buchmuller:1985jz,Grzadkowski:2010es} (see e.g., Ref. \cite{Brivio:2017vri} for a recent review), which contains all possible independent operators. Moreover, the conceptions of the SMEFT and SEFTs are totally different. The former one is a model-independent framework where all Wilson coefficients of effective operators are unknown and free parameters, whereas the latter one indeed depends on UV models (i.e., seesaw mechanisms here) and all Wilson coefficients are specifically related to a few couplings and masses of heavy fields. In other words, those Wilson coefficients in the SEFTs are highly correlated and carry effects of heavy fields on low-energy observations. 
Therefore, the SEFTs dominated by Eq. \eqref{eq:seft} provide us some consistent frameworks to discuss low-energy consequences of the heavy fields  in seesaw mechanisms. There are two other strong motivations to consider the SEFTs:
\begin{itemize}
\item In a model with large mass hierarchies, the predicted low-energy observables usually involve large logarithmic radiative corrections, which need to be resummed for precision measurements. Such a  resummation of large logarithms can be simply achieved by means of renormalization group equations (RGEs) in the corresponding EFT, as discussed in Ref. \cite{Weinberg:1980wa}. 
\item Though all canonical seesaw mechanisms can naturally explain tiny neutrino masses in a similar way, their corresponding SEFTs  have very different structures including contents of  effective operators and the associated Wilson coefficients \cite{Zhang:2021jdf,Li:2022ipc,Coy:2021hyr,Du:2022vso}.  Then, comparisons between the structures of these SEFTs may provide a way to distinguish these seesaw mechanisms \cite{Li:2022ipc,Coy:2021hyr,Du:2022vso,Li:2022tcz}.
\end{itemize}
Thus, to some extent, the  SEFTs are indispensable to consistently explore the phenomenological consequences of seesaw mechanisms and indirectly probe the underlying heavy particles.

The tree-level Lagrangian for the SEFTs can be easily obtained by integrating out the heavy degrees of freedom at the tree level. For the three canonical seesaw mechanisms, their tree-level EFTs, i.e., the tree-level part of the type-I, type-II and type-III SEFTs, have been achieved a long time ago and are quite different from one another \cite{Broncano:2002rw,Broncano:2003fq,Abada:2007ux,Elgaard-Clausen:2017xkq}. Though the dim-5 Weinberg operator $\Op^{(5)}$ exists in all the three SEFTs and generates neutrino masses after spontaneous gauge symmetry breaking (SSB) \cite{Weinberg:1979sa}, its Wilson coefficient in different SEFTs depends on different couplings and is essentially different. Moreover, two dim-6 operators $\Op^{(1)}_{H\ell}$ and $\Op^{(3)}_{H\ell}$ in the type-I and type-III SEFTs can modify the coupling of neutrinos with weak gauge bosons and then cause the unitarity violation of lepton flavor mixing, while the lepton flavor mixing matrix keeps unitary in the type-II SEFT due to the absence of these two operators \cite{Broncano:2002rw,Broncano:2003fq,Abada:2007ux,Elgaard-Clausen:2017xkq}. However, integrating out the heavy fields in these seesaw mechanisms at the one-loop level is much more complicated and more technical than that at the tree level. Consequently, matchings of the three canonical seesaw mechanisms onto the corresponding SEFTs at the one-loop level have been achieved partially or completely only in very recent times, and so have consistent discussions on low-energy phenomena involving one-loop contributions in the frameworks of these SEFTs \cite{Zhang:2021jdf,Li:2022ipc,Coy:2021hyr,Du:2022vso,Zhang:2021tsq,Ohlsson:2022hfl,Crivellin:2022cve,Coy:2018bxr}. As expected, three complete SEFTs at the one-loop level and up to dimension six are much more complicated and significantly differ from one another in the number and content of effective operators, as well in structures of operators' Wilson coefficients. More details about those differences can be found in Refs. \cite{Zhang:2021jdf,Li:2022ipc,Du:2022vso}. 

In this work, we are mainly concerned with the one-loop structure of the SEFT arising from the type-(I+II) seesaw mechanism. This issue naturally shows up under the circumstance  where the type-I and type-II SEFTs have been completely established at the one-loop level. At the ultraviolet (UV) level, the classical Lagrangian of the type-(I+II) seesaw mechanism is simply the linear combination of those of the type-I and type-II seesaw mechanisms. Then, one may ask whether the effective Lagrangian of the type-(I+II) SEFT  up to one-loop level is also the trivial linear combination of those of the type-I and type-II SEFTs. The answer is affirmative for the tree-level part of the effective Lagrangian, but negative for the one-loop part. The former is simply due to the fact that the right-handed neutrinos and the triplet Higgs in the type-(I+II) seesaw mechanism do not interact with each other at the classical level and then can be integrated out separately at the tree level. While for the one-loop matching of the type-(I+II) seesaw mechanism onto the corresponding SEFT, the effects of the right-handed neutrinos and the triplet Higgs can be entangled via one-loop processes involving both of them (e.g., the right-handed neutrinos and the triplet Higgs simultaneously appear in a loop) and then lead to additional contributions absent in both the type-I and type-II SEFTs. For simplicity, these additional contributions from the non-linear effects can be called {\it cross contributions}. Since the complete one-loop results of the type-I and type-II SEFTs have already been gained \cite{Zhang:2021jdf,Li:2022ipc,Du:2022vso}, one may only focus on the cross contributions. This is exactly what we do in this work. After carefully considering and calculating the cross contributions by means of the Feynman diagrammatic method with the assumption that all masses of heavy particles are roughly of the same order of magnitude, we achieve the complete one-loop Lagrangian of the type-(I+II) SEFT up to dimension six and find that those cross contributions do not result in any new operators. They only give additional corrections to the quartic couplings of the SM Higgs and to Wilson coefficients of the unique Weinberg operator and nine dim-6 operators.

The rest parts of this paper are organized as follows. In Sec. \ref{sec:model}, we introduce the framework of the type-(I+II) seesaw mechanism and discuss the type and origin of cross contributions. The Feynman diagrammatic approach adopted to derive cross contributions is briefly described, and we carry out the one-loop matching and then obtain all cross contributions in the Green's basis in Sec. \ref{sec:matching}. In Sec. \ref{sec:warsaw}, cross contributions arising from removing redundant operators in the Green's basis and normalizing the kinetic terms are carefully calculated and the final results in the Warsaw basis are completely obtained. Finally, we summarize our main conclusions in Sec. \ref{sec:conclusion}.

\section{The type-(I+II) seesaw mechanism}\label{sec:model}

The classical Lagrangian of the type-(I+II) seesaw mechanism is the combination of those of the type-I and type-II seesaw mechanisms, that is 
\begin{eqnarray}
\mathcal{L}^{}_{\rm UV} = \mathcal{L}^{}_{\rm SM} + \mathcal{L}^{}_{\rm I} + \mathcal{L}^{}_{\rm II} \;,
\label{eq:type-(I+II)}
\end{eqnarray}
where the SM Lagrangian $\mathcal{L}^{}_{\rm SM}$ is explicitly given by
\begin{eqnarray}
\mathcal{L}^{}_{\rm SM} &=& - \frac{1}{4} G^A_{\mu\nu} G^{A\mu\nu} - \frac{1}{4} W^I_{\mu\nu} W^{I \mu\nu} - \frac{1}{4} B^{}_{\mu\nu} B^{\mu\nu} + \left( D^{}_\mu H \right)^\dagger \left( D^\mu H \right) - m^2 H^\dagger H - \lambda \left( H^\dagger H \right)^2 
\nonumber
\\
&& + \sum^{}_f \overline{f} \rmI \slashed{D} f - \left( \overline{Q^{}_{\rm L}} Y^{}_{\rm u} \widetilde{H} U^{}_{\rm R} + \overline{Q^{}_{\rm L}} Y^{}_{\rm d} H D^{}_{\rm R} + \overline{\ell^{}_{\rm L}} Y^{}_l H E^{}_{\rm R} + {\rm h.c.} \right) \;,
\label{eq:SM}
\end{eqnarray}
$\mathcal{L}^{}_{\rm I}$ and $\mathcal{L}^{}_{\rm II}$ come from the type-I and type-II seesaw mechanisms, respectively, and separately contain the right-handed neutrinos and the triplet Higgs, viz.,
\begin{eqnarray}
\mathcal{L}^{}_{\rm I} &=& \overline{N^{}_{\rm R}} \rmI \slashed{\partial} N^{}_{\rm R} - \left( \frac{1}{2} \overline{N^{\rm c}_{\rm R}} M^{}_{\rm R} N^{}_{\rm R} + \overline{\ell^{}_{\rm L}} Y^{}_\nu \widetilde{H} N^{}_{\rm R} + {\rm h.c.} \right) \;,
\label{eq:type-I}
\nonumber
\\
\mathcal{L}^{}_{\rm II} &=& \left( D^{}_\mu \Phi \right)^\dagger \left( D^\mu \Phi \right) - M^2_\Delta \left( \Phi^\dagger \Phi \right) - \left( \frac{1}{2} \overline{\ell^{}_{\rm L}} Y^{}_\Delta \sigma^I \widetilde{\ell^{}_{\rm L}} \Phi^{}_I + \lambda^{}_\Delta M^{}_\Delta \widetilde{H}^\dagger \sigma^I H \Phi^{}_I + {\rm h.c.}   \right) - \lambda^{}_3 \left( H^\dagger H \right)  \left( \Phi^\dagger \Phi \right) 
\nonumber
\\
&& - \lambda^{}_4 \left( H^\dagger \sigma^I H \right) \left( \Phi^\dagger T^I \Phi \right) - \left( \lambda^{}_1 + \frac{\lambda^{}_2}{4} \right) \left( \Phi^\dagger \Phi \right)^2  + \frac{\lambda^{}_2}{4} \left(\Phi^\dagger T^I \Phi\right) \left( \Phi^\dagger T^I \Phi \right) \;.
\label{eq:type-II}
\end{eqnarray}
The notations adopted in Eqs. \eqref{eq:SM} and \eqref{eq:type-II} follow those in Refs. \cite{Zhang:2021jdf,Li:2022ipc}: $f=Q^{}_{\rm L}, U^{}_{\rm R}, D^{}_{\rm R}, \ell^{}_{\rm L}, E^{}_{\rm R}$ refer to the SM fermionic doublets and singlets, $\widetilde{H}$ and $\widetilde{\ell^{}_{\rm L}}$ are respectively defined as $\widetilde{H} \equiv \rmI \sigma^2 H^\ast$ and $\widetilde{\ell^{}_{\rm L}} \equiv \rmI \sigma^2 \ell^{\rm c}_{\rm L}$ with $\sigma^I$ (for $I =1,2,3$) being the Pauli matrices, $\psi^{\rm c}$ (for $\psi = \ell^{}_{\rm L}, N^{}_{\rm R}$) represents the charge-conjugation of $\psi$, i.e., $\psi^{\rm c} \equiv {\sf C} \overline{\psi}^{\rm T}$ with ${\sf C} \equiv \rmI \gamma^2 \gamma^0$, $\Phi \equiv \left( \Phi^{}_1, \Phi^{}_2, \Phi^{}_3 \right)^{\rm T}$ is the triplet Higgs, and the covariant derivative $D^{}_\mu \equiv \partial^{}_\mu - \rmI g^{}_1 Y B^{}_\mu - \rmI g^{}_2 T^I W^I_\mu - \rmI g^{}_s T^A G^A_\mu$ has been defined as usual. $Y$, $T^I$ and $T^A$ are the generators of $\rm U(1)^{}_Y$, $\rm SU(2)^{}_L$ and $\rm SU(3)^{}_c$ groups, respectively, with $I=1,2,3$ and $A=1,2,\dots,8$ being the corresponding adjoint indices of $\rm SU(2)_L$ and $\rm SU(3)_c$. $T^I$ takes the fundamental representation of $\rm SU(2)^{}_L$, i.e., $T^I = \sigma^I/2$ (for $I=1,2,3$) for $\rm SU(2)^{}_L$ doublets, or the adjoint representation $\left( T^I \right)^{}_{JK} = -\rmI \epsilon^{IJK}$ (for $I,J,K=1,2,3$) for $\rm SU(2)^{}_L$ triplets with $\epsilon^{IJK}$ being the totally antisymmetric Levi-Civita tensor. Throughout this work, we adopt the basis where the right-handed neutrino mass matrix $M^{}_{\rm R}$ is diagonal, i.e., $M^{}_{\rm R} = {\rm Diag} \{ M^{}_1, M^{}_2, M^{}_3 \}$, and assume that masses of all three right-handed neutrinos and the triplet Higgs are roughly of the same order of magnitude but not fully degenerate, i.e., $M^{}_i \sim \Mds[] \sim \mathcal{O} (M)$. Therefore, these heavy particles can be integrated out simultaneously at the matching scale $\mu = M$ and the RG-running effects between any two mass scales of heavy particles can be neglected.

From Eqs. \eqref{eq:type-(I+II)}---\eqref{eq:type-II}, one can see that there do not exist any interactions between the right-handed neutrinos $N^{}_{\rm R}$ and the triplet  Higgs $\Phi$. Thus, one may naively think that the type-(I+II) SEFT derived by integrating out $N^{}_{\rm R}$ and $\Phi$ is simply the combination of the type-I and type-II  SEFTs. This conclusion only holds at the tree level. Actually, at the one-loop level, there are additional non-trivial contributions, i.e., {\it cross contributions} from non-linearly combined effects of the right-handed neutrinos $N^{}_{\rm R}$ and the triplet Higgs $\Phi$. There are two types of cross contributions, whose origins are quite different:
\begin{itemize}
\item One-loop diagrams involving both two kinds of heavy particles, i.e., $N^{}_{\rm R}$ and $\Phi$. Though $N^{}_{\rm R}$ and $\Phi$ do not couple with each other directly, they can show up simultaneously in some one-light-particle-irreducible (1LPI) diagrams via their Yukawa interactions with lepton and Higgs doublets, i.e., $\ell^{}_{\rm L}$ and $H$. In this case, their effects are entangled with each other when they are integrated out at the one-loop level. This type of cross contributions is very clear and easy to be understood.

\item Field redefinitions acting on the SM terms and higher-dimensional operators. The implementation of field redefinitions [or equations of motion (EOMs) of fields \footnote{Strictly speaking, it is field redefinitions that should be employed to remove operator redundancy, and  implementations of field redefinitions and EOMs are not equivalent in general \cite{Criado:2018sdb}. Nevertheless, EOMs can still exploited to eliminate redundant operators as long as the discrepancy between field redefinitions and EOMs is correctly taken into account. This point will be discussed in detail later.}] to get rid of some redundant operators of dimension six can lead to additional cross contributions, as long as field redefinitions and Wilson coefficients of some relevant redundant operators contain pure contributions from integrating out $N^{}_{\rm R}$ (or $\Phi$) and $\Phi$ (or $N^{}_{\rm R}$) separately. So do the field redefinitions used to normalize the kinetic terms of some fields. This type of cross contributions is much more obscure. 
\end{itemize}

These two types of cross contributions arise in different periods of the full matching process, which will be carefully discussed and calculated in the next two sections.

\section{Matching in the Green's basis}\label{sec:matching}

Usually, there are two equivalent approaches to perform the one-loop matching of a given UV model onto the corresponding EFT. One is the functional approach \cite{Henning:2014wua,Drozd:2015rsp,Henning:2016lyp,Ellis:2016enq,Fuentes-Martin:2016uol,Zhang:2016pja,Ellis:2017jns,Kramer:2019fwz,Cohen:2019btp,Cohen:2020fcu,Dittmaier:2021fls} based on the  so-called covariant derivative expansion (CDE) \cite{Gaillard:1985uh,Chan:1986jq,Cheyette:1987qz}, and the other one is the Feynman diagrammatic method (see, e.g., Refs. \cite{Rothstein:2003mp,Skiba:2010xn,Manohar:2018aog,Cohen:2019wxr,Penco:2020kvy}). Both of these two approaches have their own advantages and disadvantages, as can be seen from the discussions in Refs. \cite{Cohen:2020fcu,Dittmaier:2021fls,Cohen:2022tir}. It is worth pointing out that several automated tools have been developed to perform the complete matching calculations recently. For example, the Mathematica packages {\sf STrEAM} \cite{Cohen:2020qvb} and {\sf SuperTracer}  \cite{Fuentes-Martin:2020udw} \footnote{Built upon {\sf SuperTracer}, the package {\sf Matchete} \cite{Wilsch:2022kfv} will facilitate the fully automatic matching of various UV models and now is still  under development.} are based on the CDE technique to evaluate a large number of supertraces in the functional approach, and the package {\sf matchmakereft} \cite{Carmona:2021xtq} automatically carries out the complete matching of arbitrary models by means of the Feynman diagrammatic method. In this paper, we adopt the intuitive diagrammatic method to calculate the aforementioned {\it cross contributions} and achieve the complete one-loop matching of the type-(I+II) seesaw mechanism. At last, we make use of the package {\sf matchmakereft} to crosscheck our results.

\subsection{The tree-level matching}

As discussed before, results for the tree-level matching of the type-(I+II) seesaw mechanism are trivially the linear combination of those in the type-I and type-II seesaw mechanisms, which have been achieved separately in the previous works \cite{Zhang:2021jdf,Li:2022ipc,Broncano:2002rw,Broncano:2003fq,Abada:2007ux}. Following the results in Refs. \cite{Zhang:2021jdf,Li:2022ipc}, the tree-level effective Lagrangian up to dimension six in the Green's basis for the type-(I+II) SEFT is found to be
\begin{eqnarray} \label{eq:tree-level}
\mathcal{L}^{\rm G}_{\rm tree} &=& \mathcal{L}^{}_{\rm SM} +  \delta \lambda^{\rm G} |^{}_{\rm tree} \left( H^\dagger H \right)^2 + \frac{1}{2} \left( G^{(5)}_{\alpha\beta} |^{}_{\rm tree} \Op^{(5)}_{\alpha\beta} + {\rm h.c.} \right) + G^{}_H |^{}_{\rm tree} \Op^{}_H + G^{}_{HD} |^{}_{\rm tree} \Op^{}_{HD}  
\nonumber
\\[0.15cm]
&& + G^{(1)}_{H\ell} |^{\alpha\beta}_{\rm tree} \Op^{(1)\alpha\beta}_{H\ell} + G^{(3)}_{H\ell} |^{\alpha\beta}_{\rm tree} \Op^{(3)\alpha\beta}_{H\ell} + G^{}_{\ell\ell} |^{\alpha\beta\gamma\lambda}_{\rm tree} \Op^{\alpha\beta\gamma\lambda}_{\ell\ell} + G^\prime_{HD} |^{}_{\rm tree} \Opr^\prime_{HD}
\nonumber
\\[0.15cm]
&& + G^{\prime(1)}_{H\ell} |^{\alpha\beta}_{\rm tree} \Opr^{\prime(1)\alpha\beta}_{H\ell} + G^{\prime(3)}_{H\ell} |^{\alpha\beta}_{\rm tree}  \Opr^{\prime(3)\alpha\beta}_{H\ell} \;,
\end{eqnarray} 
where $\Op^{\cdots}_{\cdots}$ and $\Opr^{\cdots}_{\cdots}$ stand for independent operators in the Warsaw basis \cite{Grzadkowski:2010es} and redundant operators in the Green's basis \cite{Jiang:2018pbd,Gherardi:2020det} \footnote{All dim-6 operators in the Green's basis are listed in Table \ref{tab:green-basis-1}---\ref{tab:green-basis-4}.}, respectively. In Eq. \eqref{eq:tree-level}, $G^{\cdots}_{\cdots} |^{}_{\rm tree}$ and $\delta \lambda^{\rm G} |^{}_{\rm tree}$ denote the corresponding tree-level Wilson coefficients and the tree-level correction to the quartic coupling constant $\lambda$ in the Green's basis, respectively
, which are explicitly given by
\begin{eqnarray}\label{eq:coe-tot}
&& \delta \lambda^{\rm G} |^{}_{\rm tree} = \delta \lambda^{\rm G, II} |^{}_{\rm tree} \;,\quad G^{(5)} |^{}_{\rm tree} = G^{(5),\rm I}|^{}_{\rm tree} + G^{(5),\rm II} |^{}_{\rm tree} \;,
\nonumber
\\[0.15cm]
&& G^{}_H |^{}_{\rm tree} = G^{\rm II}_H |^{}_{\rm tree} \;,\quad G^{}_{HD} |^{}_{\rm tree} = G^{\rm II}_{HD} |^{}_{\rm tree} \;,\quad G^{\prime}_{HD} |^{}_{\rm tree} = G^{\prime, \rm II}_{HD} |^{}_{\rm tree} \;,\quad G^{}_{\ell\ell} |^{}_{\rm tree} =  G^{{\rm II}}_{\ell\ell} |^{}_{\rm tree} \;,
\nonumber
\\[0.15cm]
&& G^{(1)}_{H\ell} |^{}_{\rm tree} =  G^{(1),{\rm I}}_{H\ell}  |^{}_{\rm tree} \;,\quad G^{\prime(1)}_{H\ell} |^{}_{\rm tree} =  G^{\prime (1),{\rm I}}_{H\ell}  |^{}_{\rm tree} \;,\quad G^{(3)}_{H\ell} |^{}_{\rm tree} =  G^{(3),{\rm I}}_{H\ell}  |^{}_{\rm tree} \;,\quad G^{\prime(3)}_{H\ell} |^{}_{\rm tree} =  G^{\prime (3),{\rm I}}_{H\ell}  |^{}_{\rm tree} \quad
\end{eqnarray}
with
\begin{eqnarray}\label{eq:coe-ind}
&& \delta \lambda^{\rm G,II} |^{}_{\rm tree} = 2\lambda^2_\Delta \;,\quad G^{(5),\rm I}_{} |^{\alpha\beta}_{\rm tree} = \left( \Yn M^{-1}_{\rm R} \TYn \right)^{}_{\alpha\beta} \;,\quad G^{(5),\rm II}_{} |^{\alpha\beta}_{\rm tree} = - \frac{2\lambda^{}_\Delta}{M^{}_\Delta} \left( Y^{}_\Delta \right)^{}_{\alpha\beta} \;,
\nonumber
\\
&& G^{\rm II}_H |^{}_{\rm tree} = - \frac{2 \left( \lambda^{}_3 - \lambda^{}_4 \right) \lambda^2_\Delta}{\Mds} \;,\quad G^{\rm II}_{HD} |^{}_{\rm tree} = G^{\prime, \rm II}_{HD} |^{}_{\rm tree} = \frac{ 4\lambda^2_\Delta}{\Mds} \;,\quad G^{{\rm II}}_{\ell\ell} |^{\alpha\beta\gamma\lambda}_{\rm tree} = \frac{1}{4\Mds} \left( \Ydel \right)^{}_{\alpha\gamma} \left( \DYdel \right)^{}_{\beta\lambda} \;,
\nonumber
\\
&& G^{(1),{\rm I}}_{H\ell} |^{\alpha\beta}_{\rm tree} = G^{\prime (1),{\rm I}}_{H\ell} |^{\alpha\beta}_{\rm tree} = - G^{(3),{\rm I}}_{H\ell} |^{\alpha\beta}_{\rm tree} = - G^{\prime (3),{\rm I}}_{H\ell}  |^{\alpha\beta}_{\rm tree} =  \frac{1}{4} \left( \Yn M^{-2}_{\rm R} \DYn \right)^{}_{\alpha\beta} \;.
\end{eqnarray}
The superscripts ``I" and ``II" in Eqs. \eqref{eq:coe-tot} and \eqref{eq:coe-ind} characterize pure contributions from $N^{}_{\rm R}$ and $\Phi$, respectively. As shown in Eq. \eqref{eq:tree-level}, there are three redundant operators, i.e., $\Opr^\prime_{HD}$, $\Opr^{\prime(1)}_{H\ell}$ and $\Opr^{\prime (3)}_{H\ell}$. These redundant operators will be removed to obtain final results in the Warsaw basis, but during this process, they may induce additional one-loop contributions. Therefore, we first keep above results in the Green's basis.

\subsection{The one-loop matching} \label{subsec:one-loop-matching}

To make the one-loop matching procedure more intuitive, we adopt the Feynman diagrammatic approach to match the type-(I+II) seesaw mechanism onto the corresponding low-energy EFT, i.e., the type-(I+II) SEFT. Similar to the procedure described in Ref. \cite{Zhang:2021tsq}, the general procedure to carry out the complete one-loop matching via the Feynman diagrammatic method is as below:

\begin{itemize}
\item First, one needs to generate a set of one-light-particle-irreducible diagrams at the one-loop level in the UV model, which should cover all operators up to the considered dimension in the Green's basis. A trivial strategy is to choose a set of 1LPI diagrams, whose external lines go through the field ingredients (not including the gauge fields in the covariant derivative) of all operators and the SM terms in the Green's basis. For example, to calculate Wilson coefficients of the operators in the $\psi^2 D H^2$ class shown in Table \ref{tab:green-basis-1}, one can choose the four-point 1LPI diagrams with external lines generated by the $ \left( \overline{\psi}\psi H^\dagger H \right)$ field configurations.

\item Second, one may calculate the generated 1LPI diagrams in the $d$-dimensional space-time (i.e., $d \equiv 4 - 2\varepsilon$) with the dimensional regularization and the modified minimal subtraction ($\overline{\rm MS}$) scheme. Moreover, all external particles are taken to be off-shell. Before calculating loop integrals in amplitudes and obtaining final results for the generated 1LPI diagrams, one needs to make use of the expansion-by-regions techniques \cite{Beneke:1997zp,Smirnov:2002pj,Jantzen:2011nz} to expand loop integrals up to an enough order in the hard-momentum region (i.e., $k \sim M \gg p$ with $k$, $M$ and $p$ standing for the loop momentum, heavy field masses and external momenta, respectively)~\footnote{It is easy to see that contributions from the soft-momentum region where the loop integrals are expanded with $k\sim p \ll M$ are  exactly the same as those from the tree-level EFT Lagrangian via one-loop diagrams. Thus, those two kinds of contributions from the UV model and EFT respectively cancel each other out and hence do not need to be taken into consideration during the one-loop matching.}. On the other hand, in the EFT one needs to calculate contributions from effective operators (including the threshold corrections \cite{Wells:2017vla} to the SM terms) in the Green's basis to the tree-level vertices with  external lines being the same as those of the 1LPI diagrams in the UV model \footnote{Here, the possible tree-level Wilson coefficients of effective operators (as well as possible tree-level corrections to the SM terms) in the Green's basis induced by the tree-level matching should not be taken into consideration. }.  Then the relevant unknown Wilson coefficients and threshold corrections at the one-loop level can be solved by equating results for the vertices in the EFT with those for the corresponding 1LPI diagrams in the UV model. Note that all results obtained at this stage are in the Green's basis and hence some operators are redundant.

\item Finally, one should remove the redundant operators in the Green's basis and normalize the kinetic terms with the help of field redefinitions (or fields' EOMs). Then the final results for the effective SM couplings and Wilson coefficients of operators in the Warsaw basis are achieved, which constitute the complete one-loop EFT and consistently describe the infrared behavior of the UV model. During the process of removing the redundant operators and normalizing the kinetic terms, the loop order of contributions to Wilson coefficients and of corrections  to the SM terms should be carefully considered, since the tree-level matching results may result in additional one-loop contributions via field redefinitions (or EOMs).
\end{itemize}

The complete one-loop matchings of the type-I and type-II seesaw mechanisms have already been achieved separately \cite{Zhang:2021jdf,Li:2022ipc}, as a result, for the one-loop matching of the type-(I+II) seesaw mechanism, we only need to focus on the {\it cross contributions} arising from entangled effects of $N^{}_{\rm R}$ and $\Phi$. Then the Wilson coefficients or threshold corrections ${\sf X}$ in the type-(I+II) SEFT can be written as
\begin{eqnarray}
\sf X = X^{\rm I} + X^{\rm II} + X^{\rm ct}	\;,
\end{eqnarray}
where ${\sf X}^{\rm I}$ and ${\sf X}^{\rm II}$ are respectively the corresponding results in the type-I and type-II SEFTs, and ${\sf X}^{\rm ct}$ are the aforementioned cross contributions. During the matching procedure in the Green's basis, all cross contributions ${\sf X}^{\rm ct}$ result from the 1LPI diagrams involving both $N^{}_{\rm R}$ and $\Phi$. We adopt the trivial strategy described at the beginning of this subsection to generate 1LPI diagrams. With the help of the Mathematica packages {\sf FeynRules} \cite{Christensen:2008py} and {\sf FeynArts} \cite{Hahn:2000kx}, we find that only a few of 1LPI diagrams involve $N^{}_{\rm R}$ and $\Phi$ simultaneously and their external lines are dominated by $ \left( H^\dagger H^\dagger HH \right)$, $\left( \overline{\ell} \ell^{\rm c} \widetilde{H} \widetilde{H}  \right)$, $ \left( \overline{\ell} \ell H^\dagger H \right)$, $\left( \overline{\ell} \overline{\ell} \ell\ell \right)$, $ \left( \overline{\ell} E H^\dagger HH \right)$, $\left( H^\dagger H^\dagger H^\dagger HHH \right)$ field configurations \footnote{We do not consider the Hermitian conjugates of the non-Hermitian diagrams, e.g., those with external particles given by $\left( \overline{\ell} \ell^{\rm c} \widetilde{H} \widetilde{H}  \right)^\dagger$, $\left( \overline{\ell} E H^\dagger HH \right)^\dagger$ field configuration, whose contributions are exactly the Hermitian conjugates of those already obtained by the original ones and not necessary to be calculated repeatedly.}. In the rest of this section, we calculate all of these 1LPI diagrams in the UV model and as well the corresponding tree-level vertices in the EFT. Then we match those results in the UV model and the EFT to obtain the cross contributions to Wilson coefficients of some relevant operators as well as those to threshold corrections in the Green's basis.

\subsubsection{$H^\dagger H^\dagger HH$}

\begin{figure}
	\centering
	\includegraphics[width=0.75\linewidth]{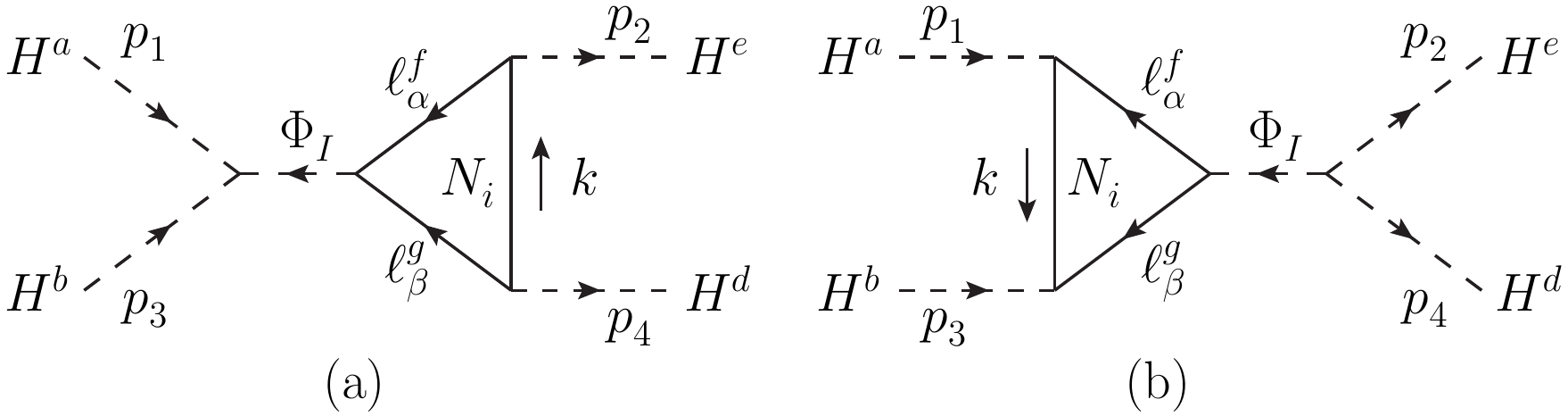}
	\caption{The 1LPI diagrams with external lines generated by the $\left( H^\dagger H^\dagger HH \right)$ field configuration in the type-(I+II) seesaw mechanism.}
	\label{fig:hhhh}
\end{figure}

We first consider the case where external lines of the 1LPI diagrams in the UV model are determined by the $\left( H^\dagger H^\dagger HH \right)$ field configuration. The relevant 1LPI diagrams are shown in Fig. \ref{fig:hhhh}. As can be seen in Fig. \ref{fig:hhhh}, the right-handed neutrino fields and the triplet Higgs field appear simultaneously in the diagrams. Note that the 1LPI Feynman diagrams in Fig. \ref{fig:hhhh} are not one-particle irreducible since they will be disconnected into two pieces by cutting the heavy triplet Higgs lines. After calculating the amplitudes of these Feynman diagrams in the $d$-dimensional space-time with the dimensional regularization and the $\overline{\rm MS}$ scheme \footnote{The expressions of the amplitudes corresponding to the two diagrams in Fig. \ref{fig:hhhh} can be found in Appendix \ref{app:B}.}, we can obtain the hard-momentum part of the total amplitude up to $\mathcal{O} \left( M^{-2} \right)$, namely
\begin{eqnarray}\label{eq:hhhh-hard}
\ampd{\rm UV} &=&  2\rmI \left( \del{ad}\del{be} + \del{ae}\del{bd} \right) \frac{ \lamd}{\left(4\pi\right)^2} \left\{ \left( \TYn \DYdel \Yn \right)^{}_{ii} \left[ - \frac{2\pM{}}{\Mds[]} \left( 1 + \lnmi \right) - \frac{2\left( 2\pM{2} - \Mds \right) }{\pM{}\Mds[3]} \left( 1 + \lnmi \right)  \right.\right.
\nonumber
\\
&& \times \left. \momp{2}\cdot\momp{4} + \frac{\Mds[2]\left( 1+ 2\lnmi \right) - 4\pM{2} \left( 1 + \lnmi \right)}{2\pM{}\Mds[3]} \left( \momp[2]{2} + \momp[2]{4} \right)\right] + \left( \DYn \Ydel \SYn \right)^{}_{ii} \left[ - \frac{2\pM{}}{\Mds[]} \left( 1 + \lnmi \right) \right. 
\nonumber
\\
&& \left. \left. + \frac{1}{\pM{}\Mds[]} \momp{3} \cdot \left( \momp{2} - \momp{3} + \momp{4} \right) + \frac{\Mds[2]\left( 1+ 2\lnmi \right) - 4\pM{2} \left( 1 + \lnmi \right)}{2\pM{}\Mds[3]} \left( \momp{2} + \momp{4} \right)^2 \right] \right\} \;,
\end{eqnarray}
in which 
\begin{eqnarray}
L^{}_{i} \equiv \displaystyle \ln \left( \frac{\mu^2}{M^{2}_{i}} \right) + \frac{1}{\varepsilon} - \gamma^{}_{\rm E} + \ln \left( 4\pi \right) \;,\quad L^{}_{\Delta} \equiv \displaystyle \ln \left( \frac{\mu^2}{M^{2}_{\Delta}} \right) + \frac{1}{\varepsilon} - \gamma^{}_{\rm E} + \ln \left( 4\pi \right)
\end{eqnarray}
have been defined for simplicity with $\mu$ and $\gamma^{}_{\rm E}$ being the renormalization scale and the Euler constant, respectively.

Now, we need to calculate the contributions from effective operators in the EFT to the corresponding tree-level vertex with external lines given by the same field configuration as the 1LPI diagrams in the UV model, i.e., $\left( H^\dagger H^\dagger HH \right)$. One can find that only the quartic coupling of the SM Higgs $\left( H^\dagger H \right)^2$, $\Op^{}_{H\square}$, $\Op^{}_{HD}$, $\Opr^\prime_{HD}$ and $\Opr^{\prime\prime}_{HD}$ contribute, i.e.,
\begin{eqnarray}
\amp{\rm EFT} &=& 2\rmI \left( \del{ad}\del{be} + \del{ae}\del{bd} \right)  \delta \lambda^{\rm G,ct}  + \rmI \left( \del{ad}\del{be} + \del{ae}\del{bd} \right) \left( G^{\rm ct}_{HD} + G^{\prime,\rm ct}_{HD} - 2\rmI G^{\prime\prime,\rm ct}_{HD}\right) \momp{2}\cdot\momp{4} 
\nonumber
\\
&& - 2 \rmI \left( \del{ad}\del{be} + \del{ae}\del{bd} \right) \left( G^{\rm ct}_{H\square} + \rmI G^{\prime\prime,\rm ct}_{HD} \right) \momp[2]{3} + \rmI \left[ \del{ad}\del{be} \left( G^{\rm ct}_{HD} - 2G^{\rm ct}_{H\square} \right) + \del{ae}\del{bd} G^{\prime,\rm ct}_{HD} \right] \momp[2]{2}
\nonumber
\\
&& - \rmI \left[ \del{ad}\del{be} \left( G^{\rm ct}_{HD} - G^{\prime,\rm ct}_{HD} - 2\rmI G^{\prime\prime,\rm ct}_{HD} - 4 G^{\rm ct}_{H\square} \right) - \del{ae}\del{bd} \left( G^{\rm ct}_{HD} - G^{\prime,\rm ct}_{HD} + 2\rmI G^{\prime\prime,\rm ct}_{HD} \right) \right] \momp{2}\cdot\momp{3}
\nonumber
\\
&& + \rmI \left[ \del{ad}\del{be} \left( G^{\rm ct}_{HD} - G^{\prime,\rm ct}_{HD} + 2 \rmI G^{\prime\prime,\rm ct}_{HD} \right) - \del{ae}\del{bd} \left( G^{\rm ct}_{HD} - G^{\prime,\rm ct}_{HD} - 2\rmI G^{\prime\prime,\rm ct}_{HD} - 4 G^{\rm ct}_{H\square} \right)  \right] \momp{3}\cdot\momp{4}
\nonumber
\\
&& + \rmI \left[ \del{ad}\del{be} G^{\prime,\rm ct}_{HD} + \del{ae}\del{bd} \left( G^{\rm ct}_{HD} - 2 G^{\rm ct}_{H\square} \right) \right] \momp[2]{4} \;.
\end{eqnarray}

Then, equating results in the EFT and the UV model, namely $\amp{\rm EFT} = \ampd{\rm UV}$ with $\rm SU(2)$ indices (i.e., $a,b,d,e=1,2$) and momenta of the external particles being arbitrary, one can achieve a set of linear equations as below:
\begin{eqnarray}\label{eq:hhhh-matching}
\begin{cases}
\displaystyle \delta \lambda^{\rm G, ct} = - \frac{2\lamd}{\left( 4\pi\right)^2} \left( \DYn \Ydel \SYn + \TYn \DYdel \Yn \right)^{}_{ii} \frac{\pM{}}{\Mds[]} \left( 1 + \lnmi \right) \;,
\\
\displaystyle G^{\rm ct}_{HD} + G^{\prime,\rm ct}_{HD} - 2\rmI G^{\prime\prime,\rm ct}_{HD} =  \frac{2\lamd}{\left( 4\pi\right)^2}  \left( \DYn \Ydel \SYn \right)^{}_{ii} \frac{\Mds \left( 1 + 2\lnmi \right) - 4 \pM{2} \left( 1 + \lnmi \right) }{\pM{}\Mds[3]} 
\\
\displaystyle \hphantom{G^{\rm ct}_{HD} + G^{\prime,\rm ct}_{HD} - 2\rmI G^{\prime\prime,\rm ct}_{HD} =} - \frac{4 \lamd}{\left(4\pi\right)^2} \left( \TYn \DYdel \Yn \right)^{}_{ii} \frac{\left( 2\pM{2} - \Mds \right)}{\pM{}\Mds[3]} \left( 1 + \lnmi \right) \;,
\\
\displaystyle G^{\rm ct}_{H\square} + \rmI G^{\prime\prime,\rm ct}_{HD} = \frac{\lamd}{\left( 4\pi\right)^2}  \left( \DYn \Ydel \SYn \right)^{}_{ii} \frac{1}{\pM{}\Mds[]} \;,
\\
\displaystyle G^{\rm ct}_{HD} - 2G^{\rm ct}_{H\square} = G^{\prime,\rm ct}_{HD} =  - \frac{\lamd}{\left( 4\pi\right)^2}  \left( \DYn \Ydel \SYn + \TYn \DYdel \Yn \right)^{}_{ii} \frac{4\pM{2} \left( 1+ \lnmi \right) - \Mds[2] \left( 1 + 2\lnmi \right)}{\pM{}\Mds[3]} \;,
\\
\displaystyle - G^{\rm ct}_{HD} + G^{\prime,\rm ct}_{HD} + 2\rmI G^{\prime\prime,\rm ct}_{HD} + 4 G^{\rm ct}_{H\square} =  G^{\rm ct}_{HD} - G^{\prime,\rm ct}_{HD} + 2\rmI G^{\prime\prime,\rm ct}_{HD} = \frac{2\lamd}{\left( 4\pi\right)^2}  \left( \DYn \Ydel \SYn \right)^{}_{ii} \frac{1}{\pM{}\Mds[]} \;.
\end{cases} 
\end{eqnarray}
Solving these linear equations in Eq. \eqref{eq:hhhh-matching}, we can obtain the cross contributions to the quartic coupling constant $\lambda$ and Wilson coefficients of $\Op^{}_{H\square}$, $\Op^{}_{HD}$, $\Opr^\prime_{HD}$ and $\Opr^{\prime\prime}_{HD}$ in the Green's basis:
\begin{eqnarray}\label{eq:hhhh-results}
\delta \lambda^{\rm G,ct} &=& - \frac{2\lamd}{\left( 4\pi\right)^2}  \left( \DYn \Ydel \SYn + \TYn \DYdel \Yn \right)^{}_{ii} \frac{\pM{}}{\Mds[]} \left( 1 + \lnmi \right) \;,
\nonumber
\\
G^{\rm ct}_{H\square} &=& \frac{\lamd}{2\left( 4\pi\right)^2}  \frac{1}{\pM{}\Mds[]} \left( \DYn \Ydel \SYn + \TYn \DYdel \Yn \right)^{}_{ii} \;,
\nonumber
\\
G^{\rm ct}_{HD} &=& - \frac{2\lamd}{\left( 4\pi\right)^2}  \left( \DYn \Ydel \SYn + \TYn \DYdel \Yn \right)^{}_{ii} \frac{2\pM{2} - \Mds}{\pM{}\Mds[3]} \left( 1 + \lnmi \right) \;,
\nonumber
\\
G^{\prime,\rm ct}_{HD} &=& - \frac{\lamd}{\left( 4\pi\right)^2}  \left( \DYn \Ydel \SYn + \TYn \DYdel \Yn \right)^{}_{ii} \frac{4\pM{2} \left( 1+ \lnmi \right) - \Mds[2] \left( 1 + 2\lnmi \right)}{\pM{}\Mds[3]}
\nonumber
\\
G^{\prime\prime,\rm ct}_{HD} &=& - \frac{\rmI \lamd}{2\left( 4\pi\right)^2}  \frac{1}{\pM{}\Mds[]} \left( \DYn \Ydel \SYn - \TYn \DYdel \Yn \right)^{}_{ii} \;.
\end{eqnarray}
As can be seen in Eq. \eqref{eq:hhhh-results}, the couplings $\Yn$ and $\Ydel$ (or $\pM{}$ and $\Mds[]$) appearing in the results are entangled, which means these cross contributions indeed contain non-linearly combined effects of $N^{}_{\rm R}$ and $\Phi$.

\subsubsection{$\overline{\ell} \ell^{\rm c} \widetilde{H} \widetilde{H}$} \label{sec:3.2.2}

\begin{figure}
\centering
\includegraphics[width=0.9\linewidth]{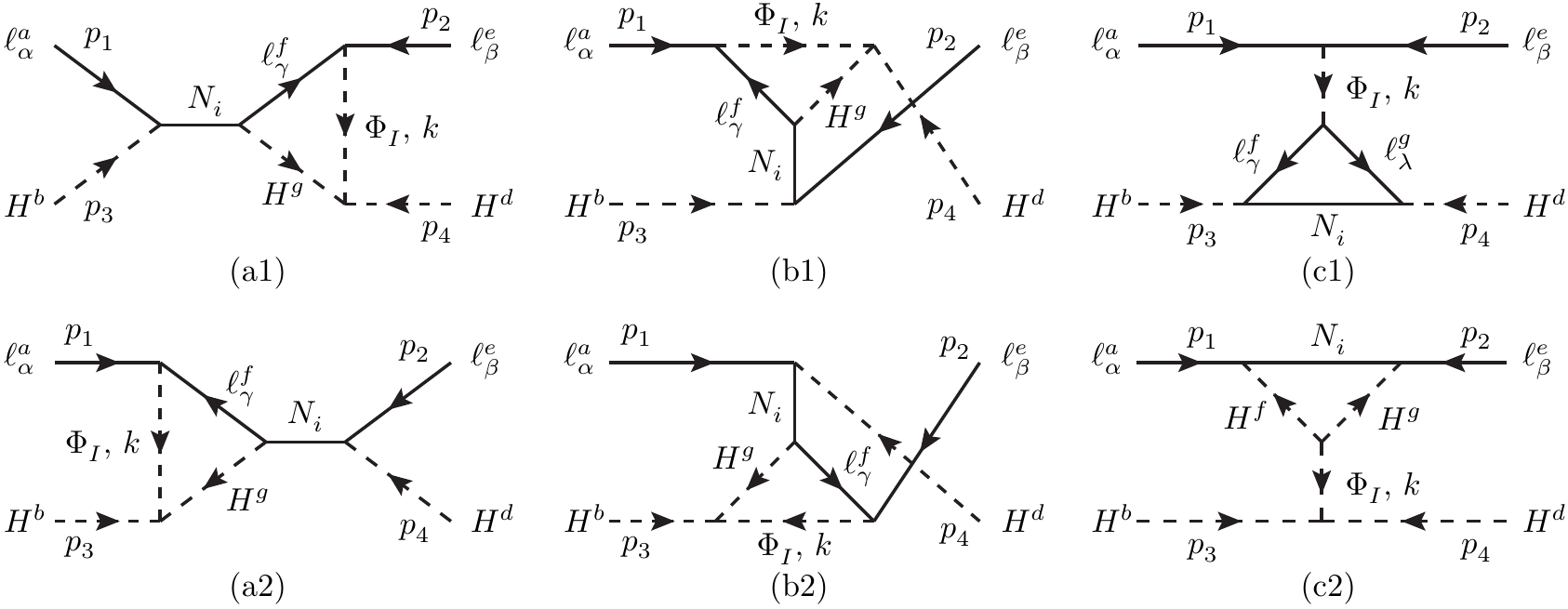}
\caption{The 1LPI diagrams with external lines generated by the $\left( \overline{\ell} \ell^{\rm c} \widetilde{H} \widetilde{H} \right)$ field configuration in the type-(I+II) seesaw mechanism.}
\label{fig:llbarhhbar}
\end{figure}

There are six 1LPI diagrams with external particles determined by the $\left( \overline{\ell} \ell^{\rm c} \widetilde{H} \widetilde{H} \right)$ field configuration in total, which are shown in Fig. \ref{fig:llbarhhbar}. The corresponding amplitudes for these Feynman diagrams can be found in Eq. \eqref{eq:llbarhhbar-amp}. With the help of the results in Eq. \eqref{eq:llbarhhbar-amp} and the expansion-by-regions technique, the hard-momentum part of the total amplitude is found to be
\begin{eqnarray}\label{eq:llbarhhbar-hard}
\ampd{\rm UV} &=& \frac{2\rmI}{\left( 4\pi \right)^2} \left( \epsilon^{}_{ad} \epsilon^{}_{be} - \epsilon^{}_{ab} \epsilon^{}_{ed} \right) \left[ 2\lamd[2] \left( \SYn \right)^{}_{\alpha i} \pM{-1} \left( \DYn \right)^{}_{i\beta} - \frac{\pM{}}{\Mds} \left( \DYdel \right)^{}_{\alpha\beta} \left( \DYn\Ydel\SYn \right)^{}_{ii}  \right]
\nonumber
\\
&& \times \left( 1 + \lnmi \right) \overline{u} \left( \momp{2} \right) \pl u \left( \momp{1} \right) \;.
\end{eqnarray}
Actually, only diagrams (c1) and (c2) in Fig. \ref{fig:llbarhhbar} have non-vanishing contributions to the result in Eq. \eqref{eq:llbarhhbar-hard}. On the other hand, the $\left( \overline{\ell} \ell^{\rm c} \widetilde{H} \widetilde{H} \right)$ field configuration breaks the lepton number by 2, which can only be achieved by the Weinberg operator $\Op^{(5)}$ among all operators up to dimension six in the Green's basis. Thus, the corresponding result in the EFT attributes to the unique Weinberg operator and is given by
\begin{eqnarray}\label{eq:llbarhhbar-eft}
\amp{\rm EFT} &=& - \rmI \left( G^{(5),{\rm ct}}_{\alpha\beta} \right)^\ast \left( \epsilon^{}_{ad} \epsilon^{}_{be} - \epsilon^{}_{ab} \epsilon^{}_{ed} \right) \overline{u} \left( \momp{2} \right) \pl u \left( \momp{1} \right) \;.
\end{eqnarray}
By equating $\amp{\rm EFT}$ in Eq. \eqref{eq:llbarhhbar-eft} with $\ampd{\rm UV}$ in Eq. \eqref{eq:llbarhhbar-hard}, we have
\begin{eqnarray}
G^{(5),\rm ct}_{\alpha\beta} = \frac{2}{\left( 4\pi \right)^2} \left[ \frac{\pM{}}{\Mds} \left( \Ydel \right)^{}_{\alpha\beta} \left( \TYn \DYdel \Yn \right)^{}_{ii} - 2 \lamd[2] \left( \Yn \right)^{}_{\alpha i} \pM{-1} \left( \TYn\right)^{}_{i\beta}  \right] \left( 1 + \lnmi \right) \;.
\end{eqnarray}

\subsubsection{$\overline{\ell}\ell H^\dagger H$}

\begin{figure}
\centering
\includegraphics[width=0.9\linewidth]{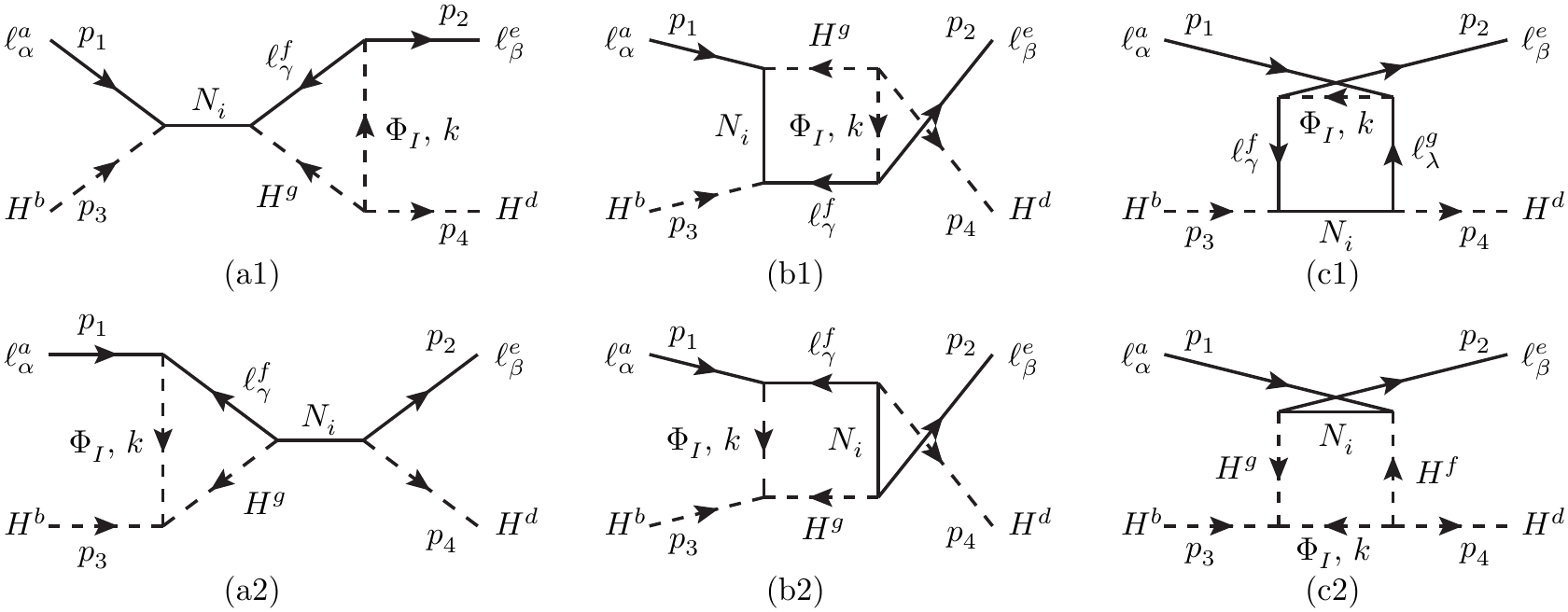}
\caption{The 1LPI diagrams with external lines generated by the $\left( \overline{\ell} \ell H^\dagger H \right)$ field configuration in the type-(I+II) seesaw mechanism.}
\label{fig:llhh}
\end{figure}

The six 1LPI diagrams with external particles generated by the $\left( \overline{\ell} \ell H^\dagger H \right)$ field configuration are shown in Fig. \ref{fig:llhh}, whose amplitudes are given in Eq. \eqref{eq:llhh-amp}. Then, the hard-momentum part of the total amplitude is 
\begin{eqnarray}\label{eq:llhh-uv}
\ampd{\rm UV} &=& \frac{\rmI}{\left( 4\pi \right)^2} \overline{u} \left( \momp{2} \right) \pr \left\{ \frac{\momps{2}}{\left(\pM{2} - \Mds \right)^2} \left[ -2 \left( 2\del{ae} \del{bd}  - \del{ad} \del{be} \right) \lamd[2] \left( \SYn \right)^{}_{\alpha i} \left( \Yn \right)^{}_{\beta i} \right.\right.
\nonumber
\\
&& \times \frac{2\pM{4} \left( 1 + \lnmd \right) - \pM{2}\Mds \left( 5 + 3\lnmi + \lnmd \right) + \Mds[4] \left( 3 + 2\lnmi \right)}{\pM{2}}   
\nonumber
\\
&& + \frac{1}{2}  \lamd  \left( \left( \DYdel \Yn \right)^{}_{\alpha i} \left( \Yn \right)^{}_{\beta i } +  \left( \SYn \right)^{}_{\alpha i} \left( \Ydel \SYn \right)^{}_{\beta i} \right)
\nonumber
\\
&& \times \left( \del{ae}\del{bd} \frac{5\pM{4} \left( 1 + 2\lnmd \right) - 2\pM{2} \Mds \left( 7 + 4\lnmi + 6\lnmd \right) + \Mds[4] \left( 9 + 4\lnmi + 6\lnmd \right) }{\pM{}\Mds[]} \right.
\nonumber
\\
&& - \left. 2 \del{ad} \del{be} \frac{2\pM{4} \left( 1 + 2\lnmd \right) - \pM{2}\Mds \left( 5 + 2\lnmi + 6\lnmd \right) + \Mds[4] \left( 3 + \lnmi + 3\lnmd \right) }{\pM{} \Mds[]} \right)
\nonumber
\\
&& + \left. \frac{1}{2} \left( 2\del{ae} \del{bd} - \del{ad} \del{be} \right) \left( \DYdel\Yn \right)^{}_{\alpha i} \left( \DYn \Ydel \right)^{}_{i \beta} \left( \pM{2} \left( 1 + \lnmi - \lnmd \right) - \Mds \right) \right] 
\nonumber
\\
&& +  \frac{\momps{3}}{\left(\pM{2} - \Mds \right)^2} \left[ \left( 2 \del{ae}\del{bd} - \del{ad}\del{be} \right) \lamd[2] \left( \SYn \right)^{}_{\alpha i} \left( \Yn \right)^{}_{\beta i} \right.
\nonumber
\\
&& \times \frac{\pM{4} \left( 1 + 2\lnmd \right) - 4\pM{2}\Mds \left( 1 + \lnmi \right) + \Mds[4] \left( 3 + 2\lnmi \right) }{\pM{2}}  
\nonumber
\\
&& +  \left(2\del{ae}\del{bd} - \del{ad}\del{be} \right) \lamd \left( \SYn \right)^{}_{\alpha i} \left( \Ydel \SYn \right)^{}_{\beta i} \frac{\Mds[] \left( \pM{2} \left( 1 + \lnmi - \lnmd \right) - \Mds \right) }{\pM{}} 
\nonumber
\\
&& + \lamd \left( \DYdel \Yn \right)^{}_{\alpha i} \left( \Yn \right)^{}_{\beta i} \left( \del{ae} \del{bd} \frac{5\pM{4} - 2\pM{2}\Mds \left( 4 - \lnmi + \lnmd \right) + 3 \Mds[4]}{\pM{}\Mds[]} \right.
\nonumber
\\
&& - \left. \del{ad}\del{be} \frac{4\pM{4} - \pM{2} \Mds \left( 7 - \lnmi + \lnmd \right) + 3\Mds[4]}{\pM{}\Mds[]} \right) 
\nonumber
\\
&& - \left. \frac{1}{2} \left( 2\del{ae}\del{bd} - \del{ad} \del{be} \right) \left( \DYdel\Yn \right)^{}_{\alpha i} \left( \DYn \Ydel \right)^{}_{i \beta} \left( \pM{2} - \Mds \left( 1 -\lnmi + \lnmd \right) \right) \right]  
\nonumber
\\
&& + \frac{\momps{4}}{\pM{2}-\Mds} \left[ - \left( 2\del{ae}\del{bd} - \del{ad} \del{be} \right) \lamd[2] \left( \SYn \right)^{}_{\alpha i} \left( \Yn \right)^{}_{\beta i} \frac{\pM{2} \left( 3 + 2\lnmd \right) - \Mds \left( 3 + 2\lnmi \right) }{\pM{2}}  \right.
\nonumber
\\
&& +  \frac{1}{2} \lamd \left( \DYdel \Yn \right)^{}_{\alpha i} \left( \Yn \right)^{}_{\beta i} \left( \del{ae}\del{bd} \frac{5\pM{2} \left( 1 + 2\lnmd \right) - \Mds \left( 5 + 4\lnmi + 6\lnmd \right) }{\pM{}\Mds[]} \right.
\nonumber
\\
&& - \left. 2\del{ad} \del{be} \frac{2\pM{2} \left(1 + 2\lnmd \right) - \Mds \left( 2 + \lnmi  + 3\lnmd \right)}{\pM{}\Mds[]} \right) + \frac{1}{2} \lamd \left( \SYn \right)^{}_{\alpha i} \left( \Ydel \SYn\right)^{}_{\beta i} 
\nonumber
\\
&& \times  \left( \del{ae}\del{bd} \frac{5\pM{2} \left( 3 + 2\lnmd \right) - \Mds \left( 15 + 4\lnmi + 6\lnmd \right)}{\pM{}\Mds[]} \right.
\nonumber
\\
&& - \left. 2\del{ad} \del{be} \frac{2\pM{2} \left( 3 + 2\lnmd \right) - \Mds \left( 6 + \lnmi + 3\lnmd \right)}{\pM{}\Mds[]} \right)  
\nonumber
\\
&& + \left.\left. \frac{1}{2} \left( 2\del{ae}\del{bd} - \del{ad} \del{be} \right) \left( \DYdel \Yn \right)^{}_{\alpha i} \left( \DYn \Ydel \right)^{}_{i \beta} \left( \lnmi - \lnmd\right)  \right] \right\} u \left( \momp{1} \right) \;,
\end{eqnarray}
to which all six diagrams in Fig. \ref{fig:llhh} contribute non-trivially. This differs from the case in Sec. \ref{sec:3.2.2}. Moreover, in the EFT there are six dim-6 operators (i.e., $\Op^{(1)}_{H\ell}$, $\Op^{(3)}_{H\ell}$, $\Opr^{\prime(1)}_{H\ell}$, $\Opr^{\prime\prime(1)}_{H\ell}$, $\Opr^{\prime (3)}_{H\ell}$ and $\Opr^{\prime\prime(3)}_{H\ell}$) in the Green's basis contributing to the corresponding amplitude, that is 
\begin{eqnarray}\label{eq:llhh-eft}
\amp{\rm EFT} &=& 2\rmI \momps{2} \left[ 2\del{ad}\del{be}  \left( G^{\prime (3), \rm ct }_{H\ell} \right)^{}_{\beta \alpha}  + \del{ae}\del{bd} \left( G^{\prime (1),\rm ct }_{H\ell} - G^{\prime (3),\rm ct}_{H\ell} \right)^{}_{\beta\alpha} \right] + \rmI \momps{3} \left[ 2\del{ad}\del{be} \left( G^{(3) ,\rm ct}_{H\ell} - G^{\prime(3),\rm ct}_{H \ell} \right.\right.
\nonumber
\\
&& - \left.\left. \rmI G^{\prime\prime (3) ,\rm ct}_{H\ell} \right)^{}_{\beta\alpha} + \del{ae} \del{bd} \left( G^{(1) ,\rm ct}_{H\ell} - G^{(3),\rm ct}_{H\ell} - G^{\prime (1),\rm ct}_{H\ell} + G^{\prime(3),\rm ct}_{H\ell} - \rmI G^{\prime\prime (1),\rm ct}_{H\ell} + \rmI G^{\prime\prime (3),\rm ct}_{H\ell} \right)^{}_{\beta\alpha} \right]
\nonumber
\\
&& + \rmI \momps{4} \left[ 2\del{ad}\del{be} \left( G^{(3),\rm ct}_{H\ell} + G^{\prime (3),\rm ct}_{H\ell} + \rmI G^{\prime\prime (3),\rm ct}_{H\ell} \right)^{}_{\beta\alpha} + \del{ae}\del{bd} \left( G^{(1),\rm ct}_{H\ell} - G^{(3),\rm ct}_{H\ell} + G^{\prime (1) ,\rm ct}_{H\ell} \right.\right.
\nonumber
\\
&& - \left.\left. G^{\prime(3) ,\rm ct}_{H\ell} + \rmI G^{\prime\prime (1),\rm ct}_{H\ell} - \rmI G^{\prime\prime (3),\rm ct}_{H\ell} \right)^{}_{\beta\alpha} \right] \;.
\end{eqnarray}
Equating the results in Eqs. \eqref{eq:llhh-eft} and \eqref{eq:llhh-uv} and solving the corresponding linear equations, one can find that the cross contributions to the Wilson coefficients of the six operators in the Green's basis are given by
\begin{eqnarray}
\left( G^{(1),\rm ct}_{H\ell} \right)^{}_{\alpha\beta} &=& \frac{1}{\left( 4\pi \right)^2}\left\{ - \frac{3}{2} \lamd[2] \Yni{i} \DYni{i} \frac{\pM{2} - \Mds \left( 1 - \lnmi + \lnmd \right)}{\left( \pM{2} - \Mds \right)^2} + \frac{3}{4} \lamd \left[ \Yni{i} \left( \DYdel \Yn \right)^{}_{\beta i} \right.\right.
\nonumber
\\
&& + \left. \left( \Ydel \SYn \right)^{}_{\alpha i} \left(\SYn\right)^{}_{\beta i} \right]  \frac{\pM{4} \left( 3 + 2\lnmd \right) - \pM{2}\Mds \left( 5 + 4\lnmd \right) + \Mds[4] \left( 2 + \lnmi + \lnmd \right)}{\pM{}\Mds[]\left(\pM{2} - \Mds \right)^2}  
\nonumber
\\
&& -  \left. \frac{3}{8} \left( \Ydel \SYn \right)^{}_{\alpha i} \left( \DYdel \Yn \right)^{}_{\beta i} \frac{\pM{2} \left( 1 - \lnmi + \lnmd \right) - \Mds \left( 1 -2\lnmi + 2\lnmd \right) }{\left( \pM{2} - \Mds \right)^2}  \right\}\;,
\nonumber
\\
\left( G^{\prime (1),\rm ct}_{H\ell} \right)^{}_{\alpha\beta} &=& \frac{1}{\left( 4\pi \right)^2} \left\{- \frac{2\pM{4} \left( 1 + \lnmd \right) - \pM{2}\Mds \left( 5 + 3\lnmi + \lnmd \right) + \Mds[4] \left( 3 + 2\lnmi \right) }{\pM{2}\left( \pM{2} - \Mds \right)^2}  \right.
\nonumber
\\
&& \times \frac{3}{2} \lamd[2] \Yni{i} \DYni{i} + \frac{3}{4} \lamd \left[ \Yni{i} \left( \DYdel \Yn \right)^{}_{\beta i} + \left( \Ydel \SYn \right)^{}_{\alpha i} \left(\SYn\right)^{}_{\beta i} \right]
\nonumber
\\
&& \times \frac{\pM{4} \left( 1 + 2\lnmd \right) - \pM{2}\Mds \left( 3 + 2\lnmi + 2\lnmd \right) + \Mds[4] \left( 2 + \lnmi + \lnmd \right)}{\pM{}\Mds[]\left(\pM{2} - \Mds \right)^2} 
\nonumber
\\
&& + \left. \frac{3}{8} \left( \Ydel \SYn \right)^{}_{\alpha i} \left( \DYdel \Yn \right)^{}_{\beta i}  \frac{\pM{2} \left( 1 + \lnmi - \lnmd \right) - \Mds  }{\left( \pM{2} - \Mds \right)^2} \right\} \;,
\nonumber
\\
\left( G^{\prime\prime(1),\rm ct}_{H\ell} \right)^{}_{\alpha\beta} &=& \frac{3\rmI \lamd}{2\left( 4\pi \right)^2} \frac{1}{\pM{}\Mds[]} \left[ \Yni{i} \left( \DYdel \Yn \right)^{}_{\beta i} - \left( \Ydel \SYn \right)^{}_{\alpha i} \left(\SYn\right)^{}_{\beta i} \right] \;,
\nonumber
\\
\left( G^{(3),\rm ct}_{H\ell} \right)^{}_{\alpha\beta} &=&  \frac{1}{\left(4\pi\right)^2} \left\{ \frac{1}{2} \lamd[2] \Yni{i} \DYni{i} \frac{\pM{2} - \Mds \left( 1 - \lnmi + \lnmd \right)}{\left( \pM{2} - \Mds \right)^2} - \frac{1}{4} \lamd \left[ \Yni{i} \left( \DYdel \Yn \right)^{}_{\beta i} \right.\right.
\nonumber
\\
&& + \left. \left( \Ydel \SYn \right)^{}_{\alpha i} \left(\SYn\right)^{}_{\beta i} \right] \frac{2\pM{4} \left( 3 + 2\lnmd \right) - \pM{2}\Mds \left( 11 + 8 \lnmd \right) + \Mds[4] \left( 5 + \lnmi + 3\lnmd \right)}{\pM{}\Mds[]\left(\pM{2} - \Mds \right)^2} 
\nonumber
\\
&& + \left. \frac{1}{8} \left( \Ydel \SYn \right)^{}_{\alpha i} \left( \DYdel \Yn \right)^{}_{\beta i}  \frac{\pM{2} \left( 1 - \lnmi + \lnmd \right) - \Mds \left( 1 -2\lnmi + 2\lnmd \right) }{\left( \pM{2} - \Mds \right)^2} \right\}\ \;,
\nonumber
\\
\left( G^{\prime (3),\rm ct}_{H\ell} \right)^{}_{\alpha\beta} &=&  \frac{1}{\left(4\pi\right)^2} \left\{ \frac{2\pM{4} \left( 1 + \lnmd \right) - \pM{2}\Mds \left( 5 + 3\lnmi + \lnmd \right) + \Mds[4] \left( 3 + 2\lnmi \right) }{\pM{2}\left( \pM{2} - \Mds \right)^2}  \right.
\nonumber
\\
&& \times \frac{1}{2} \lamd[2] \Yni{i} \DYni{i} - \frac{1}{4} \lamd \left[ \Yni{i} \left( \DYdel \Yn \right)^{}_{\beta i} + \left( \Ydel \SYn \right)^{}_{\alpha i} \left(\SYn\right)^{}_{\beta i} \right]
\nonumber
\\
&& \times \frac{2\pM{4} \left( 1 + 2\lnmd \right) - \pM{2}\Mds \left( 5 + 2\lnmi + 6\lnmd \right) + \Mds[4] \left( 3 + \lnmi + 3\lnmd \right)}{\pM{}\Mds[]\left(\pM{2} - \Mds \right)^2} 
\nonumber
\\
&& - \left. \frac{1}{8} \left( \Ydel \SYn \right)^{}_{\alpha i} \left( \DYdel \Yn \right)^{}_{\beta i}  \frac{\pM{2} \left( 1 + \lnmi - \lnmd \right) - \Mds  }{\left( \pM{2} - \Mds \right)^2}  \right\} \;,
\nonumber
\\
\left( G^{\prime\prime(3),\rm ct}_{H\ell} \right)^{}_{\alpha\beta} &=& - \frac{\rmI \lamd}{\left( 4\pi \right)^2} \frac{1}{\pM{}\Mds[]} \left[ \Yni{i} \left( \DYdel \Yn \right)^{}_{\beta i} - \left( \Ydel \SYn \right)^{}_{\alpha i} \left(\SYn\right)^{}_{\beta i} \right] \;.
\end{eqnarray}

\subsubsection{$\overline{\ell} \overline{\ell} \ell\ell$}

\begin{figure}
\centering
\includegraphics[width=0.75\linewidth]{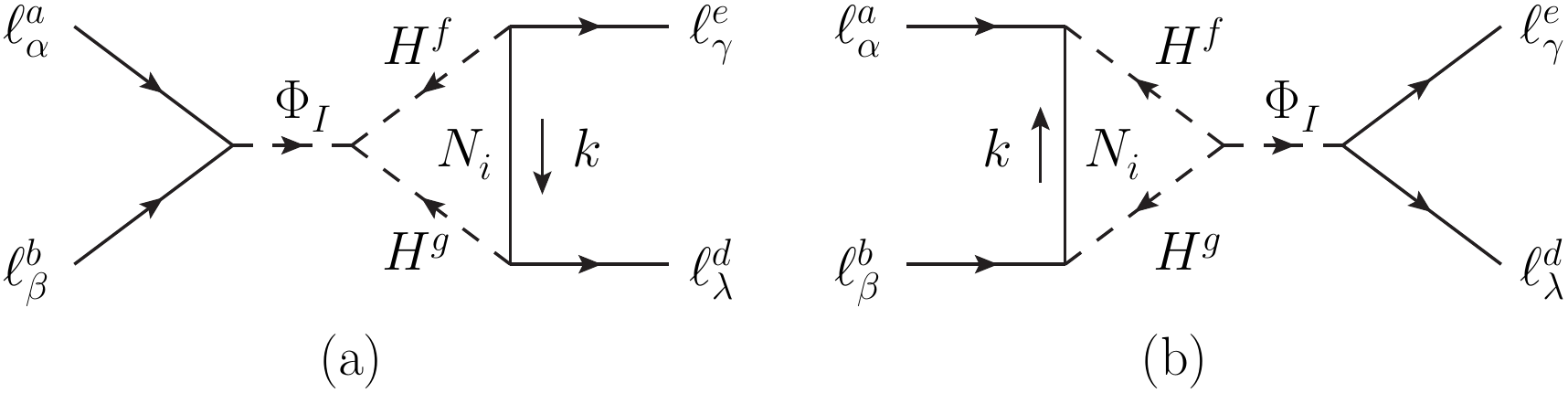}
\caption{The 1LPI diagrams with external lines generated by the $\left( \overline{\ell} \overline{\ell} \ell\ell \right)$ field configuration in the type-(I+II) seesaw mechanism.}
\label{fig:llll}
\end{figure}

Fig. \ref{fig:llll} shows the two 1LPI diagrams with external particles governed by the $\left( \overline{\ell} \overline{\ell} \ell \ell \right)$ field configuration in the UV model. The associated amplitudes for these two diagrams are given in Eq. \eqref{eq:llll-amp}, and the hard-momentum part of the total amplitude is found to be
\begin{eqnarray}\label{eq:llll-uv}
\ampd{\rm UV} &=& \frac{\rmI}{\left( 4\pi \right)^2} \left( \del{ae} \del{bd} + \del{ad} \del{be} \right) \lamd \left[  \left( \DYdel \right)^{}_{\alpha\beta} \Yni[\gamma]{i} \Yni[\lambda]{i} + \left( \SYn \right)^{}_{\alpha i} \left( \SYn \right)^{}_{\beta i} \left( \Ydel \right)^{}_{\lambda\gamma} \right] \frac{1+\lnmi}{\pM{}\Mds[]} 
\nonumber
\\
&& \times \overline{u} \left(0\right) \gamma^\mu \pl u \left( 0 \right) \overline{u} \left(0\right) \gamma^{}_\mu \pl u \left( 0 \right)  \;,
\end{eqnarray}
where Fierz transformations and $u = {\sf C} \overline{v}^{\rm T}$, $v = {\sf C} \overline{u}^{\rm T}$ have been used. Here we have taken momenta of all external particles to be zero, since the $\left( \overline{\ell} \overline{\ell} \ell \ell \right)$ field configuration already has a mass dimension of six. In the EFT, only one operator, i.e., $\Op^{}_{\ell\ell}$ in the Green's basis takes effect, and its contribution is given by
\begin{eqnarray}\label{eq:llll-eft}
\amp{\rm EFT} = 2 \rmI \overline{u} \left(0\right) \gamma^\mu \pl u \left( 0 \right) \overline{u} \left(0\right) \gamma^{}_\mu \pl u \left( 0 \right) \left[ \del{ae}\del{bd} \left( G^{\rm ct}_{\ell \ell} \right)^{}_{\gamma\alpha\lambda\beta} + \del{ad} \del{be} \left( G^{\rm ct}_{\ell \ell} \right)^{}_{\lambda\alpha \gamma\beta} \right] \;.
\end{eqnarray}
Then with Eqs. \eqref{eq:llll-uv} and \eqref{eq:llll-eft}, we have
\begin{eqnarray}
\left( G^{\rm ct}_{\ell\ell} \right)^{}_{\alpha\beta\gamma\lambda} &=& \frac{1}{2\left( 4\pi \right)^2} \lamd \left[ \left( \DYdel \right)^{}_{\beta\lambda} \Yni[\alpha]{i} \Yni[\gamma]{i}  + \left( \Ydel \right)^{}_{\alpha\gamma} \left( \SYn \right)^{}_{\beta i} \left( \SYn \right)^{}_{\lambda i} \right] \frac{1+\lnmi}{\pM{}\Mds[]} \;.
\end{eqnarray}

\subsubsection{$\overline{\ell} E H^\dagger HH$}

\begin{figure}
\centering
\includegraphics[width=1\linewidth]{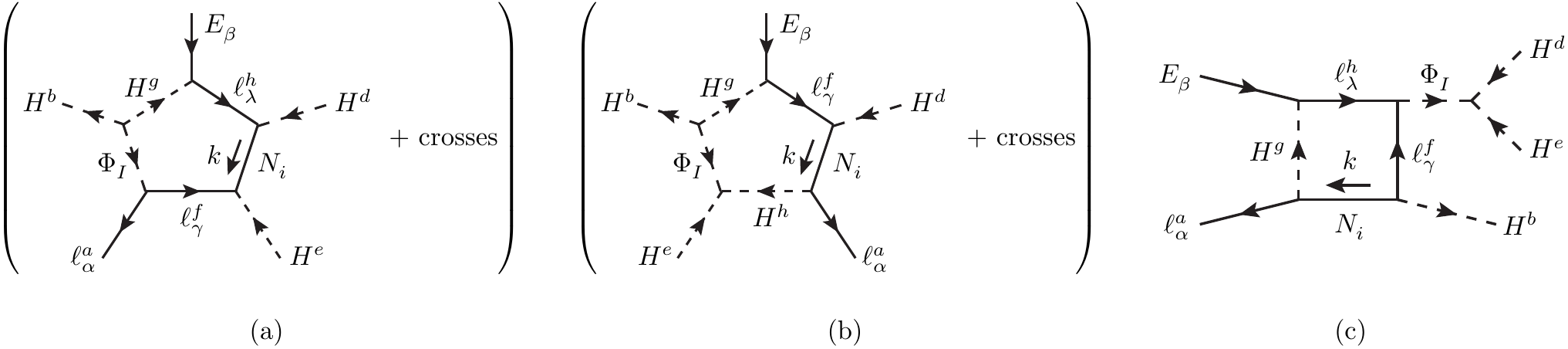}
\caption{The 1LPI diagrams with external lines generated by the $\left( \overline{\ell} E H^\dagger H H \right)$ field configuration in the type-(I+II) seesaw mechanism.}
\label{fig:hhhle}
\end{figure}

There are five 1LPI diagrams with external lines generated by the $\left( \overline{\ell} E H^\dagger H H \right)$ field configuration in the UV model, but two of them can be easily obtained by exchanging two external lines (i.e., $H^d$ and $H^e$) of the first two diagrams in Fig. \ref{fig:hhhle} and are not explicitly plotted. The corresponding amplitudes of these diagrams in Fig.~\ref{fig:hhhle} with all external momenta being vanishing are given in Eq. \eqref{eq:lehhh-amp}. Then, one can obtain the hard-momentum part of the total amplitude:
\begin{eqnarray}\label{eq:lehhh-uv}
\ampd{\rm UV} &=& \frac{2\rmI \lamd}{\left( 4\pi \right)^2} \left( \del{ad} \del{be} + \del{ae} \del{bd} \right) \overline{u} \left( 0 \right) \pr u \left( 0 \right) \left\{  \left[ 2\lamd \frac{\Mds[]}{\pM{}} \Yni{i} \left( \DYn \Yl \right)^{}_{i\beta}  -  \left( \Ydel \SYn \right)^{}_{\alpha i} \left( \DYn \Yl \right)^{}_{i \beta} \right]\right.
\nonumber
\\
&& \times \left. \frac{\pM{2} \left( 1 + \lnmd \right) - \Mds \left( 1 + \lnmi \right)}{\pM{}\Mds[] \left( \pM{2} - \Mds \right)} - \Yni{i} \left( \TYn \DYdel \Yl \right)^{}_{i\beta} \frac{1+\lnmi}{\pM{}\Mds[]}  \right\} \;.
\end{eqnarray}
In the EFT, only the operator $\Op^{}_{eH}$ contributes and the corresponding result is found to be
\begin{eqnarray}\label{eq:lehhh-eft}
\amp{\rm EFT} &=& \rmI \left( \del{ad} \del{be} + \del{ae} \del{bd} \right) \overline{u} \left( 0 \right) \pr u \left( 0 \right) \left( G^{\rm ct}_{eH} \right)^{}_{\alpha\beta} \;.
\end{eqnarray}
Equating the result in Eq. \eqref{eq:lehhh-eft} with that in Eq. \eqref{eq:lehhh-uv}, one can obtain the cross contribution to the Wilson coefficient of $\Op^{}_{eH}$ in the Green's basis, namely, 
\begin{eqnarray}
\left( G^{\rm ct}_{eH} \right)^{}_{\alpha\beta} &=& \frac{2\lamd}{\left( 4\pi \right)^2} \left\{  \left[ \frac{2\lamd}{\pM{2}} \Yni{i} \left( \DYn \Yl \right)^{}_{i\beta}  - \frac{1}{\pM{}\Mds[]} \left( \Ydel \SYn \right)^{}_{\alpha i} \left( \DYn \Yl \right)^{}_{i \beta} \right]\right.
\nonumber
\\
&& \times \left.  \frac{\pM{2} \left( 1 + \lnmd \right) - \Mds \left( 1 + \lnmi \right)}{\pM{2} - \Mds} - \Yni{i} \left( \TYn \DYdel \Yl \right)^{}_{i\beta} \frac{1+\lnmi}{\pM{}\Mds[]}  \right\} \;.
\end{eqnarray}

\subsubsection{$H^\dagger H^\dagger H^\dagger HHH$}

\begin{figure}[!t]
\centering
\includegraphics[width=1\linewidth]{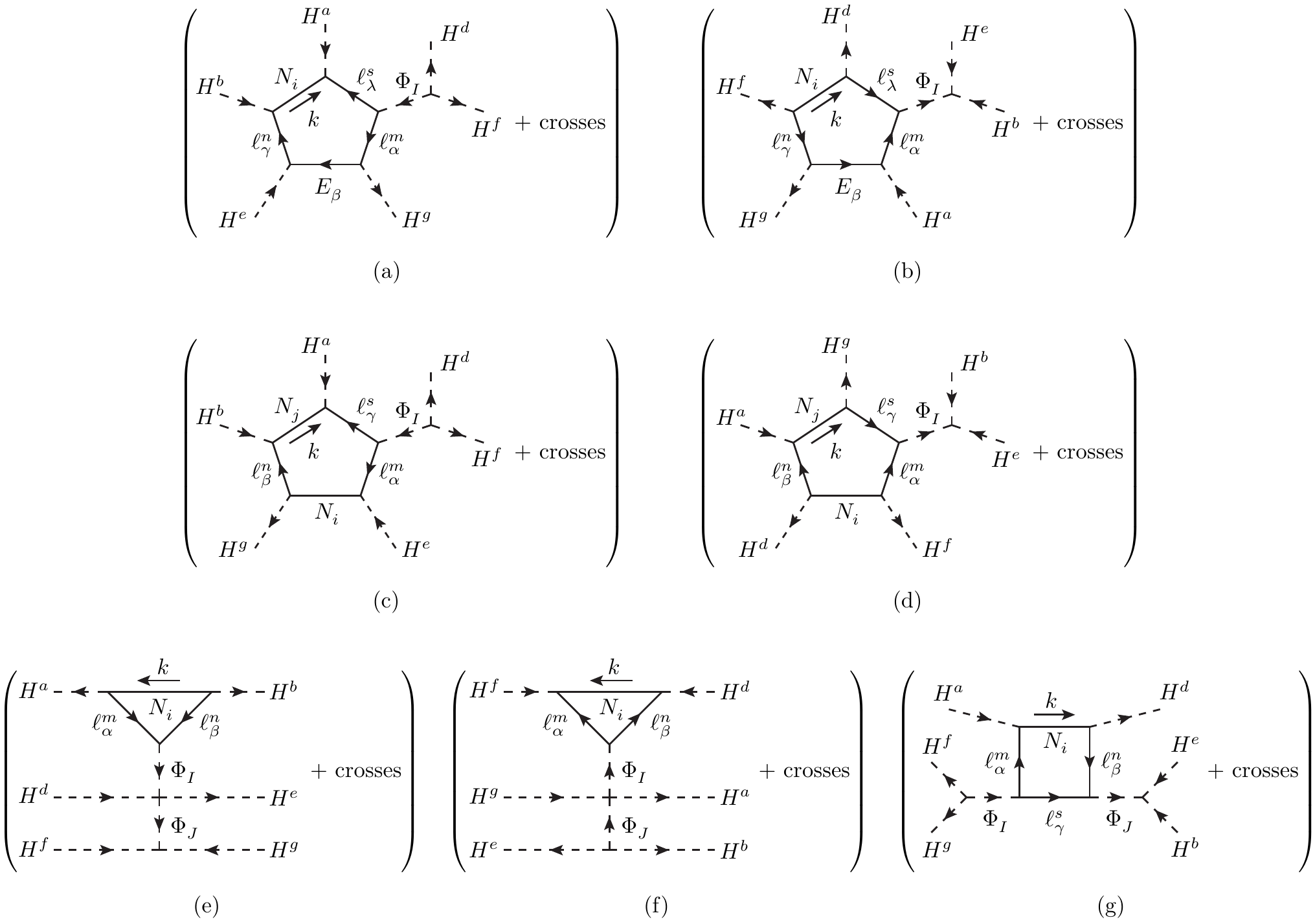}
\caption{The 1LPI diagrams with external lines generated by the $\left( H^\dagger H^\dagger H^\dagger H H H \right)$ field configuration in the type-(I+II) seesaw mechanism.}
\label{fig:hhhhhh}
\end{figure}

There are ninety-nine 1LPI diagrams with external particles given by the $\left( H^\dagger H^\dagger H^\dagger H H H \right) $ field configuration in total. We only plot seven of them explicitly in Fig. \ref{fig:hhhhhh}, where all other diagrams with crossed external lines are not shown. All amplitudes associated with these diagrams in Fig.~\ref{fig:hhhhhh} with all external momenta being zero are given in Eq. \eqref{eq:hhhhhh-amp}. Derived from Eq. \eqref{eq:hhhhhh-amp}, the hard-momentum part of the total amplitude is
\begin{eqnarray}\label{eq:hhhhhh-uv}
\ampd{\rm UV} &=& \frac{\rmI}{\left( 4\pi \right)^2} \left[ \del{ad} \left( \del{bg} \del{ef} + \del{bf} \del{eg} \right) + \del{af} \left( \del{bg} \del{ed} + \del{bd} \del{eg} \right) + \del{ag} \left( \del{bf} \del{ed} + \del{bd} \del{ef} \right) \right]
\nonumber
\\
&& \times 12 \lamd \left\{ \left(\lambda^{}_3 - \lambda^{}_4 \right) \left( \DYn \Ydel \SYn + \TYn \DYdel \Yn \right)^{}_{ii} \frac{\pM{} \left( 1 + \lnmi \right)}{\Mds[3]} - 4\lamd \left( \DYn \Ydel \DYdel \Yn \right)^{}_{ii} \frac{1+\lnmi}{\Mds[]} \right.
\nonumber
\\
&& \left. - 2 \left[ \left( \DYn\Yn \right)^{}_{ij} \left( \TYn \DYdel \Yn \right)^{}_{ij} + \left( \TYn \SYn \right)^{}_{ij} \left( \DYn \Ydel \SYn \right)^{}_{ij} \right] \frac{\pM[j]{} \left( \lnmi -\lnmj \right) }{\Mds[] \left( \pM{2} - \pM[j]{2} \right) } \right\} \;.
\end{eqnarray}
On the other hand, there is only one operator $\Op^{}_H$ in the Green's basis contributing to the corresponding amplitude in the EFT, whose contribution is given by
\begin{eqnarray}\label{eq:hhhhhh-eft}
\amp{\rm EFT} = 6 \rmI \left[ \del{ad} \left( \del{bg} \del{ef} + \del{bf} \del{eg} \right) + \del{af} \left( \del{bg} \del{ed} + \del{bd} \del{eg} \right) + \del{ag} \left( \del{bf} \del{ed} + \del{bd} \del{ef} \right) \right] G^{\rm ct}_{H} \;.
\end{eqnarray}
With the help of Eqs. \eqref{eq:hhhhhh-uv} and \eqref{eq:hhhhhh-eft}, one can easily achieve the cross contribution to the Wilson coefficient of $\Op^{}_H$ in the Green's basis, i.e.,
\begin{eqnarray}
G^{\rm ct}_H &=& \frac{2 \lamd}{\left( 4\pi \right)^2} \left\{ \left(\lambda^{}_3 - \lambda^{}_4 \right) \left( \DYn \Ydel \SYn + \TYn \DYdel \Yn \right)^{}_{ii} \frac{\pM{} \left( 1 + \lnmi \right)}{\Mds[3]} - 4\lamd \left( \DYn \Ydel \DYdel \Yn \right)^{}_{ii} \frac{1+\lnmi}{\Mds[]} \right.
\nonumber
\\
&& \left. - 2 \left[ \left( \DYn\Yn \right)^{}_{ij} \left( \TYn \DYdel \Yn \right)^{}_{ij} + \left( \TYn \SYn \right)^{}_{ij} \left( \DYn \Ydel \SYn \right)^{}_{ij} \right] \frac{\pM[j]{} \left( \lnmi -\lnmj \right) }{\Mds[] \left( \pM{2} - \pM[j]{2} \right) } \right\} \;.
\end{eqnarray}

The one-loop matching in the Green's basis only results in the first type of cross contributions (namely, induced by the one-loop diagrams with both $N^{}_{\rm R}$ and $\Phi$). In addition to the Wilson coefficients of the dim-6 operators $\{ \Op^{}_{H\square}, \Op^{}_{HD}, \Op^{(1)}_{H\ell}, \Op^{(3)}_{H\ell}, \Op^{}_{\ell\ell}, \Op^{}_{eH}, \Op^{}_H, \Opr^\prime_{HD}, \Opr^{\prime\prime}_{HD}, \Opr^{\prime(1)}_{H\ell}, \Opr^{\prime\prime(1)}_{H\ell}, \Opr^{\prime(3)}_{H\ell}, \\ \Opr^{\prime\prime(3)}_{H\ell} \}$ in the Green's basis, the Wilson coefficient of the Weinberg operator $\Op^{(5)}$ and the quartic coupling constant of the SM Higgs also acquire some cross contributions. It is worth pointing out that so far all results and discussions are given in the Green's basis. One needs to remove the redundant operators in the Green's basis and convert all results into those in the Warsaw basis via field redefinitions (or EOMs). During this procedure, additional cross contributions in the second type appear and the tree-level results in Eq. \eqref{eq:tree-level} also play an important role.

\section{Results in the Warsaw basis}\label{sec:warsaw}

For the type-(I+II) seesaw mechanism, the one-loop threshold corrections to the renormalizable terms in the Green's basis are given by
\begin{eqnarray}\label{eq:threshold-1}
\delta \mathcal{L}^{\rm G} &=& \delta Z^{\rm G}_W W^I_{\mu\nu} W^{I\mu\nu} + \delta Z^{\rm G}_B B^{}_{\mu\nu} B^{\mu\nu} + \overline{\ell^{}_{\rm L}} \delta Z^{\rm G}_\ell \rmI \slashed{D} \ell^{}_{\rm L} + \delta Z^{\rm G}_H \left( D^{}_\mu H \right)^\dagger \left( D^\mu H \right) 
\nonumber
\\
&& + \overline{\ell^{}_{\rm L}} \delta Y^{\rm G}_l H E^{}_{\rm R} + \left( \delta m^2 \right)^{\rm G} H^\dagger H + \delta \lambda^{\rm G} \left( H^\dagger H \right)^2 \;,
\end{eqnarray}
where
\begin{eqnarray}\label{eq:threshold-2}
&&\delta Z^{\rm G}_W = \delta Z^{\rm G,II}_W \;,\quad \delta Z^{\rm G}_B = \delta Z^{\rm G,II}_B \;,\quad \delta Z^{\rm G}_\ell = \delta Z^{\rm G,I}_\ell + \delta Z^{\rm G,II}_\ell \;,\quad \delta Z^{\rm G}_H = \delta Z^{\rm G,I}_H + \delta Z^{\rm G,II}_H \;,
\nonumber
\\
&&\delta Y^{\rm G}_l = \delta Y^{\rm G, I}_l \;,\quad \left( \delta m^2 \right)^{\rm G} = \left( \delta m^2 \right)^{\rm G,I} + \left( \delta m^2 \right)^{\rm G,II}  \;,\quad \delta \lambda^{\rm G} = \delta \lambda^{\rm G,I} + \delta \lambda^{\rm G,II} + \delta \lambda^{\rm G,ct} \;,
\end{eqnarray} 
and other renormalizable terms do not acquire any threshold corrections in the Green's basis. As can be seen from Eq. \eqref{eq:threshold-2}, only the threshold correction to the quartic coupling constant of the SM Higgs, i.e., $\delta \lambda^{\rm G}$ acquires the cross contribution from non-linearly combined effects of $N^{}_{\rm R}$ and $\Phi$. The contributions from $N^{}_{\rm R}$ and $\Phi$ separately in Eq. \eqref{eq:threshold-2} can be found in Ref. \cite{Zhang:2021jdf} and in Ref. \cite{Li:2022ipc}, respectively, and are not repeated here. 

To eliminate the redundant operators in the Green's basis and normalize the kinetic terms of lepton and Higgs doublets, and electroweak gauge bosons, one should redefine the relevant fields instead of applying the EOMs of fields directly in principle. As pointed out in Ref. \cite{Criado:2018sdb}, imposing EOMs is not the same as performing field redefinitions since the EOMs only capture the first-order response of the action to variations of the fields and can not recover all higher-order corrections caused by the field redefinitions. Thus, naively using the EOMs even with higher-order corrections gives in general an action that is not equivalent to the original one at the second and higher orders. However, in practice, one may impose the EOMs first and then restore the second-order (or higher-order) corrections from the second-order (or higher-order) response of the action to variations of the relevant fields. This is exactly what has been done in Ref. \cite{Li:2022ipc}. This recipe is practically simple and effective for the issue we are considering, since such higher-order contributions missed by the EOMs are very few and easy to be found. Therefore, following Ref. \cite{Li:2022ipc}, we adopt such a recipe to remove the redundant operators.

Beginning with Eqs. \eqref{eq:tree-level} and \eqref{eq:threshold-1}, one can obtain the EOMs of the relevant SM fields, namely,
\begin{eqnarray}\label{eq:eom}
\rmI \slashed{D} E^{}_{\rm R} &=& H^\dagger \left( Y^\dagger_l - \delta Y^{\rm G \dagger}_l \right) \ell^{}_{\rm L} \;,
\nonumber
\\
\rmI \slashed{D} \ell^{}_{\rm L} &=& \left( 1 - \delta Z^{\rm G}_\ell \right) \left( \Yl - \delta Y^{\rm G}_l \right) H E^{}_{\rm R} \;,
\nonumber
\\
D^\nu B^{}_{\mu\nu} &=& \frac{1}{2} g^{}_1 \left[ \left( 1 + \delta Z^{\rm G}_H + 4 \delta Z^{\rm G}_B \right) H^\dagger \rmI \Dlr H + 2 \left( 1 + 4\delta Z^{\rm G}_B \right) \sum^{}_f Y \left(f \right) \overline{f} \gamma^{}_\mu f \right] \;,
\nonumber
\\
\left( D^\nu W^{}_{\mu\nu} \right)^I &=& \frac{1}{2} g^{}_2 \left[ \left( 1 + \delta Z^{\rm G}_H + 4 \delta Z^{\rm G}_W \right) H^\dagger \rmI \Dilr H + \left( 1 + 4\delta Z^{\rm G}_W \right) \left( \overline{Q^{}_{\rm L}} \sigma^I \gamma^{}_\mu Q^{}_{\rm L} + \overline{\ell^{}_{\rm L}} \sigma^I \gamma^{}_\mu \ell^{}_{\rm L} \right) \right] \;,
\nonumber
\\
\left( D^2 H \right)^a &=& - \left[ m^2 - \left( \delta m^2 \right)^{\rm G} - m^2 \delta Z^{\rm G}_H \right] H^a - 2 \left[ \left( \lambda - 2\lamd[2] \right) \left( 1 - \delta Z^{\rm G}_H \right) - \delta \lambda^{\rm G} \right] \left( H^\dagger H \right) H^a 
\nonumber
\\
&& - \left( 1 - \delta Z^{\rm G}_H \right) \left[ \overline{E^{}_{\rm R}} \left( \DYl - \delta Y^{{\rm G}\dagger}_l \right) \ell^a_{\rm L} + \overline{D^{}_{\rm R}} \DYd Q^a_{\rm L} - \epsilon^{}_{ab} \overline{Q^b_{\rm L}} \Yu U^{}_{\rm R} \right] \;,
\end{eqnarray}
where $Y \left( f \right)$ is the hypercharge for the fermionic fields $f=Q^{}_{\rm L}, U^{}_{\rm R}, D^{}_{\rm R}, \ell^{}_{\rm L}, E^{}_{\rm R}$, and $I=1,2,3$ and $a,b=1,2$ are the SU(2) adjoint and weak isospin indices, respectively. In Eq. \eqref{eq:eom}, we have ignored the contributions from dim-5 and dim-6 operators, which are suppressed by the masses of $N^{}_{\rm R}$ or $\Phi$ and only result in operators with dimension higher than six. The implementation of the EOMs in Eq. \eqref{eq:eom} can remove all redundant operators but misses a contribution to the Wilson coefficient of $\Op^{}_H$ from the second derivative of the action with respect to the SM Higgs field, i.e.,
\begin{eqnarray}\label{eq:add-con}
\frac{1}{4} \left( G^{\prime}_{HD} |^{}_{\rm tree} \right)^2 \left( \delta m^2 \right)^{\rm G} \Op^{}_H = \frac{1}{4} \left[  \left( G^{\prime,\rm II}_{HD} |^{}_{\rm tree} \right)^2 \left( \delta m^2 \right)^{\rm G,I} + \left( G^{\prime,\rm II}_{HD} |^{}_{\rm tree} \right)^2 \left( \delta m^2 \right)^{\rm G,II} \right] \Op^{}_H \;,
\end{eqnarray}
which is induced by the field redefinition to remove the tree-level redundant operator $\Opr^\prime_{HD}$. In Eq. \eqref{eq:add-con}, the first term in square brackets is new and indeed a cross contribution, and the second term has already been obtained in the one-loop matching procedure for the type-II seesaw mechanism \cite{Li:2022ipc}. Such a second type of cross correction will be added into the final result for the Wilson coefficient of $\Op^{}_H$. On the other hand, the kinetic terms of gauge bosons $W^I_\mu,B^{}_\mu$, and the $\rm SU(2)^{}_L$ doublets $H$ and $\ell^{}_{\rm L}$ need to be normalized by
\begin{eqnarray}\label{eq:redefinition}
&&
\begin{cases}
B^{}_\mu \to \left( 1 + 2\delta Z^{\rm G}_B \right) B^{}_\mu
\\
g^{}_1 \to g^{\rm eff}_1 = \left( 1 - 2\delta Z^{\rm G}_B \right) g^{}_1
\end{cases} \;,\qquad
\begin{cases}
W^{I}_\mu \to \left( 1 + 2\delta Z^{\rm G}_W \right) W^{I}_\mu
\\
g^{}_2 \to g^{\rm eff}_2 = \left( 1 - 2\delta Z^{\rm G}_W \right) g^{}_2
\end{cases} \;,
\nonumber
\\
&& H \to \left( 1 - \frac{1}{2} \delta Z^{\rm G}_H \right) H \;,\qquad \ell^{}_{\rm L} \to \left( 1 - \frac{1}{2} \delta Z^{\rm G}_\ell \right) \ell^{}_{\rm L} \;.
\end{eqnarray}
The combined redefinitions of gauge fields and the associated gauge couplings aim at keeping the canonical form of the covariant derivative $D^{}_\mu$, and only induce the threshold corrections to the gauge couplings, which are exactly the same as those in the type-II SEFT \cite{Li:2022ipc}. The redefinitions of $H$ and $\ell^{}_{\rm L}$ in Eq. \eqref{eq:redefinition} can lead to additional cross contributions via the tree-level Wilson coefficients of some relevant operators and as well the tree-level SM terms.

With the help of Eqs. \eqref{eq:eom}---\eqref{eq:redefinition}, one can remove all redundant operators and normalize the kinetic terms, and finally gains all results in the Warsaw basis. The complete effective Lagrangian of the type-(I+II) SEFT up to dimension six and one-loop level is given by
\begin{eqnarray}\label{eq:full-eft}
\mathcal{L}^{\rm I+II}_{\rm SEFT} &=& \mathcal{L}^{}_{\rm SM} \left( m^2 \to m^2_{\rm eff}, \lambda \to \lambda^{}_{\rm eff}, Y^{}_l \to Y^{\rm eff}_l, Y^{}_{\rm u} \to Y^{\rm eff}_{\rm u}, Y^{}_{\rm d} \to Y^{\rm eff}_{\rm d}, g^{}_1 \to g^{\rm eff}_1, g^{}_2 \to g^{\rm eff}_2 \right)
\nonumber
\\
&& + \frac{1}{2} \left( C^{(5)\alpha\beta}_{\rm eff} \Op^{(5)}_{\alpha\beta} + {\rm h.c.} \right) + C^{}_H |^{}_{\rm tree} \Op^{}_H + C^{}_{H\square} |^{}_{\rm tree} \Op^{}_{H\square} + C^{}_{HD} |^{}_{\rm tree} \Op^{}_{HD} + C^{(1)}_{H\ell} |^{\alpha\beta}_{\rm tree} \Op^{(1)\alpha\beta}_{H\ell} 
\nonumber
\\
&& + C^{(3)}_{H\ell} |^{\alpha\beta}_{\rm tree} \Op^{(3)\alpha\beta}_{H\ell} + \left( C^{}_{eH} |^{\alpha\beta}_{\rm tree} \Op^{\alpha\beta}_{eH} + C^{}_{uH} |^{\alpha\beta}_{\rm tree} \Op^{\alpha\beta}_{uH} + C^{}_{dH} |^{\alpha\beta}_{\rm tree} \Op^{\alpha\beta}_{dH} + {\rm h.c.} \right)
\nonumber
\\
&& + C^{}_{\ell\ell} |^{\alpha\beta\gamma\lambda}_{\rm tree} \Op^{\alpha\beta\gamma\lambda}_{\ell\ell} + \sum^{}_i C^{}_i \Op^{}_i \;,
\end{eqnarray}
where $O^{}_i$ and $C^{}_i$ in the last term are the independent dim-6 operators shown in the gray region in Table \ref{tab:SMEFT-dim-6-Warsaw} (including Hermitian conjugations of the non-Hermitian operators) and the associated one-loop Wilson coefficients in the Warsaw basis, respectively. More details and discussions can be found as below.

\begin{table}[!t]
\centering 
\renewcommand\arraystretch{1.55}
\resizebox{\textwidth}{!}{ %
\begin{tabular}{c|c|c|c|c|c}
\hline\hline 
\multicolumn{2}{c|}{$X^{3}$} & \multicolumn{2}{c|}{$H^{6}$~and~$H^{4}D^{2}$} & \multicolumn{2}{c}{$\psi^{2}H^{3}$}
\tabularnewline
\hline 
\multicolumn{1}{c|}{$\Op_{3G}$} & $f^{ABC}G_{\mu}^{A\nu}G_{\nu}^{B\rho}G_{\rho}^{C\mu}$  & \cellcolor{gray!60}{$\Op_{H}$ } & \cellcolor{gray!60}{$\left(H^{\dagger}H\right)^{3}$ } & \cellcolor{gray!60}{$\Op_{eH}^{\alpha\beta}$ } & \cellcolor{gray!60}{$\left(\BLelli H\REi\right)\left(H^{\dagger}H\right)$}
\tabularnewline
\multicolumn{1}{c|}{$\Op_{\widetilde{3G}}$} & $f^{ABC}\widetilde{G}_{\mu}^{A\nu}G_{\nu}^{B\rho}G_{\rho}^{C\mu}$  & \cellcolor{gray!60}{$\Op_{H\square}$ } & \cellcolor{gray!60}{$\left(H^{\dagger}H\right)\square\left(H^{\dagger}H\right)$} & \cellcolor{gray!60}{$\Op_{uH}^{\alpha\beta}$ } & \cellcolor{gray!60}{$\left(\BLQi\widetilde{H}\RUi\right)\left(H^{\dagger}H\right)$}
\tabularnewline
\multicolumn{1}{c|}{\cellcolor{gray!20}{$\Op_{3W}$}} & \cellcolor{gray!20}{$\epsilon^{IJK}W_{\mu}^{I\nu}W_{\nu}^{J\rho}W_{\rho}^{K\mu}$} & \cellcolor{gray!60}{$\Op_{HD}$ } & \cellcolor{gray!60}{$\left(H^{\dagger}D^{\mu}H\right)^{\ast}\left(H^{\dagger}D_{\mu}H\right)$} & \cellcolor{gray!60}{$\Op_{dH}^{\alpha\beta}$ } & \cellcolor{gray!60}{$\left(\BLQi H\RDi\right)\left(H^{\dagger}H\right)$}
\tabularnewline
\multicolumn{1}{c|}{$\Op_{\widetilde{3W}}$} & $\epsilon^{IJK}\widetilde{W}_{\mu}^{I\nu}W_{\nu}^{J\rho}W_{\rho}^{K\mu}$  &  &  &  & 
\tabularnewline
\hline 
\multicolumn{2}{c|}{$X^{2}H^{2}$} & \multicolumn{2}{c|}{$\psi^{2}XH$} & \multicolumn{2}{c}{$\psi^{2}H^{2}D$}
\tabularnewline
\hline 
$\Op_{HG}$  & $G_{\mu\nu}^{A}G^{A\mu\nu}H^{\dagger}H$  & \cellcolor{gray!20}{$\Op_{eW}^{\alpha\beta}$ } & \cellcolor{gray!20}{$\left(\BLelli\sigma^{\mu\nu}\REi\right)\sigma^{I}HW_{\mu\nu}^{I}$} & \cellcolor{gray!60}{$\Op_{H\ell}^{(1)\alpha\beta}$ } & \cellcolor{gray!60}{$\left(\BLelli\gamma^{\mu}\Lelli\right)\left(H^{\dagger}\rmI\Dlr H\right)$}
\tabularnewline
$\Op_{H\widetilde{G}}$  & $\widetilde{G}_{\mu\nu}^{A}G^{A\mu\nu}H^{\dagger}H$  & \cellcolor{gray!20}{$\Op_{eB}^{\alpha\beta}$ } & \cellcolor{gray!20}{$\left(\BLelli\sigma^{\mu\nu}\REi\right)HB_{\mu\nu}$} & \cellcolor{gray!60}{$\Op_{H\ell}^{(3)\alpha\beta}$ } & \cellcolor{gray!60}{$\left(\BLelli\gamma^{\mu}\sigma^{I}\Lelli\right)\left(H^{\dagger}\rmI\Dilr H\right)$}
\tabularnewline
\cellcolor{gray!20}{$\Op_{HW}$} & \cellcolor{gray!20}{$W_{\mu\nu}^{I}W^{I\mu\nu}H^{\dagger}H$ } & $\Op_{uG}^{\alpha\beta}$  & $\left(\BLQi\sigma^{\mu\nu}T^{A}\RUi\right)\widetilde{H}G_{\mu\nu}^{A}$  & \cellcolor{gray!20}{$\Op_{He}^{\alpha\beta}$ } & \cellcolor{gray!20}{$\left(\BREi\gamma^{\mu}\REi\right)\left(H^{\dagger}\rmI\Dlr H\right)$}
\tabularnewline
$\Op_{H\widetilde{W}}$  & $\widetilde{W}_{\mu\nu}^{I}W^{I\mu\nu}H^{\dagger}H$  & $\Op_{uW}^{\alpha\beta}$  & $\left(\BLQi\sigma^{\mu\nu}\RUi\right)\sigma^{I}\widetilde{H}W_{\mu\nu}^{I}$  & \cellcolor{gray!20}{$\Op_{Hq}^{(1)\alpha\beta}$ } & \cellcolor{gray!20}{$\left(\BLQi\gamma^{\mu}\LQi\right)\left(H^{\dagger}\rmI\Dlr H\right)$}
\tabularnewline
\cellcolor{gray!20}{$\Op_{HB}$} & \cellcolor{gray!20}{$B_{\mu\nu}B^{\mu\nu}H^{\dagger}H$ } & $\Op_{uB}^{\alpha\beta}$  & $\left(\BLQi\sigma^{\mu\nu}\RUi\right)\widetilde{H}B_{\mu\nu}$  & \cellcolor{gray!20}{$\Op_{Hq}^{(3)\alpha\beta}$ } & \cellcolor{gray!20}{$\left(\BLQi\gamma^{\mu}\sigma^{I}\LQi\right)\left(H^{\dagger}\rmI\Dilr H\right)$}
\tabularnewline
$\Op_{H\widetilde{B}}$  & $\widetilde{B}_{\mu\nu}B^{\mu\nu}H^{\dagger}H$  & $\Op_{dG}^{\alpha\beta}$  & $\left(\BLQi\sigma^{\mu\nu}T^{A}\RDi\right)HG_{\mu\nu}^{A}$  & \cellcolor{gray!20}{$\Op_{Hu}^{\alpha\beta}$ } & \cellcolor{gray!20}{$\left(\BRUi\gamma^{\mu}\RUi\right)\left(H^{\dagger}\rmI\Dlr H\right)$}
\tabularnewline
\cellcolor{gray!20}{$\Op_{HWB}$} & \cellcolor{gray!20}{$W_{\mu\nu}^{I}B^{\mu\nu}\left(H^{\dagger}\sigma^{I}H\right)$} & $\Op_{dW}^{\alpha\beta}$  & $\left(\BLQi\sigma^{\mu\nu}\RDi\right)\sigma^{I}HW_{\mu\nu}^{I}$  & \cellcolor{gray!20}{$\Op_{Hd}^{\alpha\beta}$ } & \cellcolor{gray!20}{$\left(\BRDi\gamma^{\mu}\RDi\right)\left(H^{\dagger}\rmI\Dlr H\right)$}
\tabularnewline
$\Op_{H\widetilde{W}B}$  & $\widetilde{W}_{\mu\nu}^{I}B^{\mu\nu}\left(H^{\dagger}\sigma^{I}H\right)$  & $\Op_{dB}^{\alpha\beta}$  & $\left(\BLQi\sigma^{\mu\nu}\RDi\right)HB_{\mu\nu}$  & $\Op_{Hud}^{\alpha\beta}$  & $\rmI\left(\BRUi\gamma^{\mu}\RDi\right)\left(\widetilde{H}^{\dagger}D_{\mu}H\right)$
\tabularnewline
\hline 
\multicolumn{2}{c|}{$\rm \left(\overline{L}L\right)\left(\overline{L}L\right)$} & \multicolumn{2}{c|}{$\rm \left(\overline{R}R\right)\left(\overline{R}R\right)$} & \multicolumn{2}{c}{$\rm \left(\overline{L}L\right)\left(\overline{R}R\right)$}
\tabularnewline
\hline 
\cellcolor{gray!60}{$\Op_{\ell\ell}^{\alpha\beta\gamma\lambda}$} & \cellcolor{gray!60}{$\left(\BLelli\gamma^{\mu}\Lelli\right)\left(\BLelli[\gamma]\gamma_{\mu}\Lelli[\lambda]\right)$} & \cellcolor{gray!20}{$\Op_{ee}^{\alpha\beta\gamma\lambda}$ } & \cellcolor{gray!20}{$\left(\BREi\gamma^{\mu}\REi\right)\left(\BREi[\gamma]\gamma_{\mu}\REi[\lambda]\right)$} & \cellcolor{gray!20}{$\Op_{\ell e}^{\alpha\beta\gamma\lambda}$} & \cellcolor{gray!20}{$\left(\BLelli\gamma^{\mu}\Lelli\right)\left(\BREi[\gamma]\gamma_{\mu}\REi[\lambda]\right)$}
\tabularnewline
\cellcolor{gray!20}{$\Op_{qq}^{(1)\alpha\beta\gamma\lambda}$ } & \cellcolor{gray!20}{$\left(\BLQi\gamma^{\mu}\LQi\right)\left(\BLQi[\gamma]\gamma_{\mu}\LQi[\lambda]\right)$} & \cellcolor{gray!20}{$\Op_{uu}^{\alpha\beta\gamma\lambda}$}  & \cellcolor{gray!20}{$\left(\BRUi\gamma^{\mu}\RUi\right)\left(\BRUi[\gamma]\gamma_{\mu}\RUi[\lambda]\right)$} & \cellcolor{gray!20}{$\Op_{\ell u}^{\alpha\beta\gamma\lambda}$} & \cellcolor{gray!20}{$\left(\BLelli\gamma^{\mu}\Lelli\right)\left(\BRUi[\gamma]\gamma_{\mu}\RUi[\lambda]\right)$}
\tabularnewline
\cellcolor{gray!20}{$\Op_{qq}^{(3)\alpha\beta\gamma\lambda}$ } & \cellcolor{gray!20}{$\left(\BLQi\gamma^{\mu}\sigma^{I}\LQi\right)\left(\BLQi[\gamma]\gamma_{\mu}\sigma^{I}\LQi[\lambda]\right)$} & \cellcolor{gray!20}{$\Op_{dd}^{\alpha\beta\gamma\lambda}$ } & \cellcolor{gray!20}{$\left(\BRDi\gamma^{\mu}\RDi\right)\left(\BRDi[\gamma]\gamma_{\mu}\RDi[\lambda]\right)$}  & \cellcolor{gray!20}{$\Op_{\ell d}^{\alpha\beta\gamma\lambda}$} & \cellcolor{gray!20}{$\left(\BLelli\gamma^{\mu}\Lelli\right)\left(\BRDi[\gamma]\gamma_{\mu}\RDi[\lambda]\right)$}
\tabularnewline
\cellcolor{gray!20}{$\Op_{\ell q}^{(1)\alpha\beta\gamma\lambda}$} & \cellcolor{gray!20}{$\left(\BLelli\gamma^{\mu}\Lelli\right)\left(\BLQi[\gamma]\gamma_{\mu}^ {}\LQi[\lambda]\right)$}  & \cellcolor{gray!20}{$\Op_{eu}^{\alpha\beta\gamma\lambda}$}  & \cellcolor{gray!20}{$\left(\BREi\gamma^{\mu}\REi\right)\left(\BRUi[\gamma]\gamma_{\mu}\RUi[\lambda]\right)$} & \cellcolor{gray!20}{$\Op_{qe}^{\alpha\beta\gamma\lambda}$ } & \cellcolor{gray!20}{$\left(\BLQi\gamma^{\mu}\LQi\right)\left(\BREi[\gamma]\gamma_{\mu}\REi[\lambda]\right)$}
\tabularnewline
\cellcolor{gray!20}{$\Op_{\ell q}^{(3)\alpha\beta\gamma\lambda}$} & \cellcolor{gray!20}{$\left(\BLelli\gamma^{\mu}\sigma^{I}\Lelli\right)\left(\BLQi[\gamma]\gamma_{\mu}^ {}\sigma^{I}\LQi[\lambda]\right)$} & \cellcolor{gray!20}{$\Op_{ed}^{\alpha\beta\gamma\lambda}$}  & \cellcolor{gray!20}{$\left(\BREi\gamma^{\mu}\REi\right)\left(\BRDi[\gamma]\gamma_{\mu}\RDi[\lambda]\right)$} & \cellcolor{gray!20}{$\Op_{qu}^{(1)\alpha\beta\gamma\lambda}$} & \cellcolor{gray!20}{$\left(\BLQi\gamma^{\mu}\LQi\right)\left(\BRUi[\gamma]\gamma_{\mu}\RUi[\lambda]\right)$}
\tabularnewline
&  & \cellcolor{gray!20}{$\Op_{ud}^{(1)\alpha\beta\gamma\lambda}$}  & \cellcolor{gray!20}{$\left(\BRUi\gamma^{\mu}\RUi\right)\left(\BRDi[\gamma]\gamma_{\mu}\RDi[\lambda]\right)$} & \cellcolor{gray!20}{$\Op_{qu}^{(8)\alpha\beta\gamma\lambda}$} & \cellcolor{gray!20}{$\left(\BLQi\gamma^{\mu}T^{A}\LQi\right)\left(\BRUi[\gamma]\gamma_{\mu}T^{A}\RUi[\lambda]\right)$}
\tabularnewline
&  & $\Op_{ud}^{(8)\alpha\beta\gamma\lambda}$  & $\left(\BRUi\gamma^{\mu}T^{A}\RUi\right)\left(\BRDi[\gamma]\gamma_{\mu}T^{A}\RDi[\lambda]\right)$  & \cellcolor{gray!20}{$\Op_{qd}^{(1)\alpha\beta\gamma\lambda}$} & \cellcolor{gray!20}{$\left(\BLQi\gamma^{\mu}\LQi\right)\left(\BRDi[\gamma]\gamma_{\mu}\RDi[\lambda]\right)$}
\tabularnewline
&  &  &  & \cellcolor{gray!20}{$\Op_{qd}^{(8)\alpha\beta\gamma\lambda}$} & \cellcolor{gray!20}{$\left(\BLQi\gamma^{\mu}T^{A}\LQi\right)\left(\BRDi[\gamma]\gamma_{\mu}T^{A}\RDi[\lambda]\right)$}
\tabularnewline
\hline 
\multicolumn{2}{c|}{$\rm \left(\overline{L}R\right)\left(\overline{R}L\right)$~and~$\rm \left(\overline{L}R\right)\left(\overline{L}R\right)$} & \multicolumn{4}{c}{$B$- and $L$-number violating}
\tabularnewline
\hline 
\cellcolor{gray!20}{$\Op_{\ell edq}^{\alpha\beta\gamma\lambda}$} & \cellcolor{gray!20}{$\left(\BLelli\REi\right)\left(\BRDi[\gamma]\LQi[\lambda]\right)$} & $\Op_{duq}^{\alpha\beta\gamma\lambda}$  & \multicolumn{3}{c}{$\epsilon^{ABC}\epsilon^{ab}\left[\left(D_{\alpha{\rm R}}^{A}\right)^{T}{\sf C}U_{\beta{\rm R}}^{B}\right]\left[\left(Q_{\gamma{\rm L}}^{{C}a}\right)^{T}{\sf C}\ell_{\lambda{\rm L}}^{b}\right]$}
\tabularnewline
\cellcolor{gray!20}{$\Op_{quqd}^{(1)\alpha\beta\gamma\lambda}$} & \cellcolor{gray!20}{$\left(\overline{Q_{\alpha{\rm L}}^{a}}\RUi\right)\epsilon^{ab}\left(\overline{Q_{\gamma{\rm L}}^{b}}\RDi[\lambda]\right)$} & $\Op_{qqu}^{\alpha\beta\gamma\lambda}$  & \multicolumn{3}{c}{$\epsilon^{ABC}\epsilon^{ab}\left[\left(Q_{\alpha{\rm L}}^{Aa}\right)^{T}{\sf C}Q_{\beta{\rm L}}^{Bb}\right]\left[\left(U_{\gamma{\rm R}}^{C}\right)^{T}{\sf C}E_{\lambda{\rm R}}\right]$}
\tabularnewline
$\Op_{quqd}^{(8)\alpha\beta\gamma\lambda}$  & $\left(\overline{Q_{\alpha{\rm L}}^{a}}T^{A}\RUi\right)\epsilon^{ab}\left(\overline{Q_{\gamma{\rm L}}^{b}}T^{A}\RDi[\lambda]\right)$  & $\Op_{qqq}^{\alpha\beta\gamma\lambda}$  & \multicolumn{3}{c}{$\epsilon^{ABC}\epsilon^{ad}\epsilon^{be}\left[\left(Q_{\alpha{\rm L}}^{Aa}\right)^{T}{\sf C}Q_{\beta{\rm L}}^{Bb}\right]\left[\left(Q_{\gamma{\rm L}}^{{C}e}\right)^{T}{\sf C}\ell_{\lambda{\rm L}}^{d}\right]$}
\tabularnewline
\cellcolor{gray!20}{$\Op_{\ell equ}^{(1)\alpha\beta\gamma\lambda}$}  & \cellcolor{gray!20}{$\left(\overline{\ell_{\alpha{\rm L}}^{a}}\REi\right)\epsilon^{ab}\left(\overline{Q_{\gamma{\rm L}}^{b}}\RUi[\lambda]\right)$} & $\Op_{duu}^{\alpha\beta\gamma\lambda}$  & \multicolumn{3}{c}{$\epsilon^{ABC}\left[\left(D_{\alpha{\rm R}}^{A}\right)^{T}{\sf C}U_{\beta{\rm R}}^{B}\right]\left[\left(U_{\gamma{\rm R}}^{C}\right)^{T}{\sf C}E_{\lambda{\rm R}}\right]$}
\tabularnewline
$\Op_{\ell equ}^{(3)\alpha\beta\gamma\lambda}$  & $\left(\overline{\ell_{\alpha{\rm L}}^{a}}\sigma^{\mu\nu}\REi\right)\epsilon^{ab}\left(\overline{Q_{\gamma{\rm L}}^{b}}\sigma_{\mu\nu}\RUi[\lambda]\right)$  &  & \multicolumn{3}{c}{}
\tabularnewline
\hline\hline 
\end{tabular}
} 
\caption{The dimension-six operators in the Warsaw basis in the SMEFT. Among these operators, only the 41 operators in the gray region are contained in the type-(I+II) SEFT, as well in the type-II SEFT. Moreover, the Wilson coefficients of the 9 operators in the dark gray region acquire additional cross contributions from entangled effects of the right-handed neutrinos and the triplet Higgs in the type-(I+II) seesaw mechanism.}
\label{tab:SMEFT-dim-6-Warsaw} 
\end{table}

\subsubsection*{\bf \textbullet The renormalizable SM terms}

In Eq. \eqref{eq:full-eft}, all the SM couplings except for the $\rm SU(3)^{}_c$ gouge coupling $g^{}_s$ have been replaced by the corresponding effective couplings. The effective gauge couplings only acquire the one-loop threshold corrections from the individual effect of $\Phi$, namely,
\begin{eqnarray}
g^{\rm eff}_1 = \left[ 1 + \frac{g^2_1 \lnmd}{2\left( 4\pi \right)^2} \right] g^{}_1 \;,\qquad g^{\rm eff}_2 = \left[ 1 + \frac{g^2_2 \lnmd}{3\left( 4\pi \right)^2} \right] g^{}_2 \;.
\end{eqnarray}

The effective Yukawa couplings and quadratic coupling constant of the SM Higgs contain the one-loop contributions from the individual effects of $N^{}_{\rm R}$ and $\Phi$. They are given by
\begin{eqnarray}
Y^{\rm eff}_{f} = Y^{}_f + \delta Y^{\rm eff,I}_f + \delta Y^{\rm eff,II}_f \;,\quad m^2_{\rm eff} = m^2 + \delta m^2_{\rm eff,I} + \delta m^2_{\rm eff,II}
\end{eqnarray}
for $f=l,{\rm u},{\rm d}$, where
\begin{eqnarray}
\left( \delta Y^{\rm eff,I}_l \right)^{}_{\alpha\beta} &=& - \frac{1}{\left(4\pi\right)^2} \left\{ \Yli{\beta} \DDYni{i} \left[ \frac{1}{4} \left( 1 + 2 \lnmi \right) + \frac{m^2}{3 \pM{2}} \right] - \frac{1}{8} \Yni{i} \left( \DYn\Yl \right)^{}_{i\beta} \right.
\nonumber
\\
&& \times \left. \left[5 + \frac{6m^2}{M^2_i} + 2 \left( 3 + \frac{2m^2}{M^2_i} \right)\lnmi \right] \right\} \;,
\nonumber
\\
\left( \delta Y^{\rm eff,I}_{\rm u} \right)^{}_{\alpha\beta} &=& - \frac{1}{\left(4\pi\right)^2} \Yui{\beta} \DDYni{i} \left[ \frac{1}{4} \left( 1 + 2 \lnmi \right) + \frac{m^2}{3 \pM{2}} \right] \;,
\nonumber
\\
\left( \delta Y^{\rm eff,I}_{\rm d} \right)^{}_{\alpha\beta} &=& - \frac{1}{\left(4\pi\right)^2} \Ydi{\beta} \DDYni{i} \left[ \frac{1}{4} \left( 1 + 2 \lnmi \right) + \frac{m^2}{3\pM{2}} \right]  \;,
\nonumber
\\
\delta m^2_{\rm eff,I} &=&  - \frac{1}{\left(4\pi\right)^2} \DDYni{i} \left[ \frac{m^2}{2}  \left( 1 + 2 \lnmi \right) + \frac{m^4}{3 \pM{2}} - 2 M^2_i \left( 1+\lnmi \right) \right]  \;
\end{eqnarray}
result from the effects of $N^{}_{\rm R}$, and 
\begin{eqnarray}
\left( \delta Y^{\rm eff,II}_l \right)^{}_{\alpha\beta} &=& -\frac{1}{\left(4\pi\right)^2} \left\{ \left[ 3 + \frac{m^2}{M_\Delta^2} \left( 17 + 6 L^{}_\Delta \right) \right] \lambda_\Delta^2 \left( Y_l^{} \right)_{\alpha\beta} + \frac{3}{8}\left( Y_\Delta^{} Y_\Delta^\dag Y_l^{} \right)^{}_{\alpha\beta} \left( 1 + 2L^{}_\Delta \right) \vphantom{\frac{\lambda^4_\Delta}{M^2_\Delta}} \right\} \;,
\nonumber
\\
\left( \delta Y^{\rm eff,II}_{\rm u} \right)^{}_{\alpha\beta} &=& - \frac{1}{(4\pi)^2} \left[ 3 + \frac{m^2}{M_\Delta^2} \left( 17 + 6 L^{}_\Delta \right) \right] \lambda_\Delta^2 \left(Y^{}_{\rm u}\right)_{\alpha\beta} \;,
\nonumber
\\
\left( \delta Y^{\rm eff,II}_{\rm d} \right)^{}_{\alpha\beta} &=& - \frac{1}{(4\pi)^2} \left[ 3 + \frac{m^2}{M_\Delta^2} \left( 17 + 6 L^{}_\Delta \right) \right] \lambda_\Delta^2 \left(Y^{}_{\rm d}\right)_{\alpha\beta} \;,
\nonumber
\\
\delta m^2_{\rm eff,II} &=& -\frac{1}{(4\pi)^2} \left[ 3 M_\Delta^2 \left( 4 \lambda_\Delta^2 + \lambda^{}_3 \right) \left( 1 + L^{}_\Delta \right) + 6 m^2 \lambda^2_\Delta \left( 3 + 2 L^{}_\Delta \right) \vphantom{\frac{m^4}{M^2_\Delta}} + \frac{4 m^4}{M_\Delta^2}\lambda_\Delta^2 \left( 11 + 6L^{}_\Delta \right) \right] \;\quad\quad
\end{eqnarray}
are induced by the effects of $\Phi$. 

The effective quartic coupling constant of the SM Higgs obtain the threshold corrections not only from the separate effects of $N^{}_{\rm R}$ and $\Phi$ but also from their entangled effects, i.e.,
\begin{eqnarray}\label{eq:lambda-eff}
\lambda^{}_{\rm eff} = \lambda + \delta \lambda^{\rm I}_{\rm eff} + \delta \lambda^{\rm II}_{\rm eff} + \delta \lambda^{\rm ct}_{\rm eff} \;.
\end{eqnarray}
$\delta \lambda^{\rm I}_{\rm eff}$ and $\delta \lambda^{\rm II}_{\rm eff}$ in Eq. \eqref{eq:lambda-eff} are the threshold corrections from the individual contributions from $N^{}_{\rm R}$ and $\Phi$, and found to be
\begin{eqnarray}
\delta \lambda^{\rm I}_{\rm eff} &=& \frac{1}{\left(4\pi\right)^2} \left\{ \DDYni{i} \left[ - \lambda \left( 1 + 2 \lnmi \right) - \frac{4\lambda m^2}{3\pM{2}} + \frac{g^2_2 m^2}{18 \pM{2}} \left(5+6\lnmi\right) \right] - \frac{m^2}{2 \pM{2}} \left( \DYn \Yl \DYl \Yn \right)^{}_{ii} \right.
\nonumber
\\
&& \times \left(1+2\lnmi\right)  + \left[ m^2 \pM{2} \left( 1+2\lnmj \right) - m^2 \pM[j]{2} \left( 1+ 2\lnmi\right) - 2 \pM{2} \pM[j]{2} \lnij \right] \frac{\DDYni{j} \DDYni{j}}{2\pM{}\pM[j]{}\left( \pM{2} - \pM[j]{2} \right) }
\nonumber
\\
&&  + \left.  \left[m^2\lnij + \pM{2}\left( 1+\lnmi \right) - \pM[j]{2}\left( 1+\lnmj \right) \right] \frac{\DDYni{j} \DDYni[j]{i}}{\pM{2} - \pM[j]{2}}  \right\} \;,
\nonumber
\\
\delta \lambda^{\rm II}_{\rm eff} &=& - 2\lambda_\Delta^2 \left( 1 + \frac{2m^2}{M_\Delta^2} \right) + \frac{1}{\left(4\pi\right)^2} \left\{  - \frac{1}{2} \left( 3\lambda^2_3 + 2\lambda^2_4 \right) L^{}_\Delta + \left( g^4_2 - 20 \lambda^2_4 \right) \vphantom{\frac{\lambda^2_\Delta}{3}} \frac{m^2}{30M^2_\Delta}  \right.
\nonumber
\\
&& + \left[  3g^2_1 \left( 5+ 6L^{}_\Delta \right) + g^2_2 \left( 61 + 86L^{}_\Delta \right) - 24 \lambda\left( 59 + 34 L^{}_\Delta \right) - 48 \left( 8 \lambda^{}_1 + \lambda^{}_2 \right) \left( 1 + L^{}_\Delta \right) \right.
\nonumber
\\
&& + \left. 12 \lambda^{}_3 \left( 29 + 18 L^{}_\Delta \right) - 8 \lambda^{}_4 \left( 59 + 42L^{}_\Delta \right) \right] \frac{m^2 \lambda^2_\Delta}{6 M^2_\Delta} - 2\lambda^2_\Delta \left[ 2\lambda \left( 13 + 10L^{}_\Delta \right)  \right. 
\nonumber
\\
&& + \left. \left( 8\lambda^{}_1 + \lambda^{}_2 \right) \left( 1+ L^{}_\Delta \right) - 2\lambda^{}_3 \left( 8 + 5 L^{}_\Delta \right) + 4 \lambda^{}_4 \left( 3 + 2L^{}_\Delta \right) \right] + \frac{8\lambda^4_\Delta}{3} \left[ 3 \left( 19 + 11 L^{}_\Delta \right) \vphantom{\frac{\lambda^4_\Delta}{M^2_\Delta}} \right.
\nonumber
\\
&& + \left.\left. 20 \left( 13 + 6L^{}_\Delta \right) \frac{m^2}{M^2_\Delta} \right] \right\} \;,
\end{eqnarray}
where $\lnij \equiv \ln \left( \pM{2}/\pM[j]{2} \right)$. Note that the first term in $\delta \lambda^{\rm II}_{\rm eff}$ without a loop factor $1/\left( 4\pi\right)^2$ arises from the tree-level matching. The last term in Eq. \eqref{eq:lambda-eff}, i.e., $\delta \lambda^{\rm ct}_{\rm eff}$ contains both types of cross contributions and is given by
\begin{eqnarray}
\delta \lambda^{\rm ct}_{\rm eff} &=& 4\lamd[2] \left( 1 + 2\frac{m^2}{\Mds} \right) \delta Z^{\rm G,I}_H + 8 \lamd[2] m^2 G^{\rm I}_{DH} + \left[ \left( \delta m^2 \right)^{\rm G, I} + m^2 \delta Z^{\rm G, I}_H \right] G^{\prime, \rm II}_{HD} |^{}_{\rm tree} - m^2 G^{\prime,\rm ct}_{HD} - \delta \lambda^{\rm G, ct}
\nonumber
\\
&=& \frac{\lambda^{}_\Delta}{\left(4\pi\right)^2} \left\{ 2 \lambda^{}_\Delta \left( \DYn \Yn \right)^{}_{ii} \left[ \frac{3m^2 + M^2_\Delta}{M^2_\Delta}  \left( 1 + 2\lnmi \right) + \frac{4m^2}{3\pM{2}} - \frac{4\pM{2}}{\Mds} \left( 1 + \lnmi \right) \right] \right.
\nonumber
\\
&& + \left. \left( \DYn \Ydel Y^\ast_\nu + Y^{\rm T}_\nu \DYdel \Yn \right)^{}_{ii} \left[ \frac{2\pM{}\left( \Mds + 2m^2 \right)}{\Mds[3]} \left( 1 + \lnmi \right) - \frac{m^2}{\pM{} \Mds[]} \left( 1 + 2 \lnmi \right) \right] \right\} \;.
\end{eqnarray}

\subsubsection*{\bf \textbullet The dim-5 operator}

The Wilson coefficient $C^{(5)}_{\rm eff}$ of the Weinberg operator $\Op^{(5)}$ in Eq. \eqref{eq:full-eft} can be written as
\begin{eqnarray}\label{eq:dim-5}
C^{(5)}_{\rm eff} = C^{(5),\rm I}_{\rm eff} + C^{(5),\rm II}_{\rm eff} + C^{(5),\rm ct}_{\rm eff} \;.
\end{eqnarray}
The first two terms on the right-hand side of Eq. \eqref{eq:dim-5} contain both the tree-level and one-loop contributions induced by integrating out $N^{}_{\rm R}$ and $\Phi$, respectively:
\begin{eqnarray}
\left( C^{(5),\rm I}_{\rm eff} \right)^{}_{\alpha\beta} &=& \left( \Yn M^{-1}_{\rm R} \TYn \right)^{}_{\alpha\beta} - \frac{1}{\left(4\pi\right)^2} \left\{- \Yni{i} \pM{-1} \left( Y^{\rm T}_\nu \right)^{}_{i\beta} \left[ 2\lambda \left( 1+\lnmi \right) + \frac{g^2_1+g^2_2}{4} \left( 1+3\lnmi\right) \right] \right.
\nonumber
\\
&& + \frac{1}{2} \left( \Yn M^{-1}_{\rm R} \TYn \right)^{}_{\alpha\beta} \DDYni{i} \left( 1 + 2 \lnmi \right)  + \frac{1}{8} \Yni{i} \left( \DYn \Yn M^{-1}_{\rm R} \TYn \right)^{}_{i\beta} \left( 3+2\lnmi \right) 
\nonumber
\\
&& + \left. \frac{1}{8} \Yni[\beta]{i} \left( \DYn \Yn M^{-1}_{\rm R} \TYn \right)^{}_{i\alpha} \left( 3+2\lnmi \right) \right\} \;,
\nonumber
\\
\left( C^{(5),\rm II}_{\rm eff} \right)^{}_{\alpha\beta} &=& - \frac{2\lambda_\Delta}{\Mds[]} \left( Y^{}_\Delta \right)_{\alpha\beta} + \frac{2\lambda_\Delta}{(4\pi)^2 \Mds[]} 
\left\{ 6 \lambda^2_\Delta \left(Y^{}_\Delta\right)_{\alpha\beta} + \frac{3}{4} \left( 1 + 2L^{}_{\Delta} \right) \left( Y^{}_\Delta Y^{\dag}_\Delta Y^{}_\Delta \right)_{\alpha\beta} + \left( 1 + L_\Delta \right) \right.
\nonumber
\\
&& \times \left. \left[ \left( 2\lambda_3 - 4\lambda_4 - 8\lambda_1 - \lambda_2 \right) \left( Y_\Delta \right)^{}_{\alpha\beta} + \left( Y^{}_lY_l^\dag Y^{}_\Delta \right)_{\alpha \beta} + \left( Y^{}_\Delta Y^{\ast}_l Y_l^{\rm T} \right)_{\alpha \beta} \right]  \vphantom{\frac{\lambda^2}{4}}\right\} \;,
\end{eqnarray}
and the last term is the one-loop cross contribution from non-linearly combined effects of $N^{}_{\rm R}$ and $\Phi$, whose explicit expression is given by
\begin{eqnarray}
\left( C^{(5),\rm ct}_{\rm eff} \right)_{\alpha\beta} &=& - \left\{ G^{(5), \rm I}|^{}_{\rm tree} \delta Z^{\rm G, II}_H + G^{(5), \rm II} |^{}_{\rm tree} \delta Z^{\rm G, I}_H + \frac{1}{2} \left[ \delta Z^{\rm G, I}_\ell G^{(5),\rm II} |^{}_{\rm tree} + \delta Z^{\rm G, II}_\ell G^{(5),\rm I} |^{}_{\rm tree} \right.\right.
\nonumber
\\
&& + \left.\left. G^{(5),\rm I} |^{}_{\rm tree} \left( \delta Z^{\rm G, II}_\ell \right)^{\rm T} + G^{(5),\rm II} |^{}_{\rm tree} \left( \delta Z^{\rm G, I}_\ell \right)^{\rm T} \right]- G^{(5), \rm ct} \vphantom{\frac{1}{1}} \right\}^{}_{\alpha\beta}
\nonumber
\\
&=& \frac{1}{\left(4\pi\right)^2}  \left\{ - 2\lambda^2_\Delta \Yni{i} \pM{-1} \TYni{i} \left( 5 + 2\lnmi \right) + \frac{\lambda^{}_\Delta}{\Mds[]}  \left( \Ydel\right)^{}_{\alpha\beta} \left( \DYn\Yn\right)^{}_{ii} \left( 1 + 2 \lnmi \right)  \right.
\nonumber
\\
&& + \frac{\lambda^{}_\Delta}{4\Mds[]} \left[ \left(\Ydel Y^\ast_\nu \right)^{}_{\alpha i} \TYni{i} + \Yni{i} \left( \DYn \Ydel \right)^{}_{i\beta} \right] \left( 3 + 2 \lnmi \right)  + 2 \left( \Ydel \right)^{}_{\alpha\beta} \left( Y^{\rm T}_\nu \DYdel \Yn \right)^{}_{ii} 
\nonumber
\\
&& \times \left. \frac{\pM{}}{\Mds} \left( 1 + \lnmi \right) - \frac{3}{8} \left( \Ydel \DYdel \Yn M^{-1}_{\rm R} Y^{\rm T}_\nu + \Yn M^{-1}_{\rm R} Y^{\rm T}_\nu \DYdel \Ydel \right)^{}_{\alpha\beta} \left( 1 + 2\lnmd \right) \right\} \;.
\end{eqnarray}

\subsubsection*{\bf \textbullet The dim-6 operators}

In the type-(I+II) SEFT, there are 41 dim-6 operators in the Warsaw basis induced by integrating out both the heavy right-handed neutrinos $N^{}_{\rm R}$ and the heavy triplet Higgs $\Phi$, the number and content of which are exactly the same as those in the type-II SEFT as shown in Table \ref{tab:SMEFT-dim-6-Warsaw}, but the Wilson coefficients of some operators acquire additional cross contributions. This means that compared with the results in the type-II SEFT, the non-linear effects of $N^{}_{\rm R}$ and $\Phi$ do not lead to any new dim-6 operators and only result in some cross contributions to the Wilson coefficients of some relevant operators~\footnote{The reasons why no new dim-6 operators are generated and only parts of the dim-6 operators can obtain the cross contributions are not clear so far. We also have no idea whether this is a generic feature for models involving more than one kind of heavy field. The properties of the gauge theory may give some hints about these issues, which is very interesting and deserves further study.}. Unlike the way we deal with the unique dim-5 operator, the tree-level and one-loop contributions to the Wilson coefficients of the dim-6 operators already appearing at the tree level are not summed up in Eq. \eqref{eq:full-eft}. With the help of Eqs. \eqref{eq:tree-level} and \eqref{eq:eom}, these tree-level Wilson coefficients in Eq. \eqref{eq:full-eft} are explicitly given by
\begin{eqnarray}
&& C^{}_H |^{}_{\rm tree} = G^{}_H |^{}_{\rm tree} + 2 \left( \lambda - 2\lamd[2] \right) G^\prime_{HD} |^{}_{\rm tree} = 2(4\lambda-\lambda_3+\lambda_4)\lambda_\Delta^2-16\lambda_\Delta^4 
\;,
\nonumber
\\
&& C^{}_{H\square} |^{}_{\rm tree} = \frac{1}{2} G^\prime_{HD} |^{}_{\rm tree} = 2\lamd[2] \;,\quad
C^{}_{HD} |^{}_{\rm tree} = G^{}_{HD} |^{}_{\rm tree} = 4\lamd[2] \;,
\nonumber
\\
&& C^{(1)}_{H\ell} |^{\alpha\beta}_{\rm tree} = G^{(1)}_{H\ell} |^{\alpha\beta}_{\rm tree} = \frac{1}{4} \left( \Yn M^{-2}_{\rm R} \DYn \right)^{}_{\alpha\beta} \;,\quad C^{(3)}_{H\ell} |^{\alpha\beta}_{\rm tree} = G^{(3)}_{H\ell} |^{\alpha\beta}_{\rm tree} = - \frac{1}{4} \left( \Yn M^{-2}_{\rm R} \DYn \right)^{}_{\alpha\beta} \;,
\nonumber
\\
&& C^{}_{eH} |^{\alpha\beta}_{\rm tree} = \frac{1}{2} \Yli{\beta}  G^\prime_{HD} |^{}_{\rm tree} = 2\lamd[2] \Yli{\beta} \;,\quad C^{}_{uH} |^{\alpha\beta}_{\rm tree} = \frac{1}{2} \Yui{\beta}  G^\prime_{HD} |^{}_{\rm tree} = 2\lamd[2] \Yui{\beta} \;,
\nonumber
\\
&& C^{}_{dH} |^{\alpha\beta}_{\rm tree} = \frac{1}{2} \Ydi{\beta}  G^\prime_{HD} |^{}_{\rm tree} = 2\lamd[2] \Ydi{\beta} \;,\quad C^{}_{\ell\ell} |^{\alpha\beta\gamma\lambda}_{\rm tree} = G^{}_{\ell\ell} |^{\alpha\beta\gamma\lambda}_{\rm tree} = \frac{1}{4} \left( \Ydel \right)^{}_{\alpha\gamma} \left( \DYdel \right)^{}_{\beta\lambda} \;.
\end{eqnarray}

For the one-loop Wilson coefficients $C^{}_i$ in Eq. \eqref{eq:full-eft}, one can decompose them into
\begin{eqnarray}\label{eq:dim-6}
C^{}_i = C^{\rm I}_i + C^{\rm II}_i + C^{\rm ct}_i \;.
\end{eqnarray}
Again, the three terms on the right-hand side of Eq. \eqref{eq:dim-6} stand for the contributions from $N^{}_{\rm R}$, $\Phi$ and non-linear combinations of $N^{}_{\rm R}$ and $\Phi$, respectively. Among 41 dim-6 operators, only 9 operators, i.e., $\{ \Op^{}_{H\square}, \Op^{}_{HD}, \Op^{}_H, \Op^{(1)}_{H\ell}, \Op^{(3)}_{H\ell}, \Op^{}_{uH}, \Op^{}_{dH}, \Op^{}_{eH}, \Op^{}_{\ell\ell} \}$ shown in the dark gray region in Table \ref{tab:SMEFT-dim-6-Warsaw} can acquire cross contributions. Due to the complication of results for $C^{\rm I}_{i}$ and $C^{\rm II}_i$, we do not repeat these separate contributions from $N^{}_{\rm R}$ and $\Phi$ here. One can easily find them in Refs. \cite{Zhang:2021jdf,Li:2022ipc}. Now, we focus on the cross contributions which are new and result from the non-linear effects of $N^{}_{R}$ and $\Phi$. The cross contributions for the aforementioned 9 dim-6 operators are given as follows:
\begin{eqnarray}\label{eq:OHsquare}
C^{\rm ct}_{H\square} &=& -2 C^{}_{H\square} |^{}_{\rm tree} \delta Z^{\rm G, I}_H + G^{\rm ct}_{H \square} + \frac{1}{2} G^{\prime, \rm ct}_{HD} 
\nonumber
\\
&=& -\frac{\lamd}{\left(4\pi \right)^2 \Mds} \left[ 2\lambda^{}_\Delta \left( \DYn \Yn \right)^{}_{ii} \left( 1 + 2\lnmi \right) + \left( \DYn \Ydel Y^\ast_\nu + Y^{\rm T}_\nu \DYdel \Yn \right)^{}_{ii} \frac{2\pM{2} - \Mds}{\pM{} \Mds[] } \left( 1 + \lnmi \right) \right] \;,\qquad
\end{eqnarray}
\begin{eqnarray}
C^{\rm ct}_{HD} &=& -2 C^{}_{HD} |^{}_{\rm tree} \delta Z^{\rm G,I}_H + G^{\rm ct}_{HD}
\nonumber
\\
&=&  - \frac{2\lamd}{\left(4\pi\right)^2 \Mds} \left[ 2\lambda^{}_\Delta \left( \DYn \Yn \right)^{}_{ii} \left( 1 + 2\lnmi \right) + \left( \DYn \Ydel Y^\ast_\nu + Y^{\rm T}_\nu \DYdel \Yn \right)^{}_{ii} \frac{2\pM{2} - \Mds}{\pM{} \Mds[] } \left( 1 + \lnmi \right) \right] \;,\qquad
\end{eqnarray}
\begin{eqnarray}\label{eq:OH}
C^{\rm ct}_H &=& - 3 C^{}_H |^{}_{\rm tree} \delta Z^{\rm G, I}_H - 2G^{\prime,\rm II}_{HD} |^{}_{\rm tree} \left[ \delta \lambda^{\rm G, I} + \left( \lambda - 2\lamd[2] \right) \delta Z^{\rm G,I}_H + \delta \lambda^{\rm G, ct} \right] - 8g^{}_2 \lamd[2] G^{\rm I}_{WDH} - 4\lamd[2] G^{\prime, \rm I}_{HD} 
\nonumber
\\
&& + 16 \lamd[2] \left( \lamd[2] - \lambda \right) G^{\rm I}_{DH} + 2 \left( \lambda - 2\lamd[2] \right) G^{\prime, \rm ct}_{HD} + G^{\rm ct}_H + \frac{1}{4} \left( G^{\prime,\rm II}_{HD} |^{}_{\rm tree} \right)^2 \left( \delta m^2 \right)^{\rm G,I}
\nonumber
\\
&=& \frac{1}{\left( 4\pi\right)^2} \left\{ \lambda^2_\Delta \left( \DYn \Yn \right)^{}_{ii} \left[ \frac{16 \left(\lambda^2_\Delta - \lambda \right)}{3\pM{2}} + \frac{2g^2_2}{9\pM{2}} \left( 5 + 6\lnmi \right) - \frac{16\lambda - 3\left( \lambda^{}_3 - \lambda^{}_4 \right) - 32 \lambda^2_\Delta}{\Mds} \left( 1 + 2\lnmi \right)  \right.\right.
\nonumber
\\
&& - \left. 8\lamd[2] \frac{\pM{2}}{\Mds[4]} \left( 1 + \lnmi \right) \right] - \frac{2\lambda^2_\Delta}{\pM{2}} \left( \DYn \Yl \DYl \Yn \right)^{}_{ii} \left( 1 + 2 \lnmi \right) + 2\lambda^{}_\Delta \left( \DYn \Ydel Y^\ast_\nu + Y^{\rm T}_\nu \DYdel \Yn \right)^{}_{ii} 
\nonumber
\\
&& \times \left[ \frac{\lambda - 2\lambda^2_\Delta}{\pM{}\Mds[]} \left( 1 + 2 \lnmi \right) - \frac{\pM{} \left( 4\lambda - \lambda^{}_3 + \lambda^{}_4 - 16\lambda^2_\Delta \right)}{\Mds[3]} \left( 1 + \lnmi \right) \right] - \frac{8\lambda^2_\Delta}{\Mds} \left( \DYn\Ydel\DYdel\Yn \right)^{}_{ii} \left( 1 + \lnmi \right)  
\nonumber
\\
&& + 2\lambda^2_\Delta \left( \DYn \Yn \right)^{}_{ij} \left( \DYn \Yn \right)^{}_{ij} \frac{\Mds\left[ \pM{2} \left( 1 + 2\lnmj \right) - \pM[j]{2} \left( 1 + 2\lnmi \right) \right] - 4\pM{2}\pM[j]{2} \lnij}{\pM{} \pM[j]{} \left( \pM{2} - \pM[j]{2} \right) \Mds} 
\nonumber
\\
&& + 4\lambda^2_\Delta \left( \DYn \Yn \right)^{}_{ij} \left( \DYn \Yn \right)^{}_{ji} \frac{2\pM{2} \left( 1 + \lnmi \right) - 2 \pM[j]{2} \left( 1 + \lnmj \right) + \Mds \lnij }{\left( \pM{2} - \pM[j]{2} \right) \Mds} 
\nonumber
\\
&& + \left. 4 \lambda^{}_\Delta \left[ \left( \DYn \Yn \right)^{}_{ij} \left( Y^{\rm T}_\nu \DYdel \Yn \right)^{}_{ij} + \left( Y^{\rm T}_\nu Y^\ast_\nu \right)^{}_{ij} \left( \DYn \Ydel Y^\ast_\nu \right)^{}_{ij} \right] \frac{\pM[j]{} \lnij}{\left( \pM{2} - \pM[j]{2} \right) \Mds[]} \right\} \;,
\end{eqnarray}
\begin{eqnarray}
\left( C^{(1),{\rm ct}}_{H\ell} \right)^{}_{\alpha\beta} &=& - C^{(1)}_{H\ell} |^{\alpha\beta}_{\rm tree} \delta Z^{\rm G, II}_H - \frac{1}{2} \left( \delta Z^{\rm G, II}_\ell C^{(1)}_{H\ell} |^{}_{\rm tree} + C^{(1)}_{H\ell} |^{}_{\rm tree} \delta Z^{\rm G, II}_\ell \right)^{}_{\alpha\beta} + \left(  G^{(1),\rm ct}_{H\ell} \right)^{}_{\alpha\beta}
\nonumber
\\
&=& \frac{1}{\left( 4\pi \right)^2} \left\{- \frac{3\lambda^2_\Delta}{2} \Yni{i} \DYni[\beta]{i} \frac{ 2\pM{4} - \pM{2} \Mds \left( 3 - \lnmi + \lnmd \right) + \Mds[4]}{\pM{2} \left( \pM{2} - \Mds \right)^2} \right.
\nonumber
\\
&&  + \frac{3\lambda^{}_\Delta}{4} \left[ \Yni{i} \left( Y^{\rm T}_\nu \DYdel \right)^{}_{i\beta} + \left( \Ydel Y^\ast_\nu \right)^{}_{\alpha i} \DYni{i} \right] 
\nonumber
\\
&& \times \frac{\pM{4} \left( 3 + 2 \lnmd \right) - \pM{2}\Mds \left( 5 + 4\lnmd \right) + \Mds[4] \left( 2 + \lnmi + \lnmd \right)}{\pM{} \Mds[] \left( \pM{2} - \Mds \right)^2} 
\nonumber
\\
&&  - \frac{3}{32} \left( \Yn M^{-2}_{\rm R} \DYn \Ydel \DYdel + \Ydel \DYdel \Yn M^{-2}_{\rm R} \DYn \right)^{}_{\alpha\beta} \left( 1 + 2\lnmd \right) 
\nonumber
\\
&& - \left. \frac{3}{8} \left( \Ydel Y^\ast_\nu \right)^{}_{\alpha i} \left( Y^{\rm T}_\nu \DYdel \right)^{}_{i\beta}  \frac{\pM{2} \left( 1 - \lnmi + \lnmd \right) - \Mds \left( 1 - 2\lnmi + 2 \lnmd \right)}{\left( \pM{2} - \Mds \right)^2} \right\} \;,
\end{eqnarray}
\begin{eqnarray}
\left( C^{(3),{\rm ct}}_{H\ell} \right)^{}_{\alpha\beta} &=& -C^{(3)}_{H\ell} |^{\alpha\beta}_{\rm tree} \delta Z^{\rm G, II}_H - \frac{1}{2} \left( \delta Z^{\rm G,II}_\ell C^{(3)}_{H\ell} |^{}_{\rm tree} + C^{(3)}_{H\ell} |^{}_{\rm tree} \delta Z^{\rm G, II}_\ell \right)^{}_{\alpha\beta}  + \left(  G^{(3),\rm ct}_{H\ell} \right)^{}_{\alpha\beta}
\nonumber
\\
&=& \frac{1}{\left( 4\pi \right)^2} \left\{ \frac{\lambda^2_\Delta}{2} \Yni{i} \DYni[\beta]{i} \frac{ 4\pM{4} - \pM{2} \Mds \left( 7 - \lnmi + \lnmd \right) + 3\Mds[4]}{\pM{2} \left( \pM{2} - \Mds \right)^2} \right.
\nonumber
\\
&& - \frac{\lambda^{}_\Delta}{4} \left[ \Yni{i} \left( Y^{\rm T}_\nu \DYdel \right)^{}_{i\beta} + \left( \Ydel Y^\ast_\nu \right)^{}_{\alpha i} \DYni{i} \right]  
\nonumber
\\
&& \times \frac{2\pM{4} \left( 3 + 2 \lnmd \right) - \pM{2}\Mds \left( 11 + 8\lnmd \right) + \Mds[4] \left( 5 + \lnmi + 3\lnmd \right)}{\pM{} \Mds[] \left( \pM{2} - \Mds \right)^2} 
\nonumber
\\
&& + \frac{3}{32} \left( \Yn M^{-2}_{\rm R} \DYn \Ydel \DYdel + \Ydel \DYdel \Yn M^{-2}_{\rm R} \DYn \right)^{}_{\alpha\beta} \left( 1 + 2\lnmd \right) 
\nonumber
\\
&& + \left. \frac{1}{8} \left( \Ydel Y^\ast_\nu \right)^{}_{\alpha i} \left( Y^{\rm T}_\nu \DYdel \right)^{}_{i\beta} \frac{\pM{2} \left( 1 - \lnmi + \lnmd \right) - \Mds \left( 1 - 2\lnmi + 2 \lnmd \right)}{\left( \pM{2} - \Mds \right)^2} \right\} \;,
\end{eqnarray}
\begin{eqnarray}
\left( C^{\rm ct}_{uH} \right)^{}_{\alpha\beta} &=& - \frac{3}{2} C^{}_{uH} |^{\alpha\beta}_{\rm tree} \delta Z^{\rm G, I}_H - \frac{1}{2} \Yui{\beta} G^{\prime,\rm II}_{HD}|^{}_{\rm tree} \delta Z^{\rm G,I}_H - 4 \lamd[2] \Yui{\beta} G^{\rm I}_{DH} + \frac{1}{2} \Yui{\beta} G^{\prime, \rm ct}_{HD} 
\nonumber
\\
&& - \rmI \Yui{\beta} G^{\prime\prime,\rm ct}_{HD}
\nonumber
\\
&=& \frac{\Yui{\beta}}{\left( 4\pi \right)^2} \left\{ - \frac{\lambda^2_\Delta}{6} \left( \DYn \Yn \right)^{}_{ii} \left[ \frac{8}{\pM{2}} + \frac{15}{\Mds} \left( 1 + 2\lnmi \right) \right] - \lambda^{}_\Delta \left( Y^{\rm T}_\nu \DYdel \Yn \right)^{}_{ii} \frac{2\pM{2} - \Mds}{\pM{} \Mds[3]} \left( 1 + \lnmi \right)  \right.
\nonumber
\\
&& - \left. \lambda^{}_\Delta \left( \DYn \Ydel Y^\ast_\nu \right)^{}_{ii} \left[ \frac{2\pM{}}{\Mds} \left( 1 + \lnmi \right) - \frac{1}{\pM{}\Mds[]} \lnmi \right] \right\} \;,
\end{eqnarray}
\begin{eqnarray}
\left( C^{\rm ct}_{dH} \right)^{}_{\alpha\beta} &=& - \frac{3}{2} C^{}_{dH} |^{\alpha\beta}_{\rm tree} \delta Z^{\rm G, I}_H - \frac{1}{2} \Ydi{\beta} G^{\prime,\rm II}_{HD}|^{}_{\rm tree} \delta Z^{\rm G,I}_H - 4 \lamd[2] \Ydi{\beta} G^{\rm I}_{DH} + \frac{1}{2} \Ydi{\beta} G^{\prime, \rm ct}_{HD} 
\nonumber
\\
&& + \rmI \Ydi{\beta} G^{\prime\prime,\rm ct}_{HD}
\nonumber
\\
&=& \frac{\Ydi{\beta}}{\left( 4\pi \right)^2 } \left\{ - \frac{\lambda^2_\Delta}{6} \left( \DYn \Yn \right)^{}_{ii} \left[ \frac{8}{\pM{2}} + \frac{15}{\Mds} \left( 1 + 2\lnmi \right) \right] - \lambda^{}_\Delta \left( \DYn \Ydel Y^\ast_\nu \right)^{}_{ii} \frac{2\pM{2} - \Mds}{\pM{} \Mds[3]} \left( 1 + \lnmi \right)  \right.
\nonumber
\\
&& - \left. \lambda^{}_\Delta \left( Y^{\rm T}_\nu \DYdel \Yn \right)^{}_{ii}  \left[ \frac{2\pM{}}{\Mds} \left( 1 + \lnmi \right) - \frac{1}{\pM{}\Mds[]} \lnmi \right] \right\} \;,
\end{eqnarray}
\begin{eqnarray}
\left( C^{\rm ct}_{eH} \right)^{}_{\alpha\beta} &=& - \frac{3}{2} C^{}_{eH} |^{\alpha\beta}_{\rm tree} \delta Z^{\rm G,I}_H - \frac{1}{2} \left[ \left(\delta Z^{\rm G, I}_\ell \right)^\dagger C^{}_{eH} |^{}_{\rm tree} \right]^{}_{\alpha\beta} - \frac{1}{2}  G^{\prime,\rm II}_{HD}|^{}_{\rm tree} \left[ \Yli{\beta} \delta Z^{\rm G,I}_H + \left( \delta Y^{\rm G,I}_l \right)^{}_{\alpha\beta} \right]
\nonumber
\\
&& - 4 \lamd[2] \Yli{\beta} G^{\rm I}_{DH} + \frac{1}{2} \Yli{\beta} G^{\prime, \rm ct}_{HD} + \rmI \Yli{\beta} G^{\prime\prime,\rm ct}_{HD} + 2\lamd[2] \left[ 2 \left( G^{\rm I}_{eHD1} \right)^{}_{\alpha\beta} +  \left( G^{\rm I}_{eHD2} \right)^{}_{\alpha\beta} \right.
\nonumber
\\
&& - \left. \left( G^{\rm I}_{eHD4} \right)^{}_{\alpha\beta} \right] + \left[ \left( G^{\prime (1),\rm ct}_{H\ell} + \rmI G^{\prime\prime (1),\rm ct}_{H\ell} + G^{\prime (3),\rm ct}_{H\ell} + \rmI G^{\prime\prime (3),\rm ct}_{H\ell} \right) \Yl \right]^{}_{\alpha \gamma} + \left( G^{\rm ct}_{eH} \right)^{}_{\alpha\beta}
\nonumber
\\
&=& \frac{1}{\left( 4\pi \right)^2} \left\{ - \frac{\lambda^2_\Delta}{6} \Yli{\beta} \left( \DYn \Yn \right)^{}_{ii} \left[ \frac{15}{\Mds} \left( 1 + 2\lnmi \right) + \frac{8}{\pM{2}} \right] - \frac{\lambda^2_\Delta}{4\Mds} \Yni{i} \left( \DYn \Yl \right)^{}_{i\beta} \left( 3 + 2\lnmi \right) \right.
\nonumber
\\
&& + \lambda^2_\Delta \Yni{i} \left( \DYn \Yl \right)^{}_{i\beta}  \left[ 2 \frac{\pM{4} \left( 1 + \lnmi \right) + \pM{2} \Mds \left( 2 - \lnmi + 2\lnmd \right) - \Mds[4] \left( 3 + 2\lnmi \right)}{\pM{2}\Mds\left( \pM{2} - \Mds \right)} \right.
\nonumber
\\
&& - \left. \frac{2\pM{4} \left( 1 + \lnmd \right) - \pM{2} \Mds \left( 5 + 3\lnmi + \lnmd \right) + \Mds[4] \left( 3 + 2 \lnmi \right)}{\pM{2}\left(\pM{2} - \Mds\right)^2}   \right] - \lambda^{}_\Delta \Yli{\beta} 
\nonumber
\\
&& \times \left[ \left( \DYn \Ydel Y^\ast_\nu \right)^{}_{ii} \frac{2\pM{2}-\Mds}{\pM{}\Mds[3]} \left( 1 + \lnmi \right) + \left( Y^{\rm T}_\nu \DYdel \Yn \right)^{}_{ii} \left( \frac{2\pM{}}{\Mds[3]} \left( 1+ \lnmi \right) - \frac{1}{\pM{}\Mds[]} \lnmi \right) \right]
\nonumber
\\
&& - \frac{2\lambda^{}_\Delta}{\pM{}\Mds[]} \Yni{i} \left( Y^{\rm T}_\nu \DYdel \Yl \right)^{}_{i\beta} \left[ 1 + \lnmi + \frac{\pM{4}\left( 1 - 2\lnmd \right) + 4\pM{2}\Mds \lnmi - \Mds[4] \left( 1 + 2\lnmi \right)}{8\left( \pM{2} - \Mds[2] \right)^2} \right]
\nonumber
\\
&& - \frac{2\lambda^{}_\Delta}{\pM{}\Mds[] \left( \pM{2} - \Mds \right)} \left( \Ydel Y^\ast_\nu \right)^{}_{\alpha i} \left( \DYn \Yl \right)^{}_{i\beta}  \left[ \pM{2} \left( 1 + \lnmd \right) - \Mds \left( 1 + \lnmi \right) \right.
\nonumber
\\
&& - \left. \frac{\pM{4} \left( 3 + 2\lnmd \right) - 4\pM{2}\Mds \left( 2 + \lnmi \right) + \Mds[4] \left( 5 + 2\lnmi \right)}{8\left( \pM{2} - \Mds \right)} \right] + \frac{1}{4} \left( \Ydel Y^\ast_\nu \right)^{}_{\alpha i} \left( Y^{\rm T}_\nu \DYdel \Yl \right)^{}_{i \beta}
\nonumber
\\
&& \times \left. \frac{\pM{2} \left( 1 + \lnmi - \lnmd \right) - \Mds}{ \left( \pM{2} - \Mds \right)^2} \right\} \;,
\end{eqnarray}
\begin{eqnarray}\label{eq:Oll}
\left( C^{\rm ct}_{\ell \ell} \right)^{}_{\alpha\beta\gamma\lambda} &=& - \frac{1}{8\Mds} \left[ \left( \delta Z^{\rm G,I}_\ell \right)^{}_{\alpha\delta} C^{}_{\ell\ell} |^{\delta\beta\gamma\lambda}_{\rm tree} + \left( \delta Z^{\rm G,I}_\ell \right)^{}_{\delta\beta} C^{}_{\ell\ell} |^{\alpha\delta\gamma\lambda}_{\rm tree} + \left( \delta Z^{\rm G,I}_\ell \right)^{}_{\gamma\delta} C^{}_{\ell\ell} |^{\alpha\beta\delta\lambda}_{\rm tree} \right. 
\nonumber
\\
&& + \left. \left( \delta Z^{\rm G,I}_\ell \right)^{}_{\delta\lambda} C^{}_{\ell\ell} |^{\alpha\beta\gamma\delta}_{\rm tree} \right] + \left( G^{\rm ct}_{\ell\ell} \right)^{}_{\alpha\beta\gamma\lambda}
\nonumber
\\
&=& \frac{1}{\left(4\pi\right)^2} \left\{  \frac{\lambda^{}_\Delta}{2\pM{}\Mds[]} \left[ \Yni{i} \Yni[\gamma]{i} \left( \DYdel \right)^{}_{\beta\lambda} + \left( \Ydel \right)^{}_{\alpha \gamma} \DYni{i} \DYni[\lambda]{i} \right]\left( 1 + \lnmi \right)  \right.
\nonumber
\\
&& - \frac{1}{32\Mds} \left( 3 + 2\lnmi \right) \left[  \left(\Ydel\right)^{}_{\alpha\gamma} \left( \DYdel \Yn \right)^{}_{\lambda i} \left( Y^\ast_\nu \right)^{}_{\beta i} + \left( \DYdel \right)^{}_{\beta\lambda} \left( \Ydel Y^\ast_\nu \right)^{}_{\gamma i} \Yni{i} \right.
\nonumber
\\
&& + \left.\left.  \left(\Ydel\right)^{}_{\alpha\gamma} \left( \DYdel \Yn \right)^{}_{\beta i} \left( Y^\ast_\nu \right)^{}_{\lambda i} + \left( \DYdel \right)^{}_{\beta\lambda} \left( \Ydel Y^\ast_\nu \right)^{}_{\alpha i} \Yni[\gamma]{i}  \right] \right\} \;.
\end{eqnarray}
In each of the above equations, the first equality comes from the conversions from the Green's basis to the Warsaw basis and hence includes contributions from the implementation of the EOMs and the redefinitions of $H$  and $\ell^{}_{\rm L}$. All threshold corrections and Wilson coefficients with superscripts ``I" and ``II" are those from the separate effects of $N^{}_{\rm R}$ and $\Phi$, and can be found in Refs. \cite{Zhang:2021jdf,Li:2022ipc}. The others with the superscript ``ct" are what we calculate and show in Sec. \ref{subsec:one-loop-matching}. Together with those, one can easily achieve the second equality from the first one in each of Eqs. \eqref{eq:OHsquare}---\eqref{eq:Oll}. It is worth to point out that the last term in the second line of Eq. \eqref{eq:OH} is the cross contribution induced by the field redefinition to remove the tree-level redundant operator $\Opr^\prime_{HD}$ but missed by the direct implementation of the EOMs in Eq. \eqref{eq:eom}. Together with the one-loop contributions to Wilson coefficients of the induced dim-6 operators from the individual effects of $N^{}_{\rm R}$ and $\Phi$ achieved in Refs. \cite{Zhang:2021jdf,Li:2022ipc}, all results in Eqs. \eqref{eq:full-eft}---\eqref{eq:Oll} constitute the complete one-loop structures of the type-(I+II) SEFT. 

Now that the complete one-loop structures of the type-I, type-II and type-(I+II) SEFTs have been achieved, one can compare them and figure out the differences among them. Here we briefly discuss those results and possible phenomenological implications. The type-I SEFT contains 31 dim-6 operators in the Warsaw basis, which are all included in the 41 dim-6 operators of the type-II SEFT. The number and content of dim-6 operators in the type-(I+II) SEFT are the same as those in the type-II SEFT, but the Wilson coefficients for most operators are different from those in the type-I and type-II SEFTs. In the type-(I+II) SEFT, the Wilson coefficients of the 9 operators  $\{ \Op^{}_{H\square}, \Op^{}_{HD}, \Op^{}_H, \Op^{(1)}_{H\ell}, \Op^{(3)}_{H\ell}, \Op^{}_{uH}, \Op^{}_{dH}, \Op^{}_{eH}, \Op^{}_{\ell\ell} \}$ are not simply the linear combinations of those in the type-I and type-II SEFTs but have some cross contributions, and those of the 10 operators $\{ \Op^{}_{W}, \Op^{(1)}_{qq}, \Op^{(3)}_{qq}, \Op^{}_{ee}, \Op^{}_{uu}, \Op^{}_{dd}, \Op^{}_{eu}, \Op^{}_{ed}, \Op^{(1)}_{ud}, \Op^{}_{qe} \}$ are exactly the same as those in the type-II SEFT. The Wilson coefficients of the rest 22 dim-6 operators are directly the linear combinations of those in the type-I and type-II SEFTs. With those differences in the content and Wilson coefficients, one may make use of some low-energy observables to obtain constraints on model parameters in the SEFTs, which provide the possibility to distinguish different SEFTs from one another. For instance, the operators $\Op^{}_{eB}$ and $\Op^{}_{eW}$ can contribute to radiative decays of charged leptons, i.e., $\ell^-_\beta \to \ell^-_\alpha \gamma$ [for $(\alpha,\beta)=(e,\mu),(e,\tau),(\mu,\tau)$] and also to magnetic and electric dipole moments of charged leptons. The semi-leptonic operators, e.g., $\Op^{(1)}_{\ell q}$ and $\Op^{(3)}_{\ell q}$ can lead to $\mu$-$e$ conversion in nuclei, which is a very powerful tool to search for lepton flavor violations. The operators $\Op^{}_{\ell\ell}$ and $\Op^{}_{\ell e}$ contribute to $\mu \to eee$ and $\ell \to \ell \overline{\nu} \nu$, the latter of which can result in a shift of the extracted Fermi constant $G^{}_{\rm F}$. The operator $\Op^{}_{eH}$ contributes to $H \to \ell^-_\alpha \ell^+_\beta$, and $Z \to \ell^-_\alpha \ell^+_\beta$ can result from the operators $\Op^{(1)}_{H\ell}$, $\Op^{(3)}_{H\ell}$ and $\Op^{}_{He}$. These two decays can be lepton flavor conserving or violating, depending on whether the flavors of the charged leptons are the same or not. Some pure- and semi-leptonic operators (e.g., $\Op^{}_{\ell\ell}$ and $\Op^{(1)}_{\ell q}$) can be a source of non-standard interactions of neutrinos, which could be probed in future neutrino oscillation experiments.

Finally, we crosscheck our results with the help of the package {\sf matchmakereft} and find two minor differences. One is the Wilson coefficients of $O^{}_{H}$ in Eq. \eqref{eq:OH}, where the missed term by simply applying the EOMs is absent in the result obtained by {\sf matchmakereft}. This discrepancy has already been pointed out in Ref. \cite{Li:2022ipc}. The other one is the Wilson coefficient of $\Op^{}_{ll}$ in Eq. \eqref{eq:Oll}, i.e., all $Y^{}_\nu$ contained in the last two terms in Eq. \eqref{eq:Oll} take their conjugates in the result given by {\sf matchmakereft}, which results from typos in the expression used by the package to get contributions to $C^{}_{\ell\ell}$ from the normalization of the kinetic term of lepton doublet \footnote{We are contacting the authors to clarify these issues.}.

\section{Conclusions}\label{sec:conclusion}

Recently, the type-I, type-II and type-III seesaw effective field theories respectively describing the infrared behaviors of the three canonical seesaw mechanisms have been completely established up to one-loop level and dimension six. These SEFTs provide consistent frameworks for us to explore low-energy phenomena of the canonical seesaw mechanisms and indirectly probe the underlying heavy degrees of freedom (i.e., right-handed neutrinos, the triplet Higgs or triplet fermions). In addition, they make it possible to distinguish the canonical seesaw mechanisms by comparing the structures of SEFTs. As an appealing hybrid scenario to explain tiny neutrino masses, the type-(I+II) seesaw mechanism contains both the right-handed neutrinos and the triplet Higgs, and can naturally originate from more fundamental frameworks, such as the $\rm SO(10)$ grand unified theories. Despite the fact that there are no direct interactions between the right-handed neutrinos and the triplet Higgs in the  type-(I+II) seesaw mechanism, the corresponding type-(I+II) SEFT up to one-loop level is by no means the trivial combination of the type-I and type-II SEFTs. There are additional contributions, which result from the non-linearly combined effects of right-handed neutrinos and the triplet Higgs at the one-loop level and are absent in the type-I and type-II SEFTs. Therefore, the complete type-(I+II) SEFT up to one-loop level and dimension six needs more efforts to be constructed.

In this work, we perform the complete one-loop matching of the type-(I+II) seesaw mechanism onto the corresponding type-(I+II) SEFT with the assumption that all masses of heavy particles are roughly of the same order of magnitude. Based on the already well-established type-I and type-II SEFTs, we only need to focus on the so-called {\it cross contributions} arising from the entangled effects of the right-handed neutrinos and the triplet Higgs. Together with the individual contributions from the right-handed neutrinos and the triplet Higgs (already obtained in the type-I and type-II SEFTs), these cross contributions constitute the full one-loop structure of the type-(I+II) SEFT. We adopt the Feynman diagrammatic method to integrate out the heavy fields and match the full theory onto the low-energy EFT at the one-loop level. After gaining all these cross contributions in the Warsaw basis, we find that these cross contributions can not lead to any new operators up to dimension six, and only give additional corrections to the quartic coupling constant of the SM Higgs and the Wilson coefficients of the unique Weinberg operator as well as those of 9 dim-6 operators $\{ \Op^{}_{H\square}, \Op^{}_{HD}, \Op^{}_H, \Op^{(1)}_{H\ell}, \Op^{(3)}_{H\ell}, \Op^{}_{uH}, \Op^{}_{dH}, \Op^{}_{eH}, \Op^{}_{\ell\ell} \}$. Therefore, the number (i.e., 41 in total barring flavor structures and Hermitian conjugates) and content of dim-6 operators in the type-(I+II) SEFT are exactly the same as those in the type-II SEFT, and the one-loop type-(I+II) SEFT differs from the linear combination of the type-I and type-II SEFTs only in some cross contributions to effective quartic coupling constant of the SM Higgs and Wilson coefficients of the dim-5 operators and 9 dim-6 operators. These differences may be used to distinguish the type-(I+II) seesaw mechanism from the type-I and type-II seesaw mechanisms.

Finally, it is worth pointing out that these completely constructed SEFTs are only the first step to consistently explore low-energy consequences of seesaw mechanisms in the framework of EFT and then confront them with the low-energy precision measurements. In principle, the two-loop RGEs of the SM parameters and the Wilson coefficients should be derived and applied to run the physical parameters or Wilson coefficients from the matching scale $\mu = M$ down to the relevant low-energy scale, during which all results achieved in this work serve as the initial conditions at the matching scale. Moreover, if the masses of three right-handed neutrinos and the triplet Higgs are hierarchical, these heavy particles should be integrated out sequentially and the RGEs should be utilized to link physical parameters at any two mass scales. These issues are out of scope of this paper and left for future works.

\section*{Acknowledgements}

I am greatly indebted to Prof. Zhi-zhong Xing for his right question and encouragement, as well as for carefully reading this manuscript and giving many useful suggestions. I am also grateful to Prof. Shun Zhou for his encouragement. This work was supported by the National Natural Science Foundation of China under grant No.~11835013 and No.~12075254, and by the Alexander von Humboldt Foundation. All Feynman diagrams in this work are plotted with the help of {\sf JaxoDraw} \cite{Binosi:2003yf}, and all loop integrals are calculated by means of {\sf Package-X} \cite{Patel:2015tea,Patel:2016fam}. The Feynman rules for Majorana fields used in this work follow those described in Refs. \cite{Denner:1992me,Denner:1992vza}.

\begin{appendices}
\renewcommand{\theequation}{\thesection.\arabic{equation}}
\renewcommand{\thetable}{\thesection.\arabic{table}}
\setcounter{table}{0}
\setcounter{equation}{0}

\section{The Green's basis}

Here we list all the dim-six operators in the Green's basis in Table \ref{tab:green-basis-1}-\ref{tab:green-basis-4}. The notations and conventions follow those in Ref. \cite{Li:2022ipc}, and $\Op$ and $\Opr$ denote the independent operators in the Warsaw basis and the redundant operators, respectively.

\begin{table}
\centering
\renewcommand\arraystretch{1.4}
\resizebox{\textwidth}{!}{
\begin{tabular}{c|c|c|c|c|c}
\hline\hline 
\multicolumn{2}{c|}{$\psi^{2}D^{3}$} & \multicolumn{2}{c|}{$\psi^{2}XD$} & \multicolumn{2}{c}{$\psi^{2}DH^{2}$}
\tabularnewline
\hline 
$\Opr_{qD}^{\alpha\beta}$ & $\frac{\rmI}{2}\BLQi\left\{D_{\mu}D^{\mu},\slashed{D}\right\}\LQi$ & $\Opr_{Gq}^{\alpha\beta}$ & $\left(\BLQi T^{A}\gamma^{\mu}\LQi\right)D^{\nu}G_{\mu\nu}^{A}$ & $\Op_{Hq}^{(1)\alpha\beta}$  & $\left(\BLQi\gamma^{\mu}\LQi\right)\left(H^{\dagger}\rmI\Dlr H\right)$
\tabularnewline
$\Opr_{uD}^{\alpha\beta}$ & $\frac{\rmI}{2}\BRUi\left\{D_{\mu}D^{\mu},\slashed{D}\right\}\RUi$ & $\Opr_{Gq}^{\prime\alpha\beta}$ & $\frac{1}{2}\left(\BLQi T^{A}\gamma^{\mu}\rmI\Dlrn\LQi\right)G_{\mu\nu}^{A}$ & $\Opr_{Hq}^{\prime(1)\alpha\beta}$ & $\left(\BLQi\rmI\Dlrs\LQi\right)\left(H^{\dagger}H\right)$
\tabularnewline
$\Opr_{dD}^{\alpha\beta}$ & $\frac{\rmI}{2}\BRDi\left\{D_{\mu}D^{\mu},\slashed{D}\right\}\RDi$ & $\Opr_{\widetilde{G}q}^{\prime\alpha\beta}$ & $\frac{1}{2}\left(\BLQi T^{A}\gamma^{\mu}\rmI\Dlrn\LQi\right)\widetilde{G}_{\mu\nu}^{A}$ & $\Opr_{Hq}^{\prime\prime(1)\alpha\beta}$ & $\left(\BLQi\gamma^{\mu}\LQi\right)\partial_{\mu}\left(H^{\dagger}H\right)$
\tabularnewline
$\Opr_{\ell D}^{\alpha\beta}$ & $\frac{\rmI}{2}\BLelli\left\{D_{\mu}D^{\mu},\slashed{D}\right\}\Lelli$ & $\Opr_{Wq}^{\alpha\beta}$ & $\left(\BLQi\sigma^{I}\gamma^{\mu}\LQi\right)D^{\nu}W_{\mu\nu}^{I}$ & $\Op_{Hq}^{(3)\alpha\beta}$  & $\left(\BLQi\gamma^{\mu}\sigma^{I}\LQi\right)\left(H^{\dagger}\rmI\Dilr H\right)$
\tabularnewline
$\Opr_{eD}^{\alpha\beta}$ & $\frac{\rmI}{2}\BREi\left\{D_{\mu}D^{\mu},\slashed{D}\right\}\REi$ & $\Opr_{Wq}^{\prime\alpha\beta}$ & $\frac{1}{2}\left(\BLQi\sigma^{I}\gamma^{\mu}\rmI\Dlrn\LQi\right)W_{\mu\nu}^{I}$ & $\Opr_{Hq}^{\prime(3)\alpha\beta}$ & $\left(\BLQi\rmI\Dilrs\LQi\right)\left(H^{\dagger}\sigma^{I}H\right)$
\tabularnewline
\cline{1-2} \cline{2-2} 
\multicolumn{2}{c|}{$\psi^{2}HD^{2}$} & $\Opr_{\widetilde{W}q}^{\prime\alpha\beta}$ & $\frac{1}{2}\left(\BLQi\sigma^{I}\gamma^{\mu}\rmI\Dlrn\LQi\right)\widetilde{W}_{\mu\nu}^{I}$ & $\Opr_{Hq}^{\prime\prime(3)\alpha\beta}$ & $\left(\BLQi\sigma^{I}\gamma^{\mu}\LQi\right)D_{\mu}\left(H^{\dagger}\sigma^{I}H\right)$
\tabularnewline
\cline{1-2} \cline{2-2} 
$\Opr_{uHD1}^{\alpha\beta}$ & $\left(\BLQi\RUi\right)D_{\mu}D^{\mu}\widetilde{H}$ & $\Opr_{Bq}^{\alpha\beta}$ & $\left(\BLQi\gamma^{\mu}\LQi\right)\partial^{\nu}B_{\mu\nu}$ & $\Op_{Hu}^{\alpha\beta}$  & $\left(\BRUi\gamma^{\mu}\RUi\right)\left(H^{\dagger}\rmI\Dlr H\right)$
\tabularnewline
$\Opr_{uHD2}^{\alpha\beta}$ & $\left(\BLQi\rmI\sigma_{\mu\nu}D^{\mu}\RUi\right)D^{\nu}\widetilde{H}$ & $\Opr_{Bq}^{\prime\alpha\beta}$ & $\frac{1}{2}\left(\BLQi\gamma^{\mu}\rmI\Dlrn\LQi\right)B_{\mu\nu}$ & $\Opr_{Hu}^{\prime\alpha\beta}$ & $\left(\BRUi\rmI\Dlrs\RUi\right)\left(H^{\dagger}H\right)$
\tabularnewline
$\Opr_{uHD3}^{\alpha\beta}$ & $\left(\BLQi D_{\mu}D^{\mu}\RUi\right)\widetilde{H}$ & $\Opr_{\widetilde{B}q}^{\prime\alpha\beta}$ & $\frac{1}{2}\left(\BLQi\gamma^{\mu}\rmI\Dlrn\LQi\right)\widetilde{B}_{\mu\nu}$ & $\Opr_{Hu}^{\prime\prime\alpha\beta}$ & $\left(\BRUi\gamma^{\mu}\RUi\right)\partial_{\mu}\left(H^{\dagger}H\right)$
\tabularnewline
$\Opr_{uHD4}^{\alpha\beta}$ & $\left(\BLQi D_{\mu}\RUi\right)D^{\mu}\widetilde{H}$ & $\Opr_{Gu}^{\alpha\beta}$ & $\left(\BRUi T^{A}\gamma^{\mu}\RUi\right)D^{\nu}G_{\mu\nu}^{A}$ & $\Op_{Hd}^{\alpha\beta}$  & $\left(\BRDi\gamma^{\mu}\RDi\right)\left(H^{\dagger}\rmI\Dlr H\right)$ 
\tabularnewline
$\Opr_{dHD1}^{\alpha\beta}$ & $\left(\BLQi\RDi\right)D_{\mu}D^{\mu}H$ & $\Opr_{Gu}^{\prime\alpha\beta}$ & $\frac{1}{2}\left(\BRUi T^{A}\gamma^{\mu}\rmI\Dlrn\RUi\right)G_{\mu\nu}^{A}$ & $\Opr_{Hd}^{\prime\alpha\beta}$ & $\left(\BRDi\rmI\Dlrs\RDi\right)\left(H^{\dagger}H\right)$
\tabularnewline
$\Opr_{dHD2}^{\alpha\beta}$ & $\left(\BLQi\rmI\sigma_{\mu\nu}D^{\mu}\RDi\right)D^{\nu}H$ & $\Opr_{\widetilde{G}u}^{\prime\alpha\beta}$ & $\frac{1}{2}\left(\BRUi T^{A}\gamma^{\mu}\rmI\Dlrn\RUi\right)\widetilde{G}_{\mu\nu}^{A}$ & $\Opr_{Hd}^{\prime\prime\alpha\beta}$ & $\left(\BRDi\gamma^{\mu}\RDi\right)\partial_{\mu}\left(H^{\dagger}H\right)$
\tabularnewline
$\Opr_{dHD3}^{\alpha\beta}$ & $\left(\BLQi D_{\mu}D^{\mu}\RDi\right)H$ & $\Opr_{Bu}^{\alpha\beta}$ & $\left(\BRUi\gamma^{\mu}\RUi\right)\partial^{\nu}B_{\mu\nu}$ & $\Op_{Hud}^{\alpha\beta}$  & $\rmI\left(\BRUi\gamma^{\mu}\RDi\right)\left(\widetilde{H}^{\dagger}D_{\mu}H\right)$ 
\tabularnewline
$\Opr_{dHD4}^{\alpha\beta}$ & $\left(\BLQi D_{\mu}\RDi\right)D^{\mu}H$ & $\Opr_{Bu}^{\prime\alpha\beta}$ & $\frac{1}{2}\left(\BRUi\gamma^{\mu}\rmI\Dlrn\RUi\right)B_{\mu\nu}$ & $\Op_{H\ell}^{(1)\alpha\beta}$  & $\left(\BLelli\gamma^{\mu}\Lelli\right)\left(H^{\dagger}\rmI\Dlr H\right)$
\tabularnewline
$\Opr_{eHD1}^{\alpha\beta}$ & $\left(\BLelli\REi\right)D_{\mu}D^{\mu}H$ & $\Opr_{\widetilde{B}u}^{\prime\alpha\beta}$ & $\frac{1}{2}\left(\BRUi\gamma^{\mu}\rmI\Dlrn\RUi\right)\widetilde{B}_{\mu\nu}$ & $\Opr_{H\ell}^{\prime(1)\alpha\beta}$ & $\left(\BLelli\rmI\Dlrs\Lelli\right)\left(H^{\dagger}H\right)$
\tabularnewline
$\Opr_{eHD2}^{\alpha\beta}$ & $\left(\BLelli\rmI\sigma_{\mu\nu}D^{\mu}\REi\right)D^{\nu}H$ & $\Opr_{Gd}^{\alpha\beta}$ & $\left(\BRDi T^{A}\gamma^{\mu}\RDi\right)D^{\nu}G_{\mu\nu}^{A}$ & $\Opr_{H\ell}^{\prime\prime(1)\alpha\beta}$  & $\left(\BLelli\gamma^{\mu}\Lelli\right)\partial_{\mu}\left(H^{\dagger}H\right)$
\tabularnewline
$\Opr_{eHD3}^{\alpha\beta}$ & $\left(\BLelli D_{\mu}D^{\mu}\REi\right)H$ & $\Opr_{Gd}^{\prime\alpha\beta}$ & $\frac{1}{2}\left(\BRDi T^{A}\gamma^{\mu}\rmI\Dlrn\RDi\right)G_{\mu\nu}^{A}$ & $\Op_{H\ell}^{(3)\alpha\beta}$  & $\left(\BLelli\gamma^{\mu}\sigma^{I}\Lelli\right)\left(H^{\dagger}\rmI\Dilr H\right)$
\tabularnewline
$\Opr_{eHD4}^{\alpha\beta}$ & $\left(\BLelli D_{\mu}\REi\right)D^{\mu}H$ & $\Opr_{\widetilde{G}d}^{\prime\alpha\beta}$ & $\frac{1}{2}\left(\BRDi T^{A}\gamma^{\mu}\rmI\Dlrn\RDi\right)\widetilde{G}_{\mu\nu}^{A}$ & $\Opr_{H\ell}^{\prime(3)\alpha\beta}$  & $\left(\BLelli\rmI\Dilrs\Lelli\right)\left(H^{\dagger}\sigma^{I}H\right)$
\tabularnewline
\cline{1-2} \cline{2-2} 
\multicolumn{2}{c|}{$\psi^{2}XH$} & $\Opr_{Bd}^{\alpha\beta}$ & $\left(\BRDi\gamma^{\mu}\RDi\right)\partial^{\nu}B_{\mu\nu}$ & $\Opr_{H\ell}^{\prime\prime(3)\alpha\beta}$ & $\left(\BLelli\sigma^{I}\gamma^{\mu}\Lelli\right)D_{\mu}\left(H^{\dagger}\sigma^{I}H\right)$
\tabularnewline
\cline{1-2} \cline{2-2} 
$\Op_{uG}^{\alpha\beta}$  & $\left(\BLQi\sigma^{\mu\nu}T^{A}\RUi\right)\widetilde{H}G_{\mu\nu}^{A}$  & $\Opr_{Bd}^{\prime\alpha\beta}$ & $\frac{1}{2}\left(\BRDi\gamma^{\mu}\rmI\Dlrn\RDi\right)B_{\mu\nu}$ & $\Op_{He}^{\alpha\beta}$  & $\left(\BREi\gamma^{\mu}\REi\right)\left(H^{\dagger}\rmI\Dlr H\right)$ \tabularnewline
$\Op_{uW}^{\alpha\beta}$  & $\left(\BLQi\sigma^{\mu\nu}\RUi\right)\sigma^{I}\widetilde{H}W_{\mu\nu}^{I}$  & $\Opr_{\widetilde{B}d}^{\prime\alpha\beta}$ & $\frac{1}{2}\left(\BRDi\gamma^{\mu}\rmI\Dlrn\RDi\right)\widetilde{B}_{\mu\nu}$ & $\Opr_{He}^{\prime\alpha\beta}$  & $\left(\BREi\rmI\Dlrs\REi\right)\left(H^{\dagger}H\right)$
\tabularnewline
$\Op_{uB}^{\alpha\beta}$  & $\left(\BLQi\sigma^{\mu\nu}\RUi\right)\widetilde{H}B_{\mu\nu}$  & $\Opr_{W\ell}^{\alpha\beta}$ & $\left(\BLelli\sigma^{I}\gamma^{\mu}\Lelli\right)D^{\nu}W_{\mu\nu}^{I}$ & $\Opr_{He}^{\prime\prime\alpha\beta}$ & $\left(\BREi\gamma^{\mu}\REi\right)\partial_{\mu}\left(H^{\dagger}H\right)$
\tabularnewline
\cline{5-6} \cline{6-6} 
$\Op_{dG}^{\alpha\beta}$  & $\left(\BLQi\sigma^{\mu\nu}T^{A}\RDi\right)HG_{\mu\nu}^{A}$  & $\Opr_{W\ell}^{\prime\alpha\beta}$ & $\frac{1}{2}\left(\BLelli\sigma^{I}\gamma^{\mu}\rmI\Dlrn\Lelli\right)W_{\mu\nu}^{I}$ & \multicolumn{2}{c}{$\psi^{2}H^{3}$}
\tabularnewline
\cline{5-6} \cline{6-6} 
$\Op_{dW}^{\alpha\beta}$  & $\left(\BLQi\sigma^{\mu\nu}\RDi\right)\sigma^{I}HW_{\mu\nu}^{I}$  & $\Opr_{\widetilde{W}\ell}^{\prime\alpha\beta}$ & $\frac{1}{2}\left(\BLelli\sigma^{I}\gamma^{\mu}\rmI\Dlrn\Lelli\right)\widetilde{W}_{\mu\nu}^{I}$ & $\Op_{uH}^{\alpha\beta}$  & $\left(\BLQi\widetilde{H}\RUi\right)\left(H^{\dagger}H\right)$ \tabularnewline
$\Op_{dB}^{\alpha\beta}$  & $\left(\BLQi\sigma^{\mu\nu}\RDi\right)HB_{\mu\nu}$  & $\Opr_{B\ell}^{\alpha\beta}$ & $\left(\BLelli\gamma^{\mu}\Lelli\right)\partial^{\nu}B_{\mu\nu}$ & $\Op_{dH}^{\alpha\beta}$  & $\left(\BLQi H\RDi\right)\left(H^{\dagger}H\right)$
\tabularnewline
$\Op_{eW}^{\alpha\beta}$  & $\left(\BLelli\sigma^{\mu\nu}\REi\right)\sigma^{I}HW_{\mu\nu}^{I}$  & $\Opr_{B\ell}^{\prime\alpha\beta}$ & $\frac{1}{2}\left(\BLelli\gamma^{\mu}\rmI\Dlrn\Lelli\right)B_{\mu\nu}$ & $\Op_{eH}^{\alpha\beta}$ & $\left(\BLelli H\REi\right)\left(H^{\dagger}H\right)$
\tabularnewline
$\Op_{eB}^{\alpha\beta}$  & $\left(\BLelli\sigma^{\mu\nu}\REi\right)HB_{\mu\nu}$  & $\Opr_{\widetilde{B}\ell}^{\prime\alpha\beta}$ & $\frac{1}{2}\left(\BLelli\gamma^{\mu}\rmI\Dlrn\Lelli\right)\widetilde{B}_{\mu\nu}$ &  & 
\tabularnewline
&  & $\Opr_{Be}^{\alpha\beta}$ & $\left(\BREi\gamma^{\mu}\REi\right)\partial^{\nu}B_{\mu\nu}$ &  & 
\tabularnewline
&  & $\Opr_{Be}^{\prime\alpha\beta}$ & $\frac{1}{2}\left(\BREi\gamma^{\mu}\rmI\Dlrn\REi\right)B_{\mu\nu}$ &  & 
\tabularnewline
&  & $\Opr_{\widetilde{B}e}^{\prime\alpha\beta}$ & $\frac{1}{2}\left(\BREi\gamma^{\mu}\rmI\Dlrn\REi\right)\widetilde{B}_{\mu\nu}$ &  & 
\tabularnewline
\hline
\hline 
\end{tabular}
}
\caption{Two-fermion operators in the Green's basis in the SMEFT.}
\label{tab:green-basis-1}
\end{table}

\begin{table}
\centering
\renewcommand\arraystretch{1.4}
\resizebox{0.78\textwidth}{!}{
\begin{tabular}{c|c|c|c|c|c}
\hline\hline
\multicolumn{2}{c|}{$X^{3}$} & \multicolumn{2}{c|}{$X^{2}H^{2}$} & \multicolumn{2}{c}{$H^{2}D^{4}$}\tabularnewline
\hline 
$\Op_{3G}$ & $f^{ABC}G_{\mu}^{A\nu}G_{\nu}^{B\rho}G_{\rho}^{C\mu}$  & $\Op_{HG}$  & $G_{\mu\nu}^{A}G^{A\mu\nu}H^{\dagger}H$  & $\Opr_{DH}$ & $\left(D_{\mu}D^{\mu}H\right)^{\dagger}\left(D_{\nu}D^{\nu}H\right)$
\tabularnewline
\cline{5-6} \cline{6-6} 
$\Op_{\widetilde{3G}}$ & $f^{ABC}\widetilde{G}_{\mu}^{A\nu}G_{\nu}^{B\rho}G_{\rho}^{C\mu}$  & $\Op_{H\widetilde{G}}$  & $\widetilde{G}_{\mu\nu}^{A}G^{A\mu\nu}H^{\dagger}H$  & \multicolumn{2}{c}{$H^{4}D^{2}$}
\tabularnewline
\cline{5-6} \cline{6-6} 
$\Op_{3W}$ & $\epsilon^{IJK}W_{\mu}^{I\nu}W_{\nu}^{J\rho}W_{\rho}^{K\mu}$  & $\Op_{HW}$  & $W_{\mu\nu}^{I}W^{I\mu\nu}H^{\dagger}H$  & $\Op_{H\square}$  & $\left(H^{\dagger}H\right)\square\left(H^{\dagger}H\right)$ 
\tabularnewline
$\Op_{\widetilde{3W}}$ & $\epsilon^{IJK}\widetilde{W}_{\mu}^{I\nu}W_{\nu}^{J\rho}W_{\rho}^{K\mu}$  & $\Op_{H\widetilde{W}}$  & $\widetilde{W}_{\mu\nu}^{I}W^{I\mu\nu}H^{\dagger}H$  & $\Op_{HD}$  & $\left(H^{\dagger}D^{\mu}H\right)^{\ast}\left(H^{\dagger}D_{\mu}H\right)$ 
\tabularnewline
\cline{1-2} \cline{2-2} 
\multicolumn{2}{c|}{$X^{2}D^{2}$} & $\Op_{HB}$  & $B_{\mu\nu}B^{\mu\nu}H^{\dagger}H$  & $\Opr_{HD}^{\prime}$ & $\left(H^{\dagger}H\right)\left(D_{\mu}H\right)^{\dagger}\left(D^{\mu}H\right)$
\tabularnewline
\cline{1-2} \cline{2-2} 
$\Opr_{2G}$ & $-\frac{1}{2}\left(D_{\mu}G^{A\mu\nu}\right)\left(D^{\rho}G_{\rho\nu}^{A}\right)$ & $\Op_{H\widetilde{B}}$  & $\widetilde{B}_{\mu\nu}B^{\mu\nu}H^{\dagger}H$  & $\Opr_{HD}^{\prime\prime}$ & $\left(H^{\dagger}H\right)D_{\mu}\left(H^{\dagger}\rmI\Dlr H\right)$\tabularnewline
\cline{5-6} \cline{6-6} 
$\Opr_{2W}$ & $-\frac{1}{2}\left(D_{\mu}W^{I\mu\nu}\right)\left(D^{\rho}W_{\rho\nu}^{I}\right)$ & $\Op_{HWB}$  & $W_{\mu\nu}^{I}B^{\mu\nu}\left(H^{\dagger}\sigma^{I}H\right)$  & \multicolumn{2}{c}{$H^{6}$}
\tabularnewline
\cline{5-6} \cline{6-6} 
$\Opr_{2B}$ & $-\frac{1}{2}\left(\partial_{\mu}B^{\mu\nu}\right)\left(\partial^{\rho}B_{\rho\nu}\right)$ & $\Op_{H\widetilde{W}B}$  & $\widetilde{W}_{\mu\nu}^{I}B^{\mu\nu}\left(H^{\dagger}\sigma^{I}H\right)$  & $\Op_{H}$  & $\left(H^{\dagger}H\right)^{3}$
\tabularnewline
\cline{3-4} \cline{4-4} 
&  & \multicolumn{2}{c|}{$H^{2}XD^{2}$} &  & 
\tabularnewline
\cline{3-4} \cline{4-4} 
&  & $\Opr_{WDH}$ & $D_{\nu}W^{I\mu\nu}\left(H^{\dagger}\rmI\Dilr H\right)$ &  & 
\tabularnewline
&  & $\Opr_{BDH}$ & $\partial_{\nu}B^{\mu\nu}\left(H^{\dagger}\rmI\Dlr H\right)$ &  & 
\tabularnewline
\hline\hline 
\end{tabular}
}
\caption{Bosonic operators in the Green's basis in the SMEFT.}
\label{tab:green-basis-2}
\end{table}

\begin{table}
\centering
\renewcommand\arraystretch{1.38}
\resizebox{\textwidth}{!}{
\begin{tabular}{c|c|c|c|c|c}
\hline\hline
\multicolumn{2}{c|}{Four-quark} & \multicolumn{2}{c|}{Four-lepton} & \multicolumn{2}{c}{Semileptonic}
\tabularnewline
\hline 
$\Op_{qq}^{(1)\alpha\beta\gamma\lambda}$  & $\left(\BLQi\gamma^{\mu}\LQi\right)\left(\BLQi[\gamma]\gamma_{\mu}\LQi[\lambda]\right)$  & $\Op_{\ell\ell}^{\alpha\beta\gamma\lambda}$  & $\left(\BLelli\gamma^{\mu}\Lelli\right)\left(\BLelli[\gamma]\gamma_{\mu}\Lelli[\lambda]\right)$  & $\Op_{\ell q}^{(1)\alpha\beta\gamma\lambda}$  & $\left(\BLelli\gamma^{\mu}\Lelli\right)\left(\BLQi[\gamma]\gamma_{\mu}\LQi[\lambda]\right)$ 
\tabularnewline
$\Op_{qq}^{(3)\alpha\beta\gamma\lambda}$  & $\left(\BLQi\gamma^{\mu}\sigma^{I}\LQi\right)\left(\BLQi[\gamma]\gamma_{\mu}\sigma^{I}\LQi[\lambda]\right)$  & $\Op_{ee}^{\alpha\beta\gamma\lambda}$  & $\left(\BREi\gamma^{\mu}\REi\right)\left(\BREi[\gamma]\gamma_{\mu}\REi[\lambda]\right)$  & $\Op_{\ell q}^{(3)\alpha\beta\gamma\lambda}$  & $\left(\BLelli\gamma^{\mu}\sigma^{I}\Lelli\right)\left(\BLQi[\gamma]\gamma_{\mu}\sigma^{I}\LQi[\lambda]\right)$ 
\tabularnewline
$\Op_{uu}^{\alpha\beta\gamma\lambda}$  & $\left(\BRUi\gamma^{\mu}\RUi\right)\left(\BRUi[\gamma]\gamma_{\mu}\RUi[\lambda]\right)$  & $\Op_{\ell e}^{\alpha\beta\gamma\lambda}$  & $\left(\BLelli\gamma^{\mu}\Lelli\right)\left(\BREi[\gamma]\gamma_{\mu}\REi[\lambda]\right)$ & $\Op_{eu}^{\alpha\beta\gamma\lambda}$  & $\left(\BREi\gamma^{\mu}\REi\right)\left(\BRUi[\gamma]\gamma_{\mu}\RUi[\lambda]\right)$ 
\tabularnewline
$\Op_{dd}^{\alpha\beta\gamma\lambda}$  & $\left(\BRDi\gamma^{\mu}\RDi\right)\left(\BRDi[\gamma]\gamma_{\mu}\RDi[\lambda]\right)$  &  &  & $\Op_{ed}^{\alpha\beta\gamma\lambda}$  & $\left(\BREi\gamma^{\mu}\REi\right)\left(\BRDi[\gamma]\gamma_{\mu}\RDi[\lambda]\right)$ 
\tabularnewline
$\Op_{ud}^{(1)\alpha\beta\gamma\lambda}$  & $\left(\BRUi\gamma^{\mu}\RUi\right)\left(\BRDi[\gamma]\gamma_{\mu}\RDi[\lambda]\right)$  &  &  & $\Op_{qe}^{\alpha\beta\gamma\lambda}$  & $\left(\BLQi\gamma^{\mu}\LQi\right)\left(\BREi[\gamma]\gamma_{\mu}\REi[\lambda]\right)$ 
\tabularnewline
$\Op_{ud}^{(8)\alpha\beta\gamma\lambda}$  & $\left(\BRUi\gamma^{\mu}T^{A}\RUi\right)\left(\BRDi[\gamma]\gamma_{\mu}T^{A}\RDi[\lambda]\right)$  &  &  & $\Op_{\ell u}^{\alpha\beta\gamma\lambda}$  & $\left(\BLelli\gamma^{\mu}\Lelli\right)\left(\BRUi[\gamma]\gamma_{\mu}\RUi[\lambda]\right)$
\tabularnewline
$\Op_{qu}^{(1)\alpha\beta\gamma\lambda}$  & $\left(\BLQi\gamma^{\mu}\LQi\right)\left(\BRUi[\gamma]\gamma_{\mu}\RUi[\lambda]\right)$  &  &  & $\Op_{\ell d}^{\alpha\beta\gamma\lambda}$ & $\left(\BLelli\gamma^{\mu}\Lelli\right)\left(\BRDi[\gamma]\gamma_{\mu}\RDi[\lambda]\right)$ 
\tabularnewline
$\Op_{qu}^{(8)\alpha\beta\gamma\lambda}$  & $\left(\BLQi\gamma^{\mu}T^{A}\LQi\right)\left(\BRUi[\gamma]\gamma_{\mu}T^{A}\RUi[\lambda]\right)$  &  &  & $\Op_{\ell edq}^{\alpha\beta\gamma\lambda}$  & $\left(\BLelli\REi\right)\left(\BRDi[\gamma]\LQi[\lambda]\right)$ 
\tabularnewline
$\Op_{qd}^{(1)\alpha\beta\gamma\lambda}$  & $\left(\BLQi\gamma^{\mu}\LQi\right)\left(\BRDi[\gamma]\gamma_{\mu}\RDi[\lambda]\right)$  &  &  & $\Op_{\ell equ}^{(1)\alpha\beta\gamma\lambda}$  & $\left(\overline{\ell_{\alpha{\rm L}}^{a}}\REi\right)\epsilon^{ab}\left(\overline{Q_{\gamma{\rm L}}^{b}}\RUi[\lambda]\right)$ 
\tabularnewline
$\Op_{qd}^{(8)\alpha\beta\gamma\lambda}$  & $\left(\BLQi\gamma^{\mu}T^{A}\LQi\right)\left(\BRDi[\gamma]\gamma_{\mu}T^{A}\RDi[\lambda]\right)$  &  &  & $\Op_{\ell equ}^{(3)\alpha\beta\gamma\lambda}$  & $\left(\overline{\ell_{\alpha{\rm L}}^{a}}\sigma^{\mu\nu}\REi\right)\epsilon^{ab}\left(\overline{Q_{\gamma{\rm L}}^{b}}\sigma_{\mu\nu}\RUi[\lambda]\right)$ 
\tabularnewline
$\Op_{quqd}^{(1)\alpha\beta\gamma\lambda}$  & $\left(\overline{Q_{\alpha{\rm L}}^{a}}\RUi\right)\epsilon^{ab}\left(\overline{Q_{\gamma{\rm L}}^{b}}\RDi[\lambda]\right)$  &  &  &  &
\tabularnewline
$\Op_{quqd}^{(8)\alpha\beta\gamma\lambda}$  & $\left(\overline{Q_{\alpha{\rm L}}^{a}}T^{A}\RUi\right)\epsilon^{ab}\left(\overline{Q_{\gamma{\rm L}}^{b}}T^{A}\RDi[\lambda]\right)$  &  &  &  & 
\tabularnewline
\hline\hline
\end{tabular}
}
\caption{Baryon and lepton number conserving four-fermion operators in the Green's basis (and also in the Warsaw basis) in the SMEFT.}
\label{tab:green-basis-3}
\end{table}

\begin{table}
\centering
\renewcommand\arraystretch{1.6}
\resizebox{0.45\textwidth}{!}{
\begin{tabular}{c|c}
\hline\hline 
\multicolumn{2}{c}{$B$- and $L$-number violating}
\tabularnewline
\hline 
$\Op_{duq}^{\alpha\beta\gamma\lambda}$  & $\epsilon^{ABC}\epsilon^{ab}\left[\left(D_{\alpha{\rm R}}^{A}\right)^{T}{\sf C}U_{\beta{\rm R}}^{B}\right]\left[\left(Q_{\gamma{\rm L}}^{{C}a}\right)^{T}{\sf C}\ell_{\lambda{\rm L}}^{b}\right]$
\tabularnewline
$\Op_{qqu}^{\alpha\beta\gamma\lambda}$  & $\epsilon^{ABC}\epsilon^{ab}\left[\left(Q_{\alpha{\rm L}}^{Aa}\right)^{T}{\sf C}Q_{\beta{\rm L}}^{Bb}\right]\left[\left(U_{\gamma{\rm R}}^{C}\right)^{T}{\sf C}E_{\lambda{\rm R}}\right]$
\tabularnewline
$\Op_{qqq}^{\alpha\beta\gamma\lambda}$  & $\epsilon^{ABC}\epsilon^{ad}\varepsilon^{be}\left[\left(Q_{\alpha{\rm L}}^{Aa}\right)^{T}{\sf C}Q_{\beta{\rm L}}^{Bb}\right]\left[\left(Q_{\gamma{\rm L}}^{{C}e}\right)^{T}{\sf C}\ell_{\lambda{\rm L}}^{d}\right]$
\tabularnewline
$\Op_{duu}^{\alpha\beta\gamma\lambda}$  & $\epsilon^{ABC}\left[\left(D_{\alpha{\rm R}}^{A}\right)^{T}{\sf C}U_{\beta{\rm R}}^{B}\right]\left[\left(U_{\gamma{\rm R}}^{C}\right)^{T}{\sf C}E_{\lambda{\rm R}}\right]$
\tabularnewline
\hline\hline
\end{tabular}
}
\caption{Baryon and lepton number violating four-fermion operators in the Green's basis (and also in the Warsaw basis) in the SMEFT.}
\label{tab:green-basis-4}
\end{table}
	
\section{Amplitudes for the relevant diagrams} \label{app:B}
\subsection*{\textbullet $H^\dagger H^\dagger H H$}

The amplitudes for the diagrams in Fig. \ref{fig:hhhh} are 
\begin{eqnarray}\label{eq:hhhh} \label{eq:hhhh-amp}
\amp{a} &=& 4 \left( \del{ad}\del{be} + \del{ae}\del{bd} \right) \lamd \Mds[] \pM{} \left( \TYn \DYdel \Yn \right)^{}_{ii} 
\nonumber
\\
&& \times \lint \frac{\left(k- \momp{2} \right) \cdot \left( k + \momp{4} \right)}{\left( k - \momp{2} \right)^2 \left( k + \momp{4} \right)^2 \left( k^2 - \pM{2} \right) \left[ \left( \momp{2} + \momp{4} \right)^2 - \Mds \right]} \;,
\nonumber
\\
\amp{b} &=& 4 \left( \del{ad}\del{be} + \del{ae}\del{bd} \right) \lamd \Mds[] \pM{} \left( \DYn \Ydel \SYn \right)^{}_{ii} 
\nonumber
\\
&& \times \lint \frac{\left(k- \momp{1} \right) \cdot \left( k + \momp{3} \right)}{\left( k - \momp{1} \right)^2 \left( k + \momp{3} \right)^2 \left( k^2 - \pM{2} \right) \left[ \left( \momp{1} + \momp{3} \right)^2 - \Mds \right]} \;.
\end{eqnarray}
To obtain the hard-momentum parts of these amplitudes, one needs to expend loop integrals in the limit of $p^{}_i \ll k, M^{}_j, M^{}_\Delta$ (for $i=1,2,3,4$ and $j=1,2,3$) before calculating loop integrals. Then with Eq. \eqref{eq:hhhh}, the hard-momentum part of the total amplitude with external lines generated by $\left( H^\dagger H^\dagger HH\right)$ field configuration is given in Eq. \eqref{eq:hhhh-hard}.

\subsection*{\textbullet $\overline{\ell} \ell^{\rm c} \widetilde{H} \widetilde{H}$}

The amplitudes for the six diagrams shown in Fig. \ref{fig:llbarhhbar} are give by
\begin{eqnarray} \label{eq:llbarhhbar-amp}
\amp{a1} &=& - 6 \epsilon^{}_{ab} \epsilon^{}_{ed} \lamd \left( \DYdel \Yn \right)^{}_{\beta i} \left( \DYn \right)^{}_{i\alpha} \frac{ \Mds[] }{\left( \momp{2} + \momp{4} \right)^2 - \pM{2} }
\nonumber
\\
&& \times \overline{u} \left( p^{}_2 \right) \pl \left\{ \lint \frac{\left( \slashed{k} + \momps{2} \right) \left( \momps{2} + \momps{4} \right)}{\left( k + \momp{2} \right)^2 \left[ \left( k - \momp{4} \right)^2 - m^2 \right] \left( k^2 -\Mds[2] \right) } \right\} u\left( \momp{1} \right) \;,
\nonumber
\\
\amp{a2} &=& 6 \epsilon^{}_{ab} \epsilon^{}_{ed} \lamd \left( \DYdel \Yn \right)^{}_{\alpha i} \left( \DYn \right)^{}_{i\beta} \frac{ \Mds[] }{\left( \momp{1} + \momp{3} \right)^2 - \pM{2} }
\nonumber
\\
&& \times \overline{u} \left( p^{}_2 \right) \pl \left\{ \lint \frac{\left( \momps{1} + \momps{3} \right) \left( \slashed{k} - \momps{1} \right) }{\left( k - \momp{1} \right)^2 \left[ \left( k + \momp{3} \right)^2 - m^2 \right] \left( k^2 -\Mds[2] \right) } \right\} u\left( \momp{1} \right) \;,
\nonumber
\\
\amp{b1} &=& - 6 \epsilon^{}_{ad} \epsilon^{}_{be} \lamd \left( \DYdel \Yn \right)^{}_{\alpha i} \left( \DYn \right)^{}_{i\beta} \frac{ \Mds[] }{\left( \momp{1} - \momp{4} \right)^2 - \pM{2} }
\nonumber
\\
&& \times \overline{u} \left( p^{}_2 \right) \pl \left\{ \lint \frac{ \left( \momps{1} - \momps{4} \right) \left( \slashed{k} - \momps{1} \right) }{\left( k - \momp{1} \right)^2 \left[ \left( k - \momp{4} \right)^2 - m^2 \right] \left( k^2 -\Mds[2] \right) } \right\} u\left( \momp{1} \right) \;,
\nonumber
\\
\amp{b2} &=& 6 \epsilon^{}_{ad} \epsilon^{}_{be} \lamd \left( \DYdel \Yn \right)^{}_{\beta i} \left( \DYn \right)^{}_{i\alpha} \frac{ \Mds[] }{\left( \momp{2} - \momp{3} \right)^2 - \pM{2} }
\nonumber
\\
&& \times \overline{u} \left( p^{}_2 \right) \pl \left\{ \lint \frac{ \left( \slashed{k} + \momps{2} \right) \left( \momps{2} - \momps{3} \right) }{\left( k + \momp{2} \right)^2 \left[ \left( k + \momp{3} \right)^2 - m^2 \right] \left( k^2 -\Mds[2] \right) } \right\} u\left( \momp{1} \right) \;,
\nonumber
\\
\amp{c1} &=& 2 \left( \epsilon^{}_{ad} \epsilon^{}_{be} - \epsilon^{}_{ab} \epsilon^{}_{ed} \right) \left( \DYdel \right)^{}_{\alpha\beta} \left( \DYn\Ydel\SYn \right)^{}_{ii} \frac{\pM{}}{\left( \momp{4} - \momp{3} \right)^2 - \Mds[2]} \overline{u} \left( \momp{2} \right) \pl u \left( \momp{1} \right) 
\nonumber
\\
&& \times \lint \frac{\left( k - \momp{3} \right) \cdot \left( k - \momp{4} \right)}{ \left( k - \momp{3} \right)^2 \left( k - \momp{4} \right)^2 \left( k^2 - \pM{2} \right)} \;,
\nonumber
\\
\amp{c2} &=& -4 \left( \epsilon^{}_{ad} \epsilon^{}_{be} - \epsilon^{}_{ab} \epsilon^{}_{ed} \right) \lamd[2] \left( \SYn \right)^{}_{\alpha i} \left( \SYn \right)^{}_{\beta i} \frac{\pM{} \Mds}{ \left( \momp{1} - \momp{2} \right)^2 - \Mds} \overline{u} \left( \momp{2} \right) \pl u \left( \momp{1} \right) 
\nonumber
\\
&& \times \lint \frac{1}{\left[ \left( k - \momp{1} \right)^2 - m^2 \right] \left[ \left( k - \momp{2} \right)^2 - m^2 \right] \left( k^2 - \pM{2} \right) } \;.
\end{eqnarray}
After the expansion-by-regions technique is applied, the hard-momentum parts of the first four amplitudes in Eq. \eqref{eq:llbarhhbar-amp} are vanishing, but not those of the last two.

\subsection*{\textbullet $\overline{\ell} \ell H^\dagger H$}

The amplitudes for the diagrams with their external lines generated by the $\left( \overline{\ell} \ell H^\dagger H \right)$ field configuration in Fig. \ref{fig:llhh} are found to be
\begin{eqnarray}\label{eq:llhh-amp}
\amp{a1} &=& 6 \epsilon^{}_{ab} \epsilon^{}_{ed} \lamd \left( \SYn \right)^{}_{\alpha i} \left( \Ydel \SYn \right)^{}_{\beta i}  \frac{\pM{}\Mds[]}{\left( \momp{2} + \momp{4} \right)^2 - \pM{2}}
\nonumber
\\
&& \times \overline{u} \left( \momp{2} \right) \pr \left\{ \lint \frac{\slashed{k} - \momps{2}}{\left( k - \momp{2} \right)^2 \left[ \left(k + \momp{4} \right)^2 - m^2 \right] \left( k^2 - \Mds \right)} \right\} u \left( p^{}_1 \right) \;,
\nonumber
\\
\amp{a2} &=& 6 \epsilon^{}_{ab}\epsilon^{}_{ed} \lamd \left( \DYdel \Yn \right)^{}_{\alpha i} \left( \Yn \right)^{}_{\beta i} \frac{\pM{}\Mds[]}{\left( \momp{1} + \momp{3} \right)^2 - \pM{2}} 
\nonumber
\\
&& \times \overline{u} \left( \momp{2} \right) \pr \left\{ \lint \frac{\slashed{k} - \momps{1} }{ \left( k - \momp{1} \right)^2 \left[ \left( k + \momp{3} \right)^2 - m^2 \right] \left( k^2 - \Mds \right)}  \right\} u \left( \momp{1} \right) \;,
\nonumber
\\
\amp{b1} &=& 2 \sigma^I_{da} \sigma^I_{eb} \lamd \pM{}\Mds[] \left( \SYn \right)^{}_{\alpha i} \left( \Ydel \SYn \right)^{}_{\beta i} \overline{u} \left( \momp{2} \right) \pr
\nonumber
\\
&& \times \left\{ \lint \frac{\slashed{k} - \momps{2}}{\left( k - \momp{2} \right)^2 \left[ \left(k + \momp{4} \right)^2 - m^2 \right] \left( k^2 - \Mds \right) \left[ \left( k + \momp{4} - \momp{1} \right)^2 - \pM{2} \right]} \right\} u \left( p^{}_1 \right) \;,
\nonumber
\\
\amp{b2} &=& 2 \sigma^I_{da} \sigma^I_{eb}  \lamd \pM{}\Mds[] \left( \DYdel \Yn \right)^{}_{\alpha i} \left( \Yn \right)^{}_{\beta i} \overline{u} \left( \momp{2} \right) \pr 
\nonumber
\\
&& \times \left\{ \lint \frac{\slashed{k} - \momps{1} }{ \left( k - \momp{1} \right)^2 \left[ \left( k + \momp{3} \right)^2 - m^2 \right] \left( k^2 - \Mds \right) \left[ \left( k + \momp{3} - \momp{2} \right)^2 - \pM{2} \right] }  \right\} u \left( \momp{1} \right) \;,
\nonumber
\\
\amp{c1} &=& \sigma^I_{da} \sigma^I_{eb}  \left( \DYdel \Yn \right)^{}_{\alpha i} \left( \DYn \Ydel \right)^{}_{i \beta} \overline{u} \left( \momp{2} \right) \pr 
\nonumber
\\
&& \times \left\{ \lint \frac{ \left(\slashed{k} - \momps{2} \right) \left( \momps{2} - \momps{3} - \slashed{k} \right) \left( \slashed{k} - \momps{1} \right) }{\left( k -\momp{1} \right)^2 \left( k - \momp{2} \right)^2 \left( k^2 - \Mds \right) \left[ \left( k + \momp{3} - \momp{2} \right)^2 - \pM{2} \right] } \right\} u \left( \momp{1} \right) \;,
\nonumber
\\
\amp{c2} &=& 4 \sigma^I_{da} \sigma^I_{eb} \lamd[2] \Mds \left( \SYn \right)^{}_{\alpha i} \left( \Yn \right)^{}_{\beta i} \overline{u} \left( \momp{2} \right) \pr \left\{ \lint \left( \momps{2}-\momps{3} - \slashed{k} \right) \right.
\nonumber
\\
&& \times \left. \frac{1}{ \left[ \left( k + \momp{3} \right)^2 - m^2 \right] \left[ \left( k + \momp{4} \right)^2 - m^2 \right] \left( k^2 - \Mds \right) \left[ \left( k + \momp{3} - \momp{2} \right)^2 - \pM{2} \right] }  \right\} u \left( \momp{1} \right) \;.
\end{eqnarray}

\subsection*{\textbullet $\overline{\ell}\overline{\ell} \ell \ell$}

The amplitudes for the two diagrams in Fig. \ref{fig:llll} are give by
\begin{eqnarray} \label{eq:llll-amp}
\amp{a} &=& 2 \left( \del{ae} \del{bd} + \del{ad} \del{be} \right) \lamd \left( \DYdel \right)^{}_{\alpha\beta} \Yni[\gamma]{i} \Yni[\lambda]{i} \frac{\pM{}}{\Mds[]} \overline{u} \left(0\right) \pr v \left( 0 \right) \overline{v} \left(0\right) \pl u \left( 0 \right) 
\nonumber
\\
&& \times \lint \frac{1}{\left( k^2 -m^2\right)^2 \left( k^2 - \pM{2} \right)} \;,
\nonumber
\\
\amp{b} &=& 2 \left( \del{ae} \del{bd} + \del{ad} \del{be} \right) \lamd \left( \SYn \right)^{}_{\alpha i} \left( \SYn \right)^{}_{\beta i} \left( \Ydel \right)^{}_{\lambda\gamma} \frac{\pM{}}{\Mds[]} \overline{u} \left(0\right) \pr v \left( 0 \right) \overline{v} \left(0\right) \pl u \left( 0 \right) 
\nonumber
\\
&& \times \lint \frac{1}{\left( k^2 -m^2\right)^2 \left( k^2 - \pM{2} \right)} \;,
\end{eqnarray}
where the momenta of all external particles are taken to be zero since the $\left( \overline{\ell} \overline{\ell} \ell \ell \right)$ field configuration already has a mass dimension of six.

\subsection*{ \textbullet $\overline{\ell} E H^\dagger HH$}

The corresponding amplitudes of the diagrams in Fig. \ref{fig:hhhle} are 
\begin{eqnarray}\label{eq:lehhh-amp}
\amp{a} &=& 2 \left( \sigma^I_{ae} \sigma^I_{bd} + \sigma^I_{ad}\sigma^I_{be} \right) \lamd \Mds[] \pM{} \left( \Ydel \SYn \right)^{}_{\alpha i} \left( \DYn \Yl \right)^{}_{i \beta} \overline{u} \left( 0 \right) \pr u \left( 0 \right) 
\nonumber
\\
&& \times \lint \frac{1}{k^2 \left( k^2 - m^2 \right) \left( k^2 - \pM{2} \right) \left( k^2 - \Mds \right)} \;,
\nonumber
\\
\amp{b} &=& - 4 \left( \sigma^I_{ae} \sigma^I_{bd} + \sigma^I_{ad}\sigma^I_{be} \right) \lamd[2] \Mds \left(\Yn \right)^{}_{\alpha i} \left( \DYn \Yl \right)^{}_{i \beta} \overline{u} \left( 0 \right) \pr u \left( 0 \right) 
\nonumber
\\
&& \times \lint \frac{1}{ \left( k^2 - m^2 \right)^2 \left( k^2 - \pM{2} \right) \left( k^2 - \Mds \right)} \;,
\nonumber
\\
\amp{c} &=& - 2 \left( \del{ad} \del{be} + \del{ae} \del{bd} \right) \lamd \Yni{i} \left( \TYn \DYdel \Yl \right)^{}_{i\beta} \frac{\pM{}}{\Mds[]} \overline{u} \left( 0 \right) \pr u \left( 0 \right) 
\nonumber
\\
&& \times \lint \frac{1}{k^2 \left( k^2 - m^2 \right) \left( k^2 - \pM{2} \right) } \;,
\end{eqnarray}
where the momenta of all external particles are also taken to be vanishing.

\subsection*{\textbullet $H^\dagger H^\dagger H^\dagger H H H$}

The amplitudes for the diagrams with their external lines generated by the $\left( H^\dagger H^\dagger H^\dagger H H H \right)$ field configuration in Fig. \ref{fig:hhhhhh} are given by
\begin{eqnarray}\label{eq:hhhhhh-amp}
\amp{a} &=& 4 \epsilon^{}_{be} \sigma^I_{ga} \left( \sigma^I \epsilon \right)^{}_{df} \lamd \left( \DYn \Yl \DYl \Ydel \SYn \right)^{}_{ii} \frac{\pM{}}{\Mds[]} \lint \frac{1}{k^4 \left( k^2 - \pM{2} \right)} + {\rm crosses} = 0 \;,
\nonumber
\\
\amp{b} &=& 4 \epsilon^{}_{gf} \sigma^I_{da} \left( \epsilon \sigma^I \right)^{}_{bc} \lamd \left( \TYn \DYdel \Yl \DYl \Yn \right)^{}_{ii} \frac{\pM{}}{\Mds[]} \lint \frac{1}{k^4 \left( k^2 - \pM{2} \right)} + {\rm crosses} = 0 \;,
\nonumber
\\
\amp{c} &=&  - 4 \del{bg} \left( \del{af} \del{ed} + \del{ad} \del{ef} \right) \lamd \left( \TYn \SYn \right)^{}_{ij} \left( \DYn \Ydel \SYn \right)^{}_{ij} \frac{\pM[j]{}}{\Mds[]} 
\nonumber
\\
&& \times \lint \frac{1}{k^2 \left( k^2 - \pM{2} \right) \left( k^2 - \pM[j]{2} \right)} + {\rm crosses}
\nonumber
\\
&=& - 24 \left[ \del{ad} \left( \del{bg} \del{ef} + \del{bf} \del{eg} \right) + \del{af} \left( \del{bg} \del{ed} + \del{bd} \del{eg} \right) + \del{ag} \left( \del{bf} \del{ed} + \del{bd} \del{ef} \right) \right] 
\nonumber
\\
&& \times \lamd \left( \TYn \SYn \right)^{}_{ij} \left( \DYn \Ydel \SYn \right)^{}_{ij} \frac{\pM[j]{}}{\Mds[]}  \lint \frac{1}{k^2 \left( k^2 - \pM{2} \right) \left( k^2 - \pM[j]{2} \right)} \;,
\nonumber
\\
\amp{d} &=& - 4 \del{ad} \left( \del{bg} \del{ef} + \del{bf} \del{eg} \right) \lamd \left( \TYn \SYn \right)^{}_{ij} \left( \TYn \DYdel \Yn \right)^{}_{ij} \frac{\pM[i]{}}{\Mds[]} 
\nonumber
\\
&& \times \lint \frac{1}{k^2 \left( k^2 - \pM{2} \right) \left( k^2 - \pM[j]{2} \right)} + {\rm crosses}
\nonumber
\\
&=& - 24 \left[ \del{ad} \left( \del{bg} \del{ef} + \del{bf} \del{eg} \right) + \del{af} \left( \del{bg} \del{ed} + \del{bd} \del{eg} \right) + \del{ag} \left( \del{bf} \del{ed} + \del{bd} \del{ef} \right) \right] 
\nonumber
\\
&& \times \lamd \left( \TYn \SYn \right)^{}_{ij} \left( \TYn \DYdel \Yn \right)^{}_{ij} \frac{\pM[i]{}}{\Mds[]} \lint \frac{1}{k^2 \left( k^2 - \pM{2} \right) \left( k^2 - \pM[j]{2} \right)}\;,
\nonumber
\\
\amp{e} &=& - 4 \left( \sigma^I \epsilon \right)^{}_{ab} \left( \epsilon \sigma^J \right)^{}_{fg} \left( \lambda^{}_3 \del{ed} \delta^{IJ} + \rmI \lambda^{}_4 \sigma^K_{ed} \epsilon^{IJK} \right) \lamd \left( \TYn \DYdel \Yn \right)^{}_{ii} \frac{\pM{}}{\Mds[3]} \lint \frac{1}{k^2 \left( k^2 - \pM{2} \right)} 
\nonumber
\\
&& + {\rm crosses}
\nonumber
\\
&=& 12 \left[ \del{ad} \left( \del{bg} \del{ef} + \del{bf} \del{eg} \right) + \del{af} \left( \del{bg} \del{ed} + \del{bd} \del{eg} \right) + \del{ag} \left( \del{bf} \del{ed} + \del{bd} \del{ef} \right) \right] \lamd \left( \lambda^{}_3 - \lambda^{}_4 \right) 
\nonumber
\\
&& \times \left( \TYn \DYdel \Yn \right)^{}_{ii} \frac{\pM{}}{\Mds[3]} \lint \frac{1}{k^2 \left( k^2 - \pM{2} \right)} \;,
\nonumber
\\
\amp{f} &=& - 4 \left( \epsilon \sigma^I \right)^{}_{df} \left( \sigma^J \epsilon \right)^{}_{be} \left( \lambda^{}_3 \del{ag} \delta^{IJ} - \rmI \lambda^{}_4 \sigma^K_{ag} \epsilon^{IJK} \right) \lamd \left( \DYn \Ydel \SYn \right)^{}_{ii} \frac{\pM{}}{\Mds[3]} \lint \frac{1}{k^2 \left( k^2 - \pM{2} \right)}
\nonumber
\\
&& + {\rm crosses}
\nonumber
\\
&=& 12 \left[ \del{ad} \left( \del{bg} \del{ef} + \del{bf} \del{eg} \right) + \del{af} \left( \del{bg} \del{ed} + \del{bd} \del{eg} \right) + \del{ag} \left( \del{bf} \del{ed} + \del{bd} \del{ef} \right) \right] \lamd \left( \lambda^{}_3 - \lambda^{}_4 \right) 
\nonumber
\\
&& \times \left( \DYn \Ydel \SYn \right)^{}_{ii} \frac{\pM{}}{\Mds[3]} \lint \frac{1}{k^2 \left( k^2 - \pM{2} \right)} \;,
\nonumber
\\
\amp{g} &=& 8 \left( \sigma^I \epsilon \right)^{}_{fg} \left( \epsilon \sigma^J \right)^{}_{be} \left( \sigma^J \sigma^I \right)^{}_{da} \lamd[2] \left( \DYn \Ydel \DYdel \Yn \right)^{}_{ii} \frac{1}{\Mds} \lint \frac{1}{k^2 \left( k^2 - \pM{2} \right)} + {\rm crosses}
\nonumber
\\
&=& -48 \left[ \del{ad} \left( \del{bg} \del{ef} + \del{bf} \del{eg} \right) + \del{af} \left( \del{bg} \del{ed} + \del{bd} \del{eg} \right) + \del{ag} \left( \del{bf} \del{ed} + \del{bd} \del{ef} \right) \right]
\nonumber
\\
&& \times \lamd[2] \left( \DYn \Ydel \DYdel \Yn \right)^{}_{ii} \frac{1}{\Mds} \lint \frac{1}{k^2 \left( k^2 - \pM{2} \right)} \;.
\end{eqnarray}
In Eq. \eqref{eq:hhhhhh-amp}, we do not explicitly write out amplitudes of the diagrams with crossed external lines for conciseness, and similarly we take all the external momenta to be zero.

\end{appendices}

\end{document}